\documentclass{emulateapj}

\usepackage{natbib}
\usepackage{amsmath}

\newcommand{\about}{$\sim\!\!$~}
\newcommand{\kms}{\,km\,s$^{-1}$}

\def\lsim{\hbox{\rlap{\raise 0.425ex\hbox{$<$}}\lower 0.65ex\hbox{$\sim$}}}
\def\gsim{\hbox{\rlap{\raise 0.425ex\hbox{$>$}}\lower 0.65ex\hbox{$\sim$}}}

\def\arcdeg{\mbox{$^\circ$}}

\shorttitle{SN~Ia Velocity and Intrinsic Color}
\shortauthors{Foley et~al.}

\begin{document}

 \title{Velocity Evolution and the Intrinsic Color of Type~Ia Supernovae}

\def\cfa{1}
\def\clay{2}

\author{
{Ryan~J.~Foley}\altaffilmark{\cfa,\clay},
{Nathan~E.~Sanders}\altaffilmark{\cfa}, and
{Robert~P.~Kirshner}\altaffilmark{\cfa}
}

\altaffiltext{\cfa}{
Harvard-Smithsonian Center for Astrophysics,
60 Garden Street, 
Cambridge, MA 02138.
}
\altaffiltext{\clay}{
Clay Fellow. Electronic address rfoley@cfa.harvard.edu .
}

\begin{abstract}
To understand how best to use observations of Type Ia supernovae
(SNe~Ia) to obtain precise and accurate distances, we investigate the
relations between spectra of SNe~Ia and their intrinsic colors.  Using
a sample of 1630 optical spectra of 255 SNe, based primarily on data
from the CfA Supernova Program, we examine how the velocity evolution
and line strengths of \ion{Si}{2} $\lambda 6355$ and \ion{Ca}{2} H\&K
are related to the $B-V$ color at peak brightness.  We find that the
maximum-light velocity of \ion{Si}{2} $\lambda 6355$ and \ion{Ca}{2}
H\&K and the maximum-light pseudo-equivalent width of \ion{Si}{2}
$\lambda 6355$ are correlated with intrinsic color, with intrinsic
color having a linear relation with the \ion{Si}{2} $\lambda 6355$
measurements.  \ion{Ca}{2} H\&K does not have a linear relation
with intrinsic color, but lower-velocity SNe tend to be intrinsically
bluer.  Combining the spectroscopic measurements does not improve
intrinsic color inference.  The intrinsic color scatter is larger for
higher-velocity SNe~Ia --- even after removing a linear trend with
velocity --- indicating that lower-velocity SNe~Ia are more ``standard
crayons.''  Employing information derived from SN~Ia spectra has the
potential to improve the measurements of extragalactic distances and
the cosmological properties inferred from them.
\end{abstract}


\defcitealias{Wang09:2pop}{W09}
\defcitealias{Foley11:vel}{FK11}


\section{Introduction}\label{s:intro}

Type Ia supernovae (SNe~Ia) are very good distance indicators after
making an empirical correction based on their light-curve shape and
color \citep{Phillips93, Riess96}.  Using these relations, large
SN~Ia samples have a precision of \about 8\% for distance estimates
\citep[e.g.,][]{Hicken09:lc}.  This level of precision is adequate to 
 determine that the expansion of the Universe is accelerating
\citep{Riess98:Lambda, Perlmutter99}, and constrain the
equation-of-state parameter of dark energy \citep{Wood-Vasey07,
Riess07, Hicken09:de, Kessler09, Amanullah10, Conley11, Sullivan11}.
SNe~Ia are even better distance indicators in the rest-frame
near-infrared (NIR) \citep{Mandel09, Mandel11}; however, low-redshift NIR
samples are small and high-redshift NIR samples do not yet exist.

Decades ago, \citet{Branch87} showed that not all SNe~Ia have the same
ejecta velocity, as probed by the blueshift of spectral features.
\citet{Branch88} noted that a sample of SNe~Ia had a very broad
distribution of \ion{Si}{2} $\lambda 6355$ velocity near maximum
brightness and concluded: ``SNe~Ia are not {\it observationally}
homogeneous.  The only way to maintain that they are {\it physically}
homogeneous would be to postulate that they are identical but
asymmetrical.''  \citet{Benetti05} showed that the velocity gradient
for the \ion{Si}{2} $\lambda 6355$ feature did not correlate with
light-curve shape for spectroscopically normal SNe~Ia.  This
observation provided further proof that SNe~Ia were heterogeneous, but
also showed that some properties of a SN~Ia did {\it not} depend on
the width of its light curve.

Several studies have shown that SN~Ia spectral properties are related
to their photometric properties \citep[e.g.,][]{Nugent95}.  Some
studies examined the correlation between Hubble residuals after
light-curve shape correction and spectroscopic properties to produce a
smaller Hubble scatter \citep{Foley08:uv, Bailey09, Blondin11,
Chotard11, Nordin11:Si4000}.  However, using this simple approach with
only optical spectra provides only minor gains over photometry alone
\citep{Blondin11}.

In a previous study, \citet[hereafter FK11]{Foley11:vel} showed that
the maximum-light intrinsic color of SNe~Ia is highly correlated with
their ejecta velocity as probed by the \ion{Si}{2} $\lambda 6355$
feature, $v_{\rm Si~II}$.  Accounting for this correlation both
substantially improves the precision and reduces potential biases of
SN~Ia distance measurements.  Theoretically, this is understood as
increased line blanketing in the higher-velocity SNe, depressing the
$B$-band flux relative to the $V$-band flux and causing a redder $B-V$
color.  Another theoretical expectation is that higher-velocity SNe~Ia
have more absorption from lower excitation (e.g., \ion{Fe}{2} vs.\
\ion{Fe}{3}) lines, which are distributed such that higher-velocity
SNe~Ia should have more absorption in the near-ultraviolet.
Nonetheless, not all SN~Ia models show the faster-redder relation
found in the data \citep{Blondin11:2D}.

Although not necessarily the cause of the relation between intrinsic
color and $v_{\rm Si~II}$, the observations are naturally explained by
asymmetric explosions \citepalias{Foley11:vel}, as originally
suggested by \citet{Branch88}.  An asymmetric explosion model was
invoked to explain a striking relation between the velocity gradient
of the \ion{Si}{2} $\lambda 6355$ feature, $\dot{v}_{\rm Si~II}$, and
the velocity offset of nebular lines at late times
\citep{Maeda10:asym}.  \citet{Leonard05} found that SNe~Ia with
high-velocity features had stronger \ion{Si}{2} $\lambda 6355$ line
polarization than those without the features.  \citet{Maund10:asym}
expanded upon this work and found a correlation between the amount of
polarization in the \ion{Si}{2} feature and $\dot{v}_{\rm Si~II}$,
suggesting that $\dot{v}_{\rm Si~II}$ is an excellent probe of
asymmetry of the outer layers of the SN ejecta.  Since $v_{\rm Si~II}$
is an excellent proxy for $\dot{v}_{\rm Si~II}$ \citep[hereafter
W09]{Wang09:2pop}, one expects $v_{\rm Si~II}$, $\dot{v}_{\rm Si~II}$,
intrinsic color at maximum brightness, nebular line velocity offsets,
and \ion{Si}{2} polarization should all be highly correlated.
\citet{Maeda11} confirmed one of these assumptions, showing that
nebular line velocity offsets are correlated with maximum-light color.
Of all of these observables, $v_{\rm Si~II}$ is the easiest to obtain
at low redshift (and is always obtained if $\dot{v}_{\rm Si~II}$ or
polarization measurements are made) and is the only observable known
to correlate with intrinsic color that we can realistically obtain for
high-$z$ SNe.

Besides line velocity, which is determined by the minimum of an
absorption feature (corresponding to its maximum absorption), one
might expect that the total amount of absorption, as measured by
either the full-width at half maximum (FWHM) or pseudo-equivalent
width (pEW) of a feature, to strongly correlate with intrinsic color.
Both the FWHM and pEW are alternative kinematic probes, measuring the
span of the absorbing material in velocity space and may have more
physical motivation for a correlation with intrinsic color than
absorption velocity.

Although \ion{Si}{2} $\lambda 6355$ is a clean, isolated, strong
feature in SN~Ia spectra, there is no guarantee that its properties
are the most informative for determining intrinsic color.
Furthermore, \ion{Si}{2} $\lambda 6355$ redshifts out of the optical
window at $z \approx 0.4$.  Therefore, any relation determined
with \ion{Si}{2} $\lambda 6355$ cannot be applied to most high-$z$
SNe~Ia.  Another strong feature in SN~Ia spectra that has a large
velocity distribution and strong velocity evolution with time is
\ion{Ca}{2} H\&K.  \citetalias{Foley11:vel} suggested that this
feature, which is much bluer than \ion{Si}{2} $\lambda 6355$ and can
be observed in the optical window up to $z \approx 1.2$, may be an
alternative way to determine intrinsic color.
\citetalias{Foley11:vel} found that the model spectra of
\citet{Kasen07:asym} show a strong relation between intrinsic
color and $v_{\rm Ca~H\&K}$, similar to what was found with $v_{\rm
Si~II}$.

Current and previous high-$z$ SN~Ia samples typically have a single
near-maximum brightness observer-frame optical spectrum per SN
\citep[e.g.,][]{Zheng08, Foley09:year4, Walker11}, making measurements
of $\dot{v}$ impossible.  Many current and future surveys intend to
use photometric-only samples of SNe for cosmological studies where
only a small subsample of SNe will have spectra.  These samples will
also not have the spectral series necessary to determine velocity
gradients for high-$z$ SNe~Ia.  Moreover, if intrinsic color can be
determined by a single spectrum, telescope time can be used to make
this measurement for more SNe than one could if a time series is
required.

Since \citetalias{Foley11:vel} used the \citetalias{Wang09:2pop}
sample, which did not provide spectra or velocity measurements, but
only a designation of ``Normal'' or ``High-Velocity'' for SNe,
\citetalias{Foley11:vel} was unable to investigate exactly how
intrinsic color depends on ejecta velocity.  Similarly,
\citetalias{Wang09:2pop} did not provide any FWHM or pEW measurements
or any measurements for \ion{Ca}{2} H\&K.  In this paper, we use the
large and homogeneous CfA sample of SN~Ia spectra (Blondin et~al., in
prep.), supplemented by literature data (Section~\ref{s:data}), to
construct a homogeneous sample which reduces potential systematic
differences, determine a relation between $\dot{v}$ and $v (t)$ for
the two strongest features in a maximum-light SN~Ia spectrum,
\ion{Si}{2} $\lambda 6355$ and \ion{Ca}{2} H\&K
(Section~\ref{s:vgrad}), use the same data to determine a relation
between pEW for these features and their gradients
(Section~\ref{s:pew}), use those relations to measure maximum-light
values for a large sample of SNe, allowing a direct comparison of
these SNe (Sections~\ref{s:vgrad} and \ref{s:pew}), and reexamine the
relation between ejecta kinematics --- as determined by both the width
and velocity of the minimum of the absorption for \ion{Si}{2} $\lambda
6355$ and \ion{Ca}{2} H\&K --- and intrinsic color
(Section~\ref{s:col}).  We find that the maximum-light velocity of
\ion{Si}{2} $\lambda 6355$ and \ion{Ca}{2} H\&K and the maximum-light
pEW of \ion{Si}{2} $\lambda 6355$ are correlated with intrinsic color.
The intrinsic color scatter is larger for higher-velocity SNe~Ia ---
even after removing a linear trend with velocity --- indicating that
lower-velocity SNe~Ia are more ``standard crayons.''\footnote{See
\citet{Peek10} for a description of this nomenclature.}  We discuss
these results and conclude in Section~\ref{s:conc}.


\section{Data}\label{s:data}

For our study of how SN~Ia intrinsic color is related to kinematics,
we require a large sample of SNe~Ia with both photometry (to measure
colors) and spectra (to measure kinematic probes).  Below we describe
the data used for our sample.

Additionally, we use values reported by \citetalias{Wang09:2pop}.
Since \citetalias{Wang09:2pop} does not report velocity measurements
(only a designation of ``Normal'' or ``High-Velocity''), it is not
sufficient for use as a spectroscopic sample.  They do report some
relevant photometric measurements (specifically $M_{V}$ at peak,
$\Delta m_{15} (B)$, and $B_{\rm max} - V_{\rm max}$).  However, the
photometric data come from varied sources (\citealt{Hamuy96:lc},
\citealt{Hamuy96:lum}; CfA1; \citealt{Riess99:lc}; all compiled by
\citealt{Reindl05}; CfA2; \citealt{Jha06:lc}; CfA3;
\citealt{Hicken09:lc}; and preliminary LOSS light curves later
published by \citealt{Ganeshalingam10}), and there are potentially
systematic differences between each data set.  A compilation of
several photometric sources can have major differences, even after
placing all photometry on a standard system.  At the very least, S
corrections are typically necessary to achieve agreement at the
0.01~mag level \citep{Stritzinger02}.

Although we use the data presented by \citetalias{Wang09:2pop} for an
initial analysis (see Section~\ref{s:col}), we do not combine data
from \citetalias{Wang09:2pop} with other data for our final analysis.

\subsection{Photometry}\label{ss:phot}

Our photometric data comes from two sources: CfA3 \citep{Hicken09:lc}
and LOSS \citep{Ganeshalingam10}.  Each sample is large, and there is
substantial overlap in SNe between the two groups.  Each group derived
parameters from their light curves.  Specifically, they each measure
$B_{\rm max}$, $V_{\rm max}$, and $\Delta m_{15} (B)$.
\citet{Ganeshalingam10} used a combination of template fitting (when
the data are good and similar to normal SN~Ia light curves) and
polynomial fitting (when template fitting fails).  \citet{Hicken09:lc}
only used polynomial fitting.  The template fitting is more
sophisticated and should produce better measurements.  We therefore
use LOSS-derived parameters for SNe~Ia that have both LOSS and CfA3
data.

For 40, 42, and 29 SNe~Ia in common for CfA3\footnote{We correct the
CfA3 SN~2008bf photometry for an incorrect calibration which affected
the light curves by 0.07~mag in $B$ and 0.13~mag in $V$.  After
correction, the peak magnitudes are $B_{\rm max} = 15.65 \pm 0.04$~mag
and $V_{\rm max} = 15.81 \pm 0.04$~mag (M.\ Hicken, private comm.).}
and LOSS, we find average offsets of $\langle B_{\rm max}^{\rm CfA} -
B_{\rm max}^{\rm LOSS} \rangle = 0.020 \pm 0.010$~mag, $\langle V_{\rm
max}^{\rm CfA} - V_{\rm max}^{\rm LOSS} \rangle = -0.006 \pm
0.008$~mag, and $\langle \Delta m_{15}^{\rm CfA} - \Delta m_{15}^{\rm
LOSS} \rangle = 0.008 \pm 0.017$~mag, respectively.  Furthermore, for
40 SNe in common, we find $\langle (B_{\rm max} - V_{\rm max})^{\rm
CfA} - (B_{\rm max} - V_{\rm max})^{\rm LOSS} \rangle = 0.025 \pm
0.011$~mag.  Considering that \citetalias{Foley11:vel} found an offset
in $B_{\rm max} - V_{\rm max}$ of \about 0.1~mag for low and
high-velocity SNe~Ia, this offset could cause enough scatter to reduce
the significance of potential correlations.

\citet{Ganeshalingam10} examined the light curves in common between
the LOSS and CfA3 sample.  Interpolating LOSS light curves, they
examined differences between individual light-curve points.  They
found average offsets of $0.011 \pm 0.006$ ($0.016 \pm 0.005$) and
$-0.006 \pm 0.004$ ($-0.010 \pm 0.004$) mag for $B$ and $V$ using only
data brighter than magnitude 18 (all data), respectively.  These
measurements are consistent with our finding for the derived
parameters.  Confirmation with the authors indicates that extinction
corrections and K corrections were performed in the same way (M.\
Ganeshalingam \& M.\ Hicken, private comm.), leaving only S
corrections as a possible difference.  Since a SN spectrum changes
with time, the S corrections should also change with time.  Therefore,
making corrections to derived peak magnitudes using the offset we
found above is likely more accurate than that found by
\citet{Ganeshalingam10}, which used the full light curve.

Considering the relatively large offsets, we correct the CfA3
photometry to match the LOSS photometry, which is already the default
photometry when there are data from both samples.

All derived values presented by \citet{Hicken09:lc} and
\citet{Ganeshalingam10} are both K-correct and deredden for Milky Way
reddening.  We present the combined derived light-curve parameters in
Table~\ref{t:obj}.

\begin{center}
\begin{deluxetable*}{lcrrrrccc}
\tablewidth{0pc}
\tablecaption{Object Information\label{t:obj}}
\tablehead{
\colhead{} & \colhead{} & \colhead{$V_{\rm max}$\tablenotemark{a}} & \colhead{$\Delta m_{15}(B)$\tablenotemark{b}} & \colhead{$B_{\rm max} - V_{\rm max}$} & \colhead{$v_{\rm Si~II}^{0}$} & \colhead{pEW$_{0}$(\ion{Si}{2})} & \colhead{$v_{\rm Ca~H\&K}^{0}$\tablenotemark{c}} & \colhead{pEW$_{0}$(\ion{Ca}{2})} \\
\colhead{SN} & \colhead{$z_{\rm CMB}$} & \colhead{(mag)} & \colhead{(mag)} & \colhead{(mag)} & \colhead{($10^{3}$~\kms)} & \colhead{(\AA)} & \colhead{($10^{3}$~\kms)} & \colhead{(\AA)}
}

\startdata

 1981B\tablenotemark{d} & 0.0036 & 11.90 (0.05) & 1.08 (0.03) &  0.01 (0.06) & $-11.87$ (0.22) & 128.8 (6.0) & $-15.31$ (0.57) & 112.7 (6.9) \\
 1986G & 0.0027 & 11.12 (0.06) & 1.75 (0.00) &  0.88 (0.09) & $-10.48$ (0.22) & \nodata     & \nodata         & \nodata     \\
 1989B & 0.0035 & 11.89 (0.05) & 1.16 (0.02) &  0.32 (0.10) & $-10.56$ (0.22) & 121.1 (6.0) & $-11.65$ (0.57) & 106.4 (6.9) \\
 1990N\tablenotemark{d} & 0.0051 & 12.66 (0.04) & 1.11 (0.01) &  0.01 (0.06) & $ -9.38$ (0.22) &  84.7 (6.0) & \nodata         & \nodata     \\
 1990O & 0.0306 & \nodata      & 0.95 (0.02) & \nodata      & $-11.99$ (0.22) & \nodata     & \nodata         & \nodata     \\
 1991M & 0.0076 & \nodata      & \nodata     & \nodata      & $-12.81$ (0.22) & \nodata     & \nodata         & \nodata     \\
 1991T & 0.0069 & 11.46 (0.02) & 0.93 (0.01) &  0.12 (0.03) & $ -9.62$ (0.22) & \nodata     & \nodata         & \nodata     \\
1991bg & 0.0045 & 13.86 (0.04) & 1.94 (0.00) &  0.71 (0.06) & $-10.04$ (0.22) & \nodata     & \nodata         & \nodata     \\
 1992A & 0.0058 & 12.49 (0.01) & 1.48 (0.00) &  0.02 (0.02) & $-14.04$ (0.22) & 109.6 (6.0) & $-17.22$ (0.57) &  98.9 (6.9) \\
1992ag & 0.0259 & 16.17 (0.06) & 1.11 (0.09) &  0.01 (0.07) & $-11.74$ (0.30) & \nodata     & \nodata         & \nodata     \\
1992al & 0.0141 & 14.56 (0.04) & 1.10 (0.08) & $-0.13$ (0.04) & $-11.26$ (0.30) & \nodata     & \nodata         & \nodata     \\
1993ac & 0.0493 & \nodata      & 1.19 (0.10) & \nodata      & $-13.21$ (0.22) & 144.6 (6.0) & $-17.93$ (0.57) & 103.0 (6.9) \\
 1994D & 0.0041 & 11.83 (0.02) & 1.42 (0.00) & $-0.04$ (0.04) & $-11.24$ (0.24) & 100.6 (6.0) & $-11.53$ (0.58) &  92.4 (6.9) \\
 1994M & 0.0243 & \nodata      & 1.35 (0.03) & \nodata      & $-12.36$ (0.27) & 118.6 (6.1) & \nodata         & \nodata     \\
 1994S & 0.0160 & 14.79 (0.06) & 1.02 (0.00) & $-0.02$ (0.08) & $-10.72$ (0.24) &  85.0 (6.0) & \nodata         & 117.8 (6.9) \\
 1994T & 0.0356 & 17.15 (0.04) & 1.51 (0.08) &  0.18 (0.05) & $-13.43$ (0.30) & \nodata     & \nodata         & \nodata     \\
1994ae\tablenotemark{d} & 0.0062 & 13.00 (0.03) & 1.09 (0.01) & $-0.04$ (0.07) & $-10.98$ (0.23) &  81.6 (6.0) & $-12.67$ (0.59) & 110.0 (6.8) \\
 1995D & 0.0078 & 13.26 (0.05) & 0.98 (0.01) & $-0.10$ (0.07) & $-10.33$ (0.23) & \nodata     & \nodata         & \nodata     \\
 1995E & 0.0121 & 15.98 (0.05) & 1.18 (0.01) &  0.70 (0.07) & $-11.16$ (0.24) &  98.9 (6.0) & $-15.79$ (1.08) & 103.0 (7.8) \\
1995al\tablenotemark{d} & 0.0067 & 13.20 (0.05) & 0.93 (0.03) &  0.13 (0.07) & $-13.01$ (0.23) & \nodata     & \nodata         & \nodata     \\
 1996C & 0.0275 & \nodata      & 0.97 (0.01) & \nodata      & $-10.69$ (0.22) & \nodata     & \nodata         & \nodata     \\
 1996X & 0.0078 & 13.03 (0.04) & 1.24 (0.00) & $-0.05$ (0.05) & $-11.17$ (0.22) &  90.1 (6.0) & $-11.70$ (0.57) &  80.7 (6.9) \\
 1996Z & 0.0085 & \nodata      & 1.22 (0.10) & \nodata      & $-12.14$ (0.23) & 110.6 (6.0) & $-16.53$ (0.57) &  61.6 (6.9) \\
1996ai & 0.0037 & \nodata      & 0.96 (0.02) & \nodata      & $-10.82$ (0.22) & \nodata     & \nodata         & \nodata     \\
1996bk & 0.0070 & \nodata      & 1.78 (0.01) & \nodata      & $-12.16$ (0.22) & \nodata     & \nodata         & \nodata     \\
1996bl & 0.0348 & \nodata      & 1.10 (0.08) & \nodata      & $-12.34$ (0.24) &  89.0 (6.1) & \nodata         & \nodata     \\
1996bo & 0.0163 & 15.51 (0.04) & 1.20 (0.01) &  0.31 (0.06) & $-12.25$ (0.24) & 133.5 (6.1) & \nodata         & \nodata     \\
 1997E & 0.0133 & 15.08 (0.07) & 1.46 (0.02) &  0.03 (0.09) & $-12.01$ (0.23) & 120.6 (6.0) & \nodata         & \nodata     \\
 1997Y & 0.0165 & \nodata      & 1.17 (0.02) & \nodata      & $-11.03$ (0.22) & 101.1 (6.0) & \nodata         & \nodata     \\
1997bp & 0.0094 & 13.74 (0.03) & 0.96 (0.03) &  0.10 (0.04) & $-15.62$ (0.22) & \nodata     & \nodata         & \nodata     \\
1997bq & 0.0095 & \nodata      & 1.01 (0.05) & \nodata      & $-14.24$ (0.23) & 156.0 (6.0) & \nodata         & \nodata     \\
1997br & 0.0081 & 13.43 (0.08) & 1.10 (0.03) &  0.16 (0.13) & $-11.60$ (0.22) & \nodata     & $-15.53$ (0.57) &  51.8 (6.9) \\
1997cn & 0.0176 & \nodata      & 1.90 (0.05) & \nodata      & $ -9.61$ (0.22) & \nodata     & \nodata         & \nodata     \\
1997do & 0.0105 & \nodata      & 0.94 (0.04) & \nodata      & $-13.38$ (0.25) & \nodata     & \nodata         & \nodata     \\
1997dt & 0.0061 & \nodata      & 1.04 (0.15) & \nodata      & $-11.33$ (0.22) &  82.6 (6.0) & \nodata         & \nodata

\enddata

\tablecomments{As described in the text, all photometry has been places
on the LOSS system.  Uncertainties are in parantheses.}

\tablenotetext{a}{Corrected for Milky Way extinction using the
reddening maps of \citet{Schlegel98}, as performed by
\citet{Hicken09:lc} and \citet{Ganeshalingam10}.}

\tablenotetext{b}{When color inforamation is available, $\Delta
m_{15}(B)$ has been corrected for host-galaxy extinction using the
method similar to that of \citet{Phillips99}; $\Delta m_{15} (B) =
\Delta m_{15, {\rm ~obs}} (B) + 0.1 \times ((B_{\rm max} - V_{\rm
max}) + 0.081)$ (see Section~\ref{ss:w09}.  The values for $B_{\rm
max} - V_{\rm max}$ include a correction for Milky Way extinction, so
that correction is inherent in this process.}

\tablenotetext{c}{Values given for the ``red'' component.}

\tablenotetext{d}{Cepheid object with distance modulus from
\citet{Riess11}.  The redshifts listed for these SNe yield appropriate
distance moduli using $H_{0} = 70.5$ km~s$^{-1}$~Mpc$^{-1}$.}

\end{deluxetable*}
\end{center}

\subsection{Spectroscopy}

\subsubsection{CfA Spectral Sample}

Over the last two decades, the CfA Supernova Program has observed
hundreds of SNe~Ia, mostly with the FAST spectrograph
\citep{Fabricant98} mounted on the 1.5~m telescope at the F.~L.\
Whipple Observatory.  The data have been reduced in a consistent
manner (\citealt{Matheson08}; \citealt{Blondin11}; Blondin et~al., in
prep.), producing well-calibrated spectra.

For SNe~Ia in the sample with a measured time of maximum brightness
from light curves, $v_{\rm Si~II}$ and $v_{\rm Ca~H\&K}$ have been
measured (Blondin et~al., in prep.).  Briefly, this is achieved by
first generating a smoothed spectrum using an inverse-variance
Gaussian filter \citep{Blondin06}, and the wavelength of maximum
absorption in the smoothed spectrum is used to determine the velocity
(see \citealt{Blondin11} for details).  When possible the spectra are
used to measure the FWHM and pEW for the features as well (see
\citealt{Blondin11} for details).  The measurements for each spectrum
are listed in Table~\ref{t:spec}.  The reference for each spectrum is
listed in Table~\ref{t:spec}, but measurements in all cases were
obtained by Blondin et~al., in prep.

\begin{center}
\begin{deluxetable*}{lrrrrrrr}
\tablewidth{0pc}
\tablecaption{Spectral Properties\label{t:spec}}
\tablehead{
\colhead{} & \colhead{Phase} & \colhead{$v_{\rm Si~II}$} & \colhead{pEW(\ion{Si}{2})} & \colhead{$v_{\rm Ca~II,~blue}$} & \colhead{$v_{\rm Ca~II,~red}$} & \colhead{pEW(\ion{Ca}{2})} \\
\colhead{SN} & \colhead{(days)} & \colhead{($10^{3}$~\kms)} & \colhead{(\AA)} & \colhead{($10^{3}$~\kms)} & \colhead{($10^{3}$~\kms)} & \colhead{(\AA)} & \colhead{Ref.}
}
 
\startdata
 
 1981B & $  -1.89$ & $-12.05$ (0.06) & 128.6 (0.1) & \nodata         & $-15.60$ (0.06) & 115.8 (0.1) & 1 \\
 1981B & $  15.00$ & $-10.83$ (0.06) & 218.2 (0.1) & \nodata         & $-13.27$ (0.06) &  79.6 (0.1) & 1 \\
 1981B & $  17.98$ & $-10.39$ (0.06) & 169.4 (0.1) & \nodata         & $-13.10$ (0.06) &  74.4 (0.1) & 1 \\
 1981B & $  21.96$ & $-10.00$ (0.06) & 252.0 (0.1) & \nodata         & \nodata         &  78.7 (0.1) & 1 \\
 1986G & $  -4.82$ & $-11.38$ (0.05) & 140.1 (0.1) & \nodata         & $-16.26$ (0.05) & 100.1 (0.1) & 1 \\
 1986G & $  -3.82$ & $-11.55$ (0.05) & 131.0 (0.1) & \nodata         & $-15.89$ (0.05) & 110.1 (0.1) & 1 \\
 1986G & $  -3.12$ & $-11.30$ (0.05) & 134.0 (0.1) & \nodata         & \nodata         & \nodata     & 2 \\
 1986G & $  -2.82$ & $-11.15$ (0.05) & 131.2 (0.1) & \nodata         & $-16.68$ (0.05) & 109.2 (0.1) & 1 \\
 1986G & $  -2.32$ & $-11.20$ (0.05) & 126.4 (0.1) & \nodata         & $-17.17$ (0.05) &  92.3 (0.1) & 2 \\
 1986G & $  -1.33$ & $-10.91$ (0.05) & 124.8 (0.1) & \nodata         & $-17.32$ (0.05) &  90.6 (0.1) & 2 \\
 1986G & $  -0.83$ & $-10.86$ (0.05) & 133.2 (0.1) & \nodata         & \nodata         & \nodata     & 1 \\
 1986G & $  -0.33$ & $-10.13$ (0.05) & 122.7 (0.1) & \nodata         & $-16.45$ (0.05) &  97.1 (0.1) & 2 \\
 1986G & $   0.17$ & $-10.47$ (0.05) & 137.2 (0.1) & \nodata         & $-15.26$ (0.05) & 115.0 (0.1) & 1 \\
 1986G & $   0.67$ & $-10.51$ (0.05) & 117.7 (0.1) & \nodata         & $-14.77$ (0.05) & 103.7 (0.1) & 2 \\
 1986G & $   1.17$ & $-10.47$ (0.05) & 131.2 (0.1) & \nodata         & $-15.10$ (0.05) & 111.7 (0.1) & 1 \\
 1989B & $  -7.21$ & $-11.40$ (0.05) & 121.7 (0.1) & $-17.88$ (0.05) & $-12.48$ (0.05) & 126.1 (0.1) & 1 \\
 1989B & $  -1.37$ & $-10.64$ (0.05) & 120.6 (0.1) & $-16.85$ (0.05) & $-11.60$ (0.05) & 108.5 (0.1) & 1 \\
 1989B & $   2.60$ & $-10.65$ (0.05) & 119.4 (0.1) & $-15.09$ (0.05) & $-11.93$ (0.50) & 129.1 (0.1) & 1 \\
 1989B & $   4.56$ & $-10.06$ (0.05) & 131.6 (0.1) & $-15.45$ (0.05) & $-11.92$ (0.50) &  99.7 (0.1) & 1 \\
 1989B & $   6.57$ & $-10.17$ (0.05) & 130.8 (0.1) & $-15.48$ (0.05) & $-12.11$ (0.50) &  99.1 (0.1) & 1 \\
 1989B & $   8.57$ & $-10.04$ (0.05) & 136.1 (0.1) & $-14.49$ (0.05) & $-11.72$ (0.50) &  95.4 (0.1) & 1 \\
 1989B & $  10.56$ & $-10.04$ (0.05) & 144.5 (0.1) & $-15.29$ (0.05) & $-12.31$ (0.05) &  84.7 (0.1) & 1 \\
 1989B & $  11.54$ & $ -9.56$ (0.05) & 156.1 (0.1) & $-15.09$ (0.05) & $-11.92$ (0.05) &  79.4 (0.1) & 1 \\
 1989B & $  12.52$ & $ -9.56$ (0.05) & 130.6 (0.1) & $-15.09$ (0.05) & $-11.14$ (0.05) &  68.8 (0.1) & 1 \\
 1989B & $  13.53$ & $ -9.43$ (0.05) & 156.0 (0.1) & \nodata         & \nodata         & \nodata     & 1 \\
 1989B & $  15.50$ & $ -9.19$ (0.05) & 163.1 (0.1) & $-14.89$ (0.05) & $-11.92$ (0.05) &  73.1 (0.1) & 1 \\
 1989B & $  16.47$ & $ -9.43$ (0.05) & 163.5 (0.1) & $-14.49$ (0.05) & $-11.72$ (0.05) &  69.7 (0.1) & 1 \\
 1989B & $  17.47$ & $ -9.31$ (0.05) & 177.4 (0.1) & $-14.90$ (0.05) & $-12.12$ (0.05) &  66.9 (0.1) & 1 

\enddata
 
\tablerefs{(1) \citet{Wells94}; (2) \citet{Cristiani92}; (3)
\citet{Leibundgut91}; (4) \citet{Mazzali93}; (5) \citet{Blondin07};
(6) \citet{Gomez98}; (7) \citet{Jeffery92}; (8) \citet{Mazzali95}; (9)
\citet{Turatto96}; (10) \citet{Leibundgut93}; (11) \citet{Kirshner93};
(12) \citet{Patat96}; (13) SUSPECT, no reference; (14) Blondin et~al.,
in prep.; (15) \citet{Salvo01}; (16) \citet{Li99}; (17)
\citet{Turatto98}; (18) \citet{Matheson08}; (19) \citet{Branch03};
(20) \citet{Jha99:98bu}; (21) \citet{Blondin11}; (22)
\citet{Garavini04}; (23) \citet{Garavini05}; (24) Unknown; (25)
\citet{Garnavich04}; (26) \citet{Hamuy02}; (27) \citet{Valentini03};
(28) \citet{Krisciunas11}; (29) \citet{Sauer08}; (30) \citet{Wang03};
(31) \citet{Benetti04}; (32) \citet{Pignata08:02dj}; (33)
\citet{Kotak05}; (34) \citet{Elias-Rosa06}; (35)
\citet{Stanishev07:03du}; (36) \citet{Anupama05}; (37)
\citet{Leloudas09}; (38) \citet{Krisciunas07}; (39)
\citet{Altavilla07}; (40) \citet{Pastorello07:04eo}; (41)
\citet{Taubenberger08}; (42) \citet{Wang09:05cf}; (43)
\citet{Garavini07:05cf}; (44) \citet{Quimby06:05cg}; (45)
\citet{Kasliwal08}}

\end{deluxetable*}
\end{center}

\subsubsection{Literature Spectral Sample}

Many SN~Ia spectra obtained by various groups using various
telescopes, instruments, and reduction methods have been published
over time.  Although the fidelity of the spectra may not be high or
uniform as for spectra from the CfA sample, these data are useful for
expanding our sample.  When publicly available, usually through the
SUSPECT database\footnote{http://suspect.nhn.ou.edu/\textasciitilde
suspect/}, and if the spectra cover the appropriate wavelength ranges,
$v_{\rm Si~II}$, pEW(\ion{Si}{2}), $v_{\rm Ca~H\&K}$, and
pEW(\ion{Ca}{2}) have been measured for these SNe.  Since the flux
calibration is not critical for the velocity measurements, we expect
the largest errors to be from poor wavelength calibration; however,
the accuracy necessary for our measurements is easily obtained even if
mistakes are made while observing or in the reduction process.
Comparison of multiple spectra from the same SN both within a dataset
and with the CfA sample show that the measurements are consistent
within the measurement uncertainty (typically 100~\kms).  The pEW
measurements are more affected by poor flux calibration.  We have
examined the individual spectra and have rejected spectra where we
believe that the flux calibration is poor.  The measurements for each
spectrum are listed in Table~\ref{t:spec}

Between the CfA sample and Literature sample, there are 1630 $v_{\rm
Si~II}$, 1630 pEW(\ion{Si}{2}), 1192 $v_{\rm Ca~H\&K}$, and 1234
pEW(\ion{Ca}{2}) measurements for 255, 255, 192, and 211 SNe~Ia,
respectively.  For the majority of this study, we will restrict our
sample to SNe with $1 \le \Delta m_{15}(B) \le 1.5$~mag to match the
final sample criteria of \citetalias{Foley11:vel}.  With this
restriction, there are 939, 939, 685, and 708 measurements for 141,
141, 109, and 119 SNe, respectively.

\subsubsection{Telegraphic Spectral Sample}\label{sss:circ}

Despite the CfA sample being quite large, there are several SNe in the
CfA3, LOSS, and \citetalias{Wang09:2pop} photometry samples that do
not have a CfA or publicly available spectrum (or one with $-6 \le t
\le 10$~d).  However, every SN in these samples was spectroscopically
classified in the IAU Circulars or CBETs.  In these reports, the date
of observation is always reported, and for most SNe~Ia classified near
maximum brightness, $v_{\rm Si~II}$ is reported.  Although these
measurements may not be of the same quality --- and are certainly not
systematically measured in the same way, as we have done with the CfA
and public spectra --- the uncertainty in the velocity measurement
should be relatively small.  These additional measurements from the
Circulars and Telegrams can significantly expand our final sample.

For most SNe, we are able to determine the phase of each spectrum
using the time of maximum reported by
\citet{Hicken09:de}\footnote{These values are reported at
\url{http://www.cfa.harvard.edu/supernova/CfA3/sn.tBmax.mlcs17.txt}
.} and \citet{Ganeshalingam10}.  We present the data for these
SNe/spectra in Table~\ref{t:lit}.

\begin{deluxetable*}{lrlllcrr}
\tablewidth{0pc}
\tablecaption{Data for Telegraphic Supernovae\label{t:lit}}
\tablehead{
\colhead{} & \colhead{$v_{\rm Si~II}$} & \colhead{} & \colhead{Spec.} & \colhead{MJD} & \colhead{$t_{B_{\rm max}}$} & \colhead{Phase} & \colhead{$v_{\rm Si~II}^{0}$\tablenotemark{c}} \\
\colhead{SN} & \colhead{(\kms)} & \colhead{UT Date} & \colhead{Ref.\tablenotemark{a}} & \colhead{of $t_{B_{\rm max}}$} & \colhead{Ref.\tablenotemark{b}} & \colhead{(days)} & \colhead{(\kms)}
}

\startdata

 1992ag & $-11.8$ & 19920703.042 &  1 & 48807.15 & 1 &  $-0.6$ & $-11.7$ \\
 1992al & $-11.7$ & 19920729.3   &  2 & 48838.24 & 1 &  $-5.7$ & $-11.3$ \\
 1999cp & $-14.1$ & 19990619.2   &  3 & 51363.27 & 1 & $-15.2$ & \nodata \\
 1999dk & $-16.2$ & 19990815.06  &  4 & 51413.33 & 1 &  $-9.4$ & \nodata \\
 2000ca & $-11.0$ & 20000429.3   &  5 & 51666.63 & 1 &  $-3.1$ & $-10.8$ \\
 2000dr & $-12.5$ & 20001008.1   &  6 & 51833.98 & 2 &  $-8.3$ & \nodata \\
 2001ba & $-10.9$ & 20010430.17  &  7 & 52034.36 & 1 &  $-4.7$ & $-10.6$ \\
 2001cj & $-10.5$ & 20010601     &  8 & 52066.04 & 2 &  $-4.4$ & $-10.4$ \\
 2001dl & $-10.0$ & 20010808     &  9 & 52131.47 & 2 &  $-1.9$ &  $-9.9$  \\
 2003gt & $-11.0$ & 20030805     & 10 & 52862.17 & 2 &  $-5.6$ & $-10.7$ \\
 2005de & $-13.3$ & 20050804.34  & 11 & 53598.85 & 2 & $-12.2$ & \nodata \\
 2005ms & $-13.3$ & 20051231     & 12 & 53743.45 & 1 &  $-7.7$ & \nodata \\
 2006dm & $-12.0$ & 20060705.1   & 13 & 53928.96 & 2 &  $-7.3$ & \nodata \\
 2006ef & $-12.0$ & 20060824     & 14 & 53967.95 & 1 &    1.5  & $-12.2$ \\
 2006ej & $-13.0$ & 20060824     & 14 & 53973.09 & 1 &  $-4.7$ & $-12.5$ \\
 2006en & $-10.2$ & 20060828.9   & 15 & 53971.66 & 1 &    3.7  & $-10.4$ \\
 2006os & $-12.7$ & 20061122.15  & 16 & 54061.68 & 1 &  $-0.2$ & $-12.7$ \\
 2006qo & $-11.1$ & 20061201     & 17 & 54081.96 & 1 & $-11.1$ & \nodata \\
 2006td & $-10.9$ & 20061228.77  & 18 & 54098.92 & 1 &  $-1.4$ & $-10.8$ \\
 2007O  & $-10.0$ & 20070122     & 19 & 54123.80 & 1 &  $-1.3$ & $-10.0$ \\
 2007sr & $-12.1$ & 20071220.22  & 20 & 54448.34 & 1 &    6.1  & $-12.9$ \\
 2008dt & $-14.0$ & 20080701.21  & 21 & 54646.73 & 2 &    1.7  & $-14.3$ \\
 2008ec & $-12.6$ & 20080716.08  & 22 & 54674.28 & 2 & $-10.6$ & \nodata

\enddata

\tablenotetext{a}{Spectroscopy references: (1) \citet{Maza92}; (2)
\citet{McNaught92}; (3) \citet{Jha99}; (4) \citet{Salvo99}; (5)
\citet{Aldering00}; (6) \citet{Suntzeff00}; (7) \citet{Nugent01}; (8)
\citet{Wang01}; (9) \citet{Patat01}; (10) \citet{Filippenko03:03gt};
(11) \citet{Foley05:05de}; (12) \citet{Leonard05:05ms};
(13)\citet{Selj06}; (14) \citet{Foley06:06ef}; (15)
\citet{Elias-Rosa06:06en}; (16) \citet{Quimby06:06os}; (17)
\citet{Silverman06}; (18) \citet{Gurugubelli06}; (19)
\citet{Silverman07}; (20) \citet{Umbriaco07}; (21)
\citet{Blondin08:08dt}; (22) \citet{Harutyunyan08}.}

\tablenotetext{b}{Photometry references: (1) \citet{Hicken09:lc}; (2)
\citet{Ganeshalingam10}.}

\tablenotetext{c}{Uncertainty set to 300~\kms, which is slightly larger than
our expected uncertainty for Literature SNe, and larger than the
median uncertainty for the F11 sample.}

\end{deluxetable*}

\subsubsection{F11 Sample}

The intersection of the CfA, Literature, and Telegraphic spectral
samples with the LOSS and CfA3 photometric samples is the sample used
for the analysis here.  Similar to \citetalias{Foley11:vel}, we
exclude SN~2006bt because of its peculiar nature \citep{Foley10:06bt}.
We also exclude from our final analysis SNe with $E(B-V)_{\rm MW} >
0.5$~mag, corresponding to SNe~1999ek ($E(B-V)_{\rm MW} = 0.561$~mag)
and 2006lf ($E(B-V)_{\rm MW} = 0.954$~mag).  Both
\citetalias{Wang09:2pop} and \citetalias{Foley11:vel} excluded
SN~2006lf from their analyses.  We call the final sample of SNe with
photometric and spectroscopic measurements the ``F11'' sample.


\section{Velocity Gradients \& Maximum-Light Ejecta Velocity}\label{s:vgrad}

The temporal coverage of the F11 spectra (in particular, the CfA data)
provide an ideal sample for measuring velocity evolution with time.
Previous studies have examined relatively small samples of nearby
SNe~Ia and their spectra to determine velocity gradients for those SNe
\citep[e.g.,][]{Benetti05, Hachinger06}.  Additionally, ejecta
velocities have been measured for some high-$z$ SNe~Ia
\citep{Hook05, Blondin06, Bronder08, Nordin11:sdss, Konishi11}.
\citetalias{Wang09:2pop} separated their sample into two subsamples
based on ejecta velocity near maximum brightness.  Since SNe~Ia with
high velocities near maximum brightness tend to have high velocity
gradients, this separation is similar to separating based on velocity
gradient.

Although most studies of velocity gradients have focused on
\ion{Si}{2} $\lambda 6355$, other features have been examined to some
degree.  While certain features appear to have little variation from
SN to SN (e.g., \ion{S}{2}; \citealt{Pignata08:02dj}), \ion{Ca}{2}
H\&K velocity near maximum brightness varies significantly amongst
SNe~Ia.  Additionally, \ion{Ca}{2} H\&K is a strong, blue feature with
a large velocity gradient.  Using the model spectra and light curves
of \citet{Kasen07:asym}, \citetalias{Foley11:vel} found that the
maximum-brightness velocity of \ion{Ca}{2} H\&K strongly correlated
with maximum-light color.  \ion{Ca}{2} H\&K is an excellent candidate
for an alternative way to determine intrinsic color.

\subsection{\ion{Si}{2} $\lambda 6355$}\label{ss:sivel}

\begin{figure*}
\begin{center}
\epsscale{1.18}
\rotatebox{90}{
\plotone{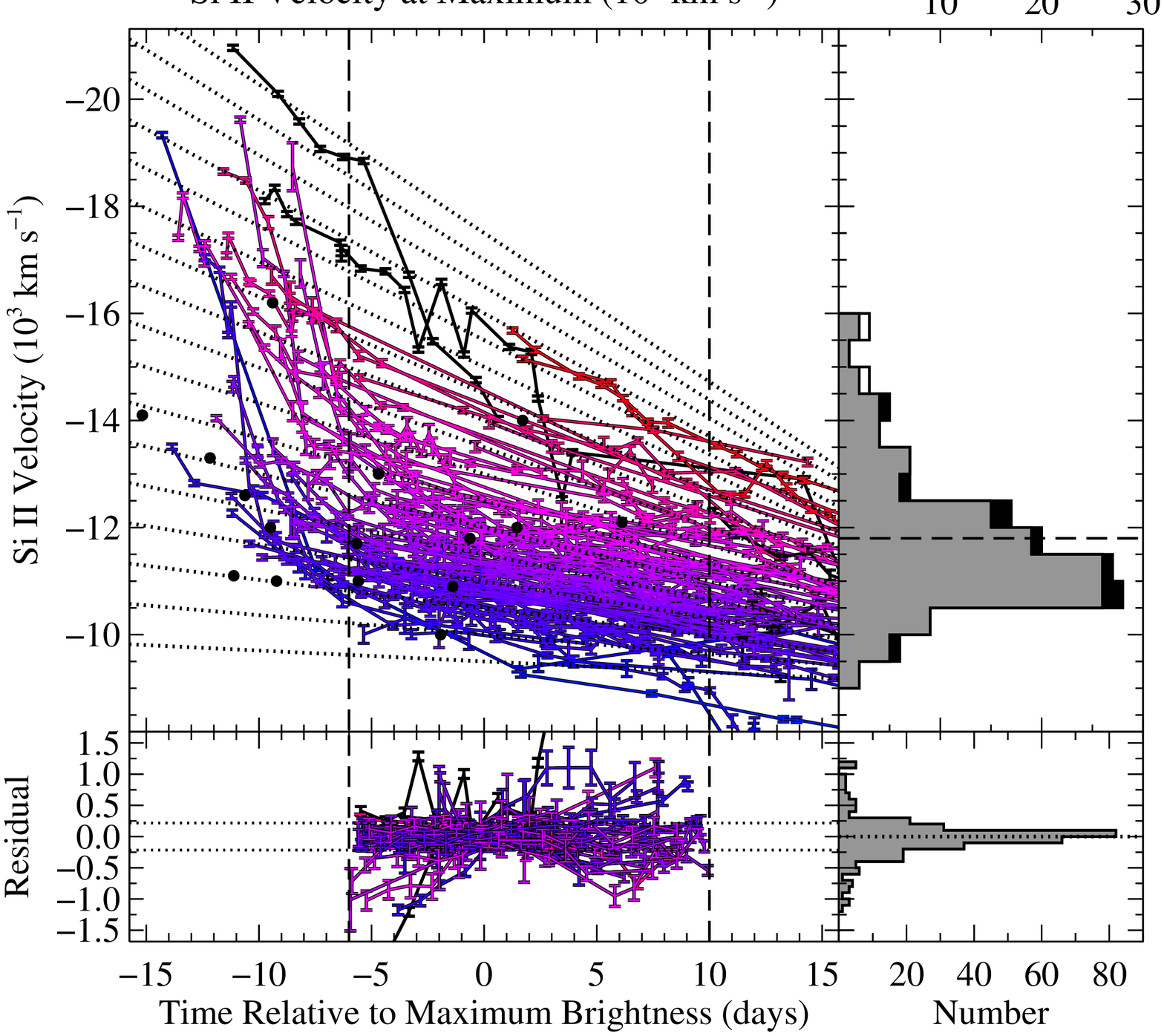}}
\caption{Top left: Temporal evolution of \ion{Si}{2} $\lambda 6355$
velocity, $v_{\rm Si~II}$, for the F11 sample of SN~Ia with $1 \le
\Delta m_{15} (B) \le 1.5$~mag.  Measurements for the same SN are
connected by solid lines.  The vertical dashed lines mark the time
interval over which we fit the velocity evolution.  The dotted lines
represent a subset of the family of functions that describe the
velocity evolution (Equation~\ref{e:grad}).  Each SN is color-coded
based on the estimate of $v_{\rm Si~II}^{0}$ (using
Equation~\ref{e:v0}) from the measurement closest to $t = 0$~d.  The
color bar at the top shows how the colors correspond to $v_{\rm
Si~II}^{0}$.  SNe with no measurement in $-6 \le t \le 10$~d are
plotted in black.  SNe~2003W and 2004dt are plotted in black and are
excluded from further analysis.  Individual measurements from the
Telegraphic sample are presented as black points (see
Section~\ref{sss:circ}).  Top right: Histograms of $v_{\rm
Si~II}^{0}$.  The grey histogram is for SNe where we have measured
velocities from spectra.  The empty histogram represents SNe~2003W and
2004dt.  The black histogram represents the SNe from the Telegraphic
sample.  The horizontal dashed line represents the velocity which
roughly separated the \citetalias{Wang09:2pop} sample into ``Normal''
and ``High-Velocity'' SNe.  Bottom-left: Residuals of $v_{\rm
Si~II}^{0}$ estimates relative to measured velocity within one day of
$t = 0$~d.  The dotted horizontal lines represent the standard
deviation of 220~km~s$^{-1}$.  Bottom-right: Histogram of the
residuals.  The dotted horizontal line is at 0.}\label{f:vel_grad}
\end{center}
\end{figure*}

In Figure~\ref{f:vel_grad}, the \ion{Si}{2} velocity is shown as a
function of phase relative to $B$ maximum for the 141 SNe~Ia with a
velocity measurement, a time of maximum, and $1 \le \Delta m_{15} (B)
\le 1.5$~mag.  The SNe with higher velocity near maximum
brightness also have larger velocity gradients.  However, at very
early times ($t \lesssim -10$~d), some low velocity gradient SNe have
similar velocities to high velocity gradient SNe.  Similarly, at later
times ($t \gtrsim 15$~d), it is difficult to distinguish SNe with
different velocity gradients with a single velocity measurement.  Near
maximum brightness, the velocity evolution is close to linear in time.

There are two SNe, SNe~2003W and SN~2004dt, that are outliers in both
early-time velocity and velocity gradient.  SN~2004dt has very high
early-time $v_{\rm Si~II}$.  The velocity gradient at early times is
consistent with other SNe, but also changes dramatically.  It declines
by 2750\kms\ in only 1.4~d, which was previously noted
\citep{Altavilla07}.  Additionally, the \ion{Si}{2} $\lambda 6355$
feature is wide, flat-bottomed, and potentially contains two
components \citep{Wang06, Altavilla07}.  SN~2004dt is also a
significant outlier in the relation between velocity gradient and
nebular velocity shifts \citep{Maeda10:asym}.  Finally, SN~2004dt has
extremely high line polarization \citep{Leonard05, Wang06}.  In
summary, it appears that SN~2004dt is spectroscopically distinct from
the majority of SNe~Ia and potentially different from even very high
velocity SNe.

SN~2003W has both the highest $v_{\rm Si~II}$ measurements of any SN
in the F11 sample, exceeding $-20,000$~\kms, and a dramatic velocity
decline near maximum light, declining by over 4000\kms\ over 5~d
(where the typical value is 100--200~\kms~d$^{-1}$).  Its \ion{Si}{2}
profile appears to be flat-bottomed and probably is the result of two
strong components.  Unfortunately, there never appears to be two {\it
distinct} components in any of our spectra.  Perhaps additional
spectroscopy at slightly later times would reveal this.
\citet{Wang08:specpol} reports that there is an unpublished epoch of
spectropolarimetry for SN~2003W.  These data may be extremely
informative to understanding the nature of this SN, particularly given
the extraordinary polarization of SN~2004dt.  Although SN~2003W is
clearly a high-velocity and high velocity gradient SN, we have
excluded it from further analysis because of its peculiar velocity
evolution.  It is unclear if SN~2003W is a truly peculiar SN or if the
velocity evolution of other SNe~Ia with extremely high $v_{\rm Si~II}$
at early times is similar to that of SN~2003W.

\begin{figure}
\begin{center}
\epsscale{1.}
\rotatebox{90}{
\plotone{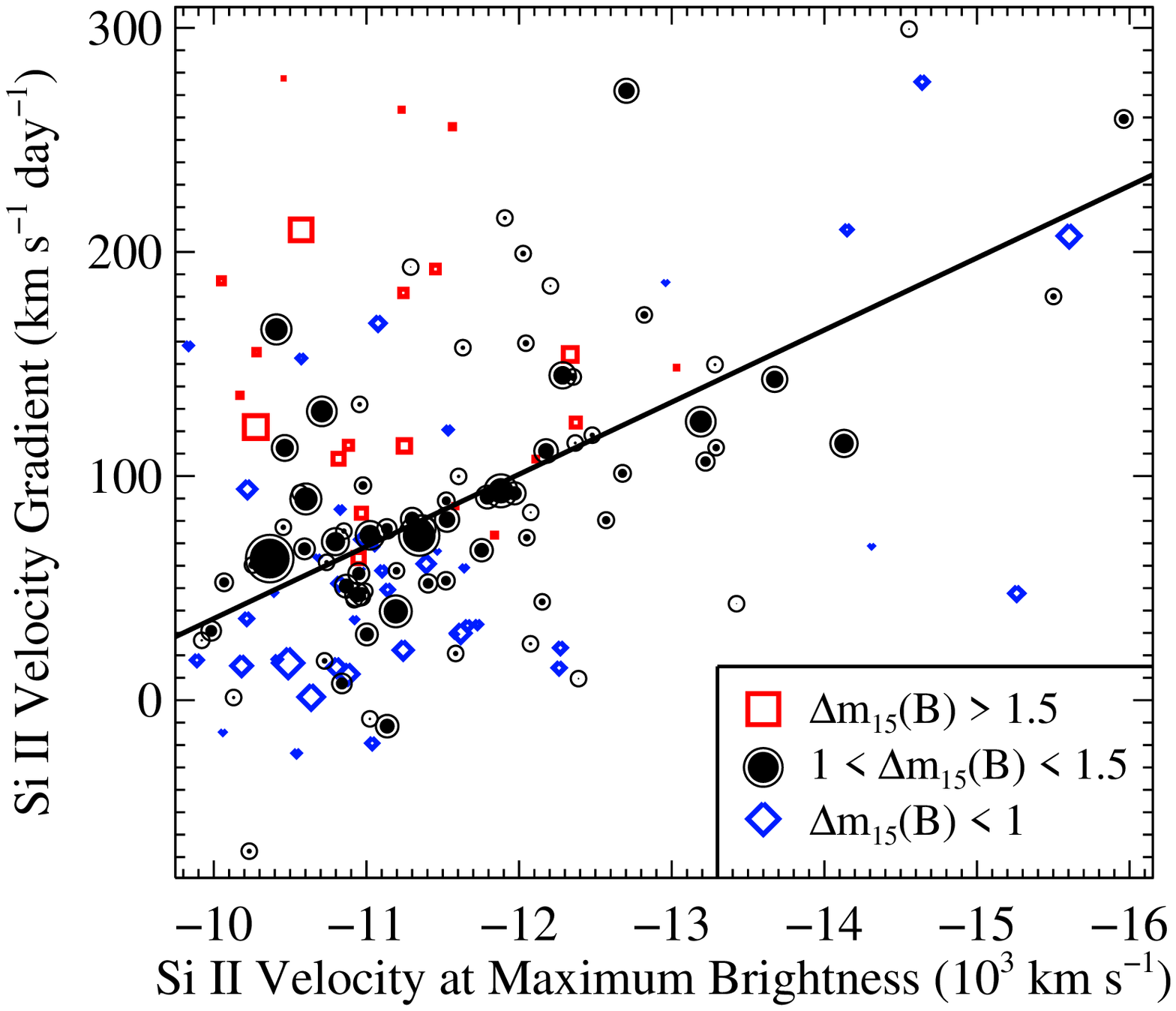}}
\caption{Best-fit values to the slope and offset for a line
(Equation~\ref{e:vt}) describing the $v_{\rm Si~II}$ evolution of the
F11 sample.  SNe with $\Delta m_{15} < 1$, $1 \le \Delta m_{15} \le
1.5$, and $\Delta m_{15} > 1.5$~mag, are represented by blue diamonds,
black circles, and red squares, respectively.  The size of the points
is inversely proportional to the size of the uncertainty.  Each point
with $1 \le \Delta m_{15} \le 1.5$~mag has a circle surrounding it,
with a minimum size to help the reader see the smallest points
(corresponding to SNe with the largest uncertainty).  The solid line
is the best-fit line for the $1 \le \Delta m_{15} \le 1.5$~mag
subsample.}\label{f:family}
\end{center}
\end{figure}

A line was fit to the velocity data for each SN (except for SNe~2003W
and 2004dt) with $\ge$3 measurements in the window $-6 \le t
\le 10$~d regardless of $\Delta m_{15} (B)$.  We fit for
\begin{equation}\label{e:vt}
  v(t) = \dot{v} t + v^{0},
\end{equation}
where $v(t)$ is the velocity, $t$ is the time relative to $B$ maximum,
and $\dot{v}$ and $v^{0}$ are a slope and offset for the line,
corresponding to the velocity gradient and velocity at time of $B$
maximum ($t = 0$), respectively.  The resulting fit values for
$\dot{v}$ and $v^{0}$ are shown in Figure~\ref{f:family}.

SNe~Ia from the different light-curve shape subsamples have different
relations between \ion{Si}{2} $\lambda 6355$ velocity and velocity
gradient.  For SNe~Ia with $1 \le \Delta m_{15} (B) \le 1.5$~mag,
there is a clear relation between the two values such that SNe
with a higher velocity at maximum also have a higher velocity
gradient.  This is quantitative confirmation of the previously noted
correlation.  The fast decliners tend to have low velocities, but high
velocity gradients relative to the other SNe, confirming previous
results \citep[e.g.,][]{Benetti05}.  The slow decliners have a large
scatter, but also have a similar trend to the moderate decliners
(although they may have a slightly lower velocity gradient for a given
velocity).

For the moderate decliners, a line was then fit to the values of
$\dot{v}$ and $v^{0}$,
\begin{equation}
  \dot{v} = \alpha v^{0} + \beta.
\end{equation}
Substituting for $\dot{v}$, we can construct a family of functions
that describe the velocity evolution near maximum brightness and only
depend on $v^{0}$,
\begin{align}
  v(t) &= \dot{v}t + v^{0} \\ \notag
       &= (\alpha v^{0} + \beta) t + v^{0} \\ \notag
       &= (1 + \alpha t) v^{0} + \beta t
\end{align}
Using the fit values, the family of functions correspond to
\begin{equation}\label{e:grad}
  v_{\rm Si~II} \left ( v_{\rm Si~II}^{0}, t \right ) = v_{\rm Si~II}^{0} (1 - 0.0322 t) - 0.285 t,
\end{equation}
where $v_{\rm Si~II}^{0}$ is the velocity at maximum, $t$ is measured
in days relative to maximum brightness, and $v_{\rm Si~II}$ and
$v_{\rm Si~II}^{0}$ are measured in $10^{3}$~km~s$^{-1}$.  This family
of functions describes the velocity evolution of SNe~Ia near maximum
brightness, and a number of the functions are plotted in
Figure~\ref{f:vel_grad}.

A test of how well this family of functions performs over our chosen
time interval ($-6 \le t \le 10$~d) is to examine the residuals of the
functions relative to the data.  Every velocity measurement within our
time interval provides an estimate of $v_{\rm Si~II}^{0}$ for that SN.
For all SNe with a velocity measurement within a day of maximum
brightness, we measure the ``residual'': the difference between the
predicted $v_{\rm Si~II}^{0}$ from a single measurement and the
measured velocity within 1 day of maximum brightness.  We call this
the residual, but it is not exactly the residual since the
measurements were not obtained exactly at $t = 0$~d.  The residuals
are shown in the bottom-left panel of Figure~\ref{f:vel_grad}.

The residuals are small and centered on zero.  The average for all
residuals is $-10$~km~s$^{-1}$ and the standard deviation is
220~km~s$^{-1}$.  There are few residuals that are more than
1000~km~s$^{-1}$ from zero.  The residuals are slightly larger at the
extremes of our time interval than near maximum brightness, but the
difference is still relatively small.  The typical measurement
uncertainty is $\lesssim$100~\kms.  Velocity uncertainty from
host-galaxy rotation is $\lesssim$300~\kms.  Therefore, the scatter in
the derived relation is not the dominant source of uncertainty for
$v_{\rm Si~II}^{0}$.  Combining these sources of uncertainty, the
total uncertainty is $\lesssim$400~\kms.

Using Equation~\ref{e:grad} and a single velocity measurement in the
time interval $-6 \le t \le 10$~d, an estimate of $v_{\rm Si~II}^{0}$
can be made for a SN.  This allows for a direct comparison of all SNe
with such measurements.  Solving Equation~\ref{e:grad} for $v_{\rm
Si~II}^{0}$, one finds
\begin{equation}\label{e:v0}
  v_{\rm Si~II}^{0} = (v_{\rm Si~II} + 0.285 t) / (1 - 0.0322 t).
\end{equation}

\begin{figure}
\begin{center}
\epsscale{1.}
\rotatebox{90}{
\plotone{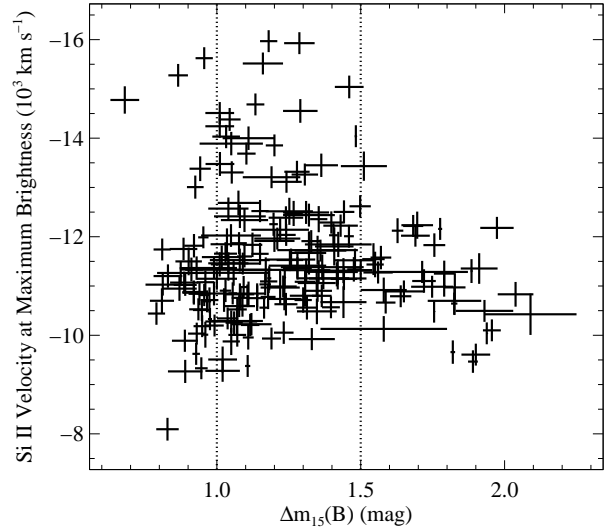}}
\caption{\ion{Si}{2} velocity at maximum brightness ($v_{\rm
Si~II}^{0}$) as a function of $\Delta m_{15}(B)$ for the F11 sample.
The vertical dotted lines mark the region of our full analysis ($1 \le
\Delta m_{15} (B) \le 1.5$~mag).}\label{f:vel_dm15}
\end{center}
\end{figure}

Table~\ref{t:obj} lists $v_{\rm Si~II}^{0}$ for each SN in the F11
sample with a measurement in $-6 \le t \le 10$~d.  The value given is
for the measurement closest to $t = 0$~d.  In Figure~\ref{f:vel_dm15},
we plot $v_{\rm Si~II}^{0}$ as a function of $\Delta m_{15} (B)$ for
the full F11 sample.  We note that these values are derived from
Equation~\ref{e:v0}, and the velocity estimated for SNe with $\Delta
m_{15} (B) < 1$~mag or $\Delta m_{15} (B) > 1.5$~mag may have
systematic offsets (but typically $< 1000$~\kms) from their true
value.  Regardless, there appears to be a dearth of
high-velocity/fast-declining SNe~Ia, but there are not many SNe with
large $\Delta m_{15} (B)$ in the F11 sample.  Slow-declining SNe~Ia
can have high-velocity ejecta, but there also appear to be a
significant population of low-velocity/slow-declining SNe~Ia.

The lack of high-velocity/fast-declining SNe~Ia is not unexpected.
SNe~Ia with $\Delta m_{15} (B) \gtrsim 1.5$~mag are spectroscopically
distinct from slower-declining SNe~Ia, being spectroscopically similar
to SN~1991bg \citep{Filippenko92:91bg, Leibundgut93}.  No
high-velocity SN~1991bg-like SN~Ia has yet been discovered.  It is
worth noting that despite having a low velocity near maximum
brightness, SNe~Ia with $\Delta m_{15} (B) > 1.5$~mag tend to have
high-velocity gradients (\citealt{Benetti05}; Figure~\ref{f:family}).
Therefore, Equation~\ref{e:grad} may not properly describe the
velocity evolution of these SNe.

Since fast-declining SNe~Ia have ejecta velocities similar to or lower
than slower-declining SNe~Ia at maximum brightness, one cannot
attribute their fast post-maximum decline to a rapidly expanding
photosphere \citep[e.g.,][]{Hoflich96, Pinto01}.  Rather, the
post-maximum decline is likely set by the amount of Fe-group elements
in the ejecta (and thus linked to $^{56}$Ni production in the
explosion and the SN peak luminosity)
\citep{Kasen07:wlr}.

\citet{Benetti05} also showed a lack of high-velocity gradient SNe~Ia
with $\Delta m_{15} (B) < 1$~mag.  Many of these SNe are
spectroscopically similar to SN~1991T \citep{Filippenko92:91T}, which
have a low velocity near maximum brightness \citepalias{Wang09:2pop}.
Although the relation between velocity at maximum brightness and
velocity gradient for SN~1991T-like SNe appears to be consistent with
that found in Equation~\ref{e:grad}, there are larger residuals for
these SNe.

Within the limited range of $1 \le \Delta m_{15} (B) \le
1.5$~mag, there is no clear trend between $v_{\rm Si~II}^{0}$ and
$\Delta m_{15} (B)$; however, the SNe with the slowest ejecta are
found among the slowest decliners.

\begin{figure}
\begin{center}
\epsscale{1.}
\rotatebox{90}{
\plotone{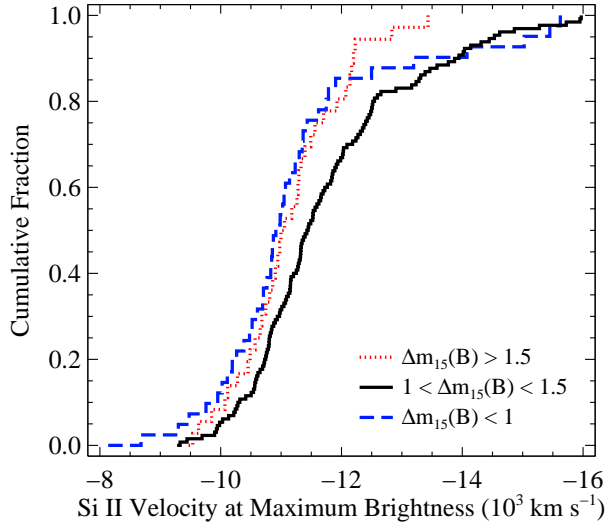}}
\caption{\ion{Si}{2} velocity at maximum brightness ($v_{\rm
Si~II}^{0}$) CDFs for the F11 sample with $\Delta m_{15} (B) < 1$, $1
\le \Delta m_{15} (B) \le 1.5$, and $\Delta m_{15} (B) > 1.5$~mag
(dashed blue, solid black, and dotted red lines,
respectively).}\label{f:cdf}
\end{center}
\end{figure}

In Figure~\ref{f:cdf}, we show the $v_{\rm Si~II}^{0}$ cumulative
distribution functions (CDFs) for SNe~Ia in the F11 sample with
$\Delta m_{15} (B) < 1$, $1 \le \Delta m_{15} (B) \le 1.5$, and
$\Delta m_{15} (B) > 1.5$~mag.  Using these broad groups, it is easy
to see the trends described above: there is a lack of
high-velocity/fast-declining SNe~Ia, there are a significant number of
high-velocity/slow-declining SNe~Ia, and there are a significant
number of low-velocity/slow-declining SNe~Ia.  Additionally, the
figure shows that SNe~Ia with $1 \le \Delta m_{15} (B) \le 1.5$~mag
tend to have a higher $v_{\rm Si~II}^{0}$ than SNe~Ia in the other
groups.  Performing Kolmogorov-Smirnov (K-S) tests, we find that it is
unlikely that SNe~Ia with $1 \le \Delta m_{15} (B) \le 1.5$~mag and
those with $\Delta m_{15} (B) < 1$~mag and those with $\Delta m_{15}
(B) > 1.5$~mag come from the same parent population ($p = 0.0065$ and
0.04, respectively).

Because of the spectroscopic difference between SNe~Ia with $1 \le
\Delta m_{15} (B) \le 1.5$~mag and many of the SNe~Ia outside that
range, we find it prudent to restrict our analysis to this range.
Additionally, the differences in velocity populations for the three
groups above and the different relation between velocity and
velocity gradient for SN~1991bg-like SNe~Ia (see
Figure~\ref{f:family}) give further reason to focus on the limited
range in light-curve shape.  Finally, \citetalias{Wang09:2pop} found
that ``Normal'' and ``High-Velocity'' SNe~Ia have similar light-curve
shape and host-galaxy morphology distributions over this range.

\begin{figure}
\begin{center}
\epsscale{1.}
\rotatebox{90}{
\plotone{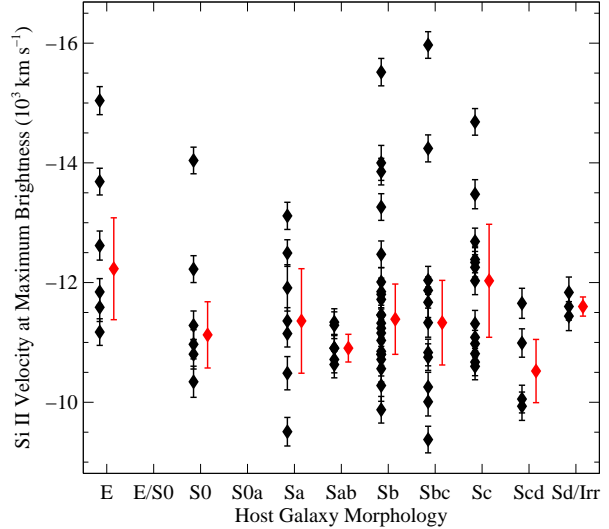}}
\caption{\ion{Si}{2} velocity at maximum brightness ($v_{\rm
Si~II}^{0}$) as a function of host-galaxy morphology for the F11
sample with $1 \le \Delta m_{15} (B) \le 1.5$~mag.  The black circles
represent individual SNe, while the red diamonds represent the median
value for each morphology bin with the error bars representing the
median absolute deviation for each morphology bin.}\label{f:vel_gal}
\end{center}
\end{figure}

\citet{Hicken09:lc} compiled morphology classifications for most of
the host galaxies of SNe~Ia in the F11 sample.  Figure~\ref{f:vel_gal}
presents $v_{\rm Si~II}^{0}$ as a function of host-galaxy morphology
for the F11 sample with $1 \le \Delta m_{15} (B) \le 1.5$~mag.  As
found by \citet{Wang09:2pop}, but now using full velocity information,
we do not see any significant differences in the \ion{Si}{2}
velocities observed in host galaxies of different morphologies.
Although there are few very low velocity ($v_{\rm Si~II}^{0} \gtrsim
-10,000$~\kms) SNe~Ia in elliptical galaxies, and no high-velocity
($v_{\rm Si~II}^{0} \lesssim -11,800$~\kms) SNe~Ia in galaxies later
than Sc, the number of SNe in these morphology bins is small.  This
may come from the combination of a slight trend between $v_{\rm
Si~II}^{0}$ and $\Delta m_{15} (B)$ (Figure~\ref{f:vel_dm15}) as well
as faster declining SNe~Ia having a higher frequency in early-type
galaxies
\citep[e.g.,][]{Howell01}.

\begin{figure}
\begin{center}
\epsscale{1.}
\rotatebox{90}{
\plotone{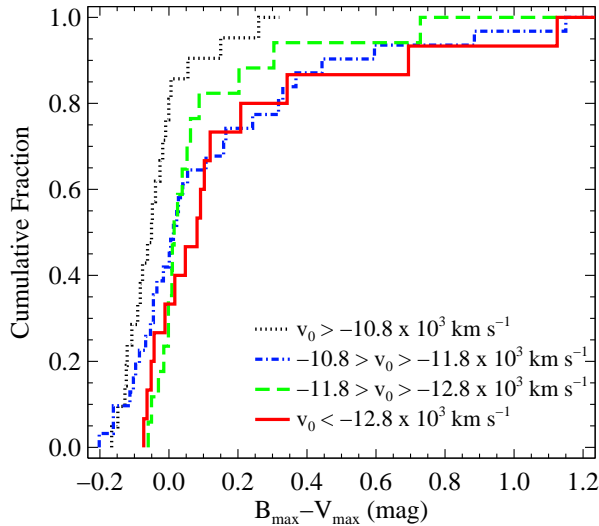}}
\caption{$B_{\rm max} - V_{\rm max}$ CDFs for the F11 sample with
$v_{\rm Si~II}^{0} > -10.8$, $-10.8 > v_{\rm Si~II}^{0} > -11.8$,
$-11.8 > v_{\rm Si~II}^{0} > -12.8$, and $v_{\rm Si~II}^{0} < -12.8
\times 10^{3}$~\kms\ (dotted black, dot-dashed blue, dashed green, and
solid red lines, respectively).  \citetalias{Wang09:2pop} set
$-11,800$~\kms\ as the dividing line between the two velocity groups
outlined in that work.}\label{f:col_cdf}
\end{center}
\end{figure}

We also examine the $B_{\rm max} - V_{\rm max}$ CDF for separate
$v_{\rm Si~II}^{0}$ bins.  Figure~\ref{f:col_cdf} shows these CDFs for
four $v_{\rm Si~II}^{0}$ bins ($v_{\rm Si~II}^{0} < -10,800$, $-10,800
> v_{\rm Si~II}^{0} > -11,800$, $-11,800 > v_{\rm Si~II}^{0} >
-12,800$, and $v_{\rm Si~II}^{0} > -12,800$~\kms, corresponding to 21,
31, 17, and 15 SNe, respectively).  There is a general trend for the
median color to become redder with velocity.  While the highest and
lowest velocity bins have similar shapes and appear to simply be
offset in color space, the middle two velocity bins have different
shapes.  The second velocity bin ($-10,800 > v_{\rm Si~II}^{0} >
-11,800$~\kms) has a broad distribution of colors, containing the
bluest SNe and a significant population of highly reddened SNe.
Meanwhile, the third velocity bin ($-11,800 > v_{\rm Si~II}^{0} >
-12,800$~\kms) has a relatively narrow range of colors and not as many
very red SNe as the second velocity bin.  Performing K-S tests, we
find that the parent population of the lowest-velocity bin is
significantly different from the other velocity bins (all with
probability $<$0.04).  Despite the apparently different shapes, K-S
tests do not find a statistically significant difference between any
of the other velocity bins.

\subsection{\ion{Ca}{2} H\&K}

In this section, we repeat much of the same analysis presented in
Section~\ref{ss:sivel}, but for the \ion{Ca}{2} H\&K feature.  As
noted in Section~\ref{s:intro}, \ion{Ca}{2} H\&K is a very strong
feature that is relatively blue and thus visible in optical spectra at
$z > 1$ and shows significant velocity evolution and diversity.  These
qualities make \ion{Ca}{2} H\&K appealing for use as an intrinsic
color indicator at high $z$.

\begin{figure}
\begin{center}
\epsscale{1.6}
\rotatebox{90}{
\plotone{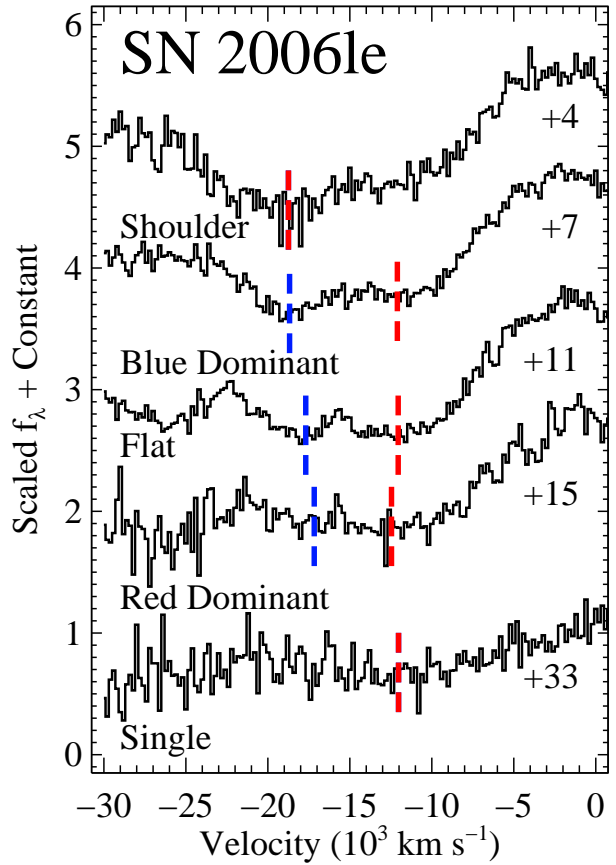}}
\caption{Optical spectra of SN~2006le showing the \ion{Ca}{2} H\&K
feature in velocity space relative to 3945~\AA.  Their phase is marked
to the right.  The blue and red absorption velocities are shown as
dashed blue and red vertical lines, respectively.  The spectra show
the range of possible spectral shapes (from top to bottom): a red
shoulder that the automatic procedure does not detect, causing the
blue feature to be classified as red, a double-absorption profile
where the blue component is stronger, a double-absorption profile
where the two components are of comparable strength, a
double-absorption profile where the red component is stronger, and a
single component where no blue absorption is
found.}\label{f:ca_profile}
\end{center}
\end{figure}

However, unlike \ion{Si}{2} $\lambda 6355$, \ion{Ca}{2} H\&K is in a
complicated spectral region.  Whereas \ion{Si}{2} $\lambda 6355$ is a
single, isolated feature, other lines are blended with \ion{Ca}{2}
H\&K, and the pseudo-continuum near the feature is set by a
combination of line blanketing and nearby strong features.
Furthermore, \ion{Ca}{2} H\&K often shows a complicated absorption
profile, with a flat bottom, shoulders, and/or multiple distinct
absorption minima, complicating any analysis
(Figure~\ref{f:ca_profile}).

\begin{figure}
\begin{center}
\epsscale{1.2}
\rotatebox{90}{
\plotone{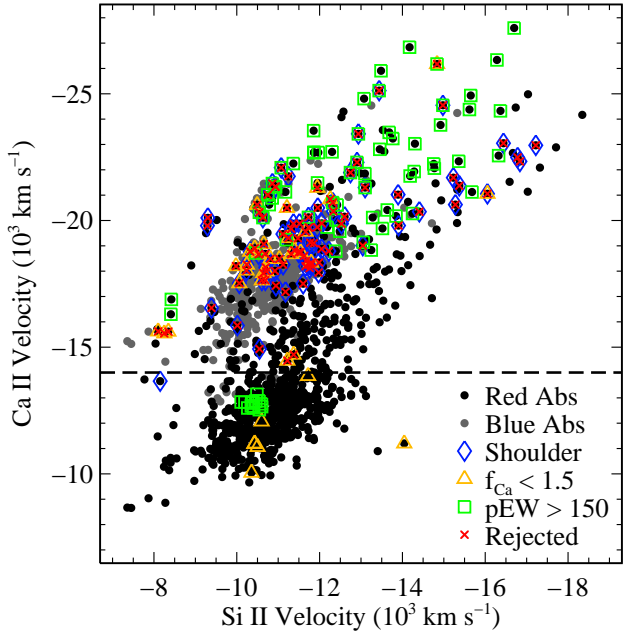}}
\caption{$v_{\rm Ca~H\&K}$ as a function of $v_{\rm Si~II}$.  The grey
and black points correspond to the blue and red absorption velocities,
respectively.  The green squares, gold triangles, and blue diamonds
correspond to spectra with pEW(\ion{Ca}{2}) $>$ 150~\AA, $f_{\rm Ca} <
1.5$, and having a visible red shoulder (e.g.,
Figure~\ref{f:ca_profile}), respectively.  The dashed line represents
a velocity of $-14,000$~\kms.  Red absorption measurements with a
visible red shoulder or $f_{\rm Ca} < 1.5$ and with $v_{\rm Ca~H\&K} <
-14,000$~\kms\ have been rejected from our analysis and are marked
with a red 'X.'}\label{f:ca_rb}
\end{center}
\end{figure}

The velocity of each absorption minimum in the \ion{Ca}{2} H\&K
feature of the smoothed spectra is automatically recorded (Blondin
et~al., in prep.).  We only examine spectra with one or two minima.
If two minima are found, the higher/lower velocities are classified as
``blue''/``red.''  If only one minimum is found, it is categorized as
the red or lower-velocity component.  In Figure~\ref{f:ca_rb}, $v_{\rm
Ca~H\&K}$ found from the blue and red absorption minima are shown as a
function of $v_{\rm Si~II}$.  Examples can be seen in
Figure~\ref{f:ca_profile}.  The two absorptions show distinct
``clouds'' representing a clear correlation between $v_{\rm Si~II}$
and $v_{\rm Ca~H\&K}$ for each component independently.  However,
there are several red velocity measurements which overlap with the
blue cloud of points.  The most likely explanation is that these
measurement correspond to a higher velocity component that was
incorrectly identified as a red component.  The obvious reason for
this is if the blue component dominates over the red component to the
point where there is no distinct minimum for the redder component.

Our goal of connecting velocities and colors will be undermined if we
misidentify the absorption components.  When $v_{\rm Si~II}$ has been
measured, there is no ambiguity, but for $z > 0.4$, $v_{\rm Si~II}$ is
rarely measurable.  To remove interlopers from our \ion{Ca}{2} H\&K
velocities, we examine 4 categories of objects with the properties:
$f_{\rm Ca} = f_{\lambda} (v_{\rm Ca~H\&K} + 9000~{\rm km~s}^{-1}) /
f_{\lambda} (v_{\rm Ca~H\&K}) < 1.5$, \ion{Ca}{2} H\&K feature has a
visible red shoulder, pEW(\ion{Ca}{2}) $>$ 150~\AA, and $v_{\rm
Ca~H\&K} > -14,000$~\kms.

A ratio of fluxes in the spectra at $v_{\rm Ca~H\&K}$ and offset by
$+9000$~\kms, which we call $f_{\rm Ca}$, is a strong discriminant.
If $v_{\rm Ca~H\&K}$ is matched to a blue feature, the flux ratio may
be small if there is an additional absorption component to the red.
Conversely, a red feature will generally have much higher flux
9000~\kms\ redward of the absorption.  One of us (RJF) visually
inspected (multiple inspections were performed for all spectra, with
at least 3 classifications per spectrum, and requiring consistent
results) all spectra to determine qualitative characteristics of the
spectra.  After making all qualitative assessments, it was found that
those with a noticeable red shoulder on the feature were exclusively
in the blue cloud (Figure~\ref{f:ca_rb}).  These spectra correspond to
the case where the blue component is dominant and the automatic
procedure could not identify the other component.  Additionally, we
found that spectra with a large pEW(\ion{Ca}{2}) generally overlapped
with the blue cloud; however, many of these spectra were at the
highest velocities (for both \ion{Ca}{2} and \ion{Si}{2}), and their
true association was ambiguous.  Since these spectra were also
typically at the earliest phases ($t \lesssim -7$~d), excluding them
will make little difference to our final analysis.  We therefore chose
to include these spectra.  Finally, very few blue absorption features
are at $v_{\rm Ca~H\&K} > 14,000$~\kms.  We therefore declare any red
$v_{\rm Ca~H\&K}$ with a velocity below this threshold to be a true
measurement of the red ejecta velocity regardless of other
measurements.  Ultimately, spectra with $v_{\rm Ca~H\&K} <
-14,000$~\kms\ and either if $f_{\rm Ca} < 1.5$ or if there was a
visible red shoulder, are assumed to be a misidentification and were
rejected.  Graphically, these samples can be seen in
Figure~\ref{f:ca_rb}.

\begin{figure*}
\begin{center}
\epsscale{1.18}
\rotatebox{90}{
\plotone{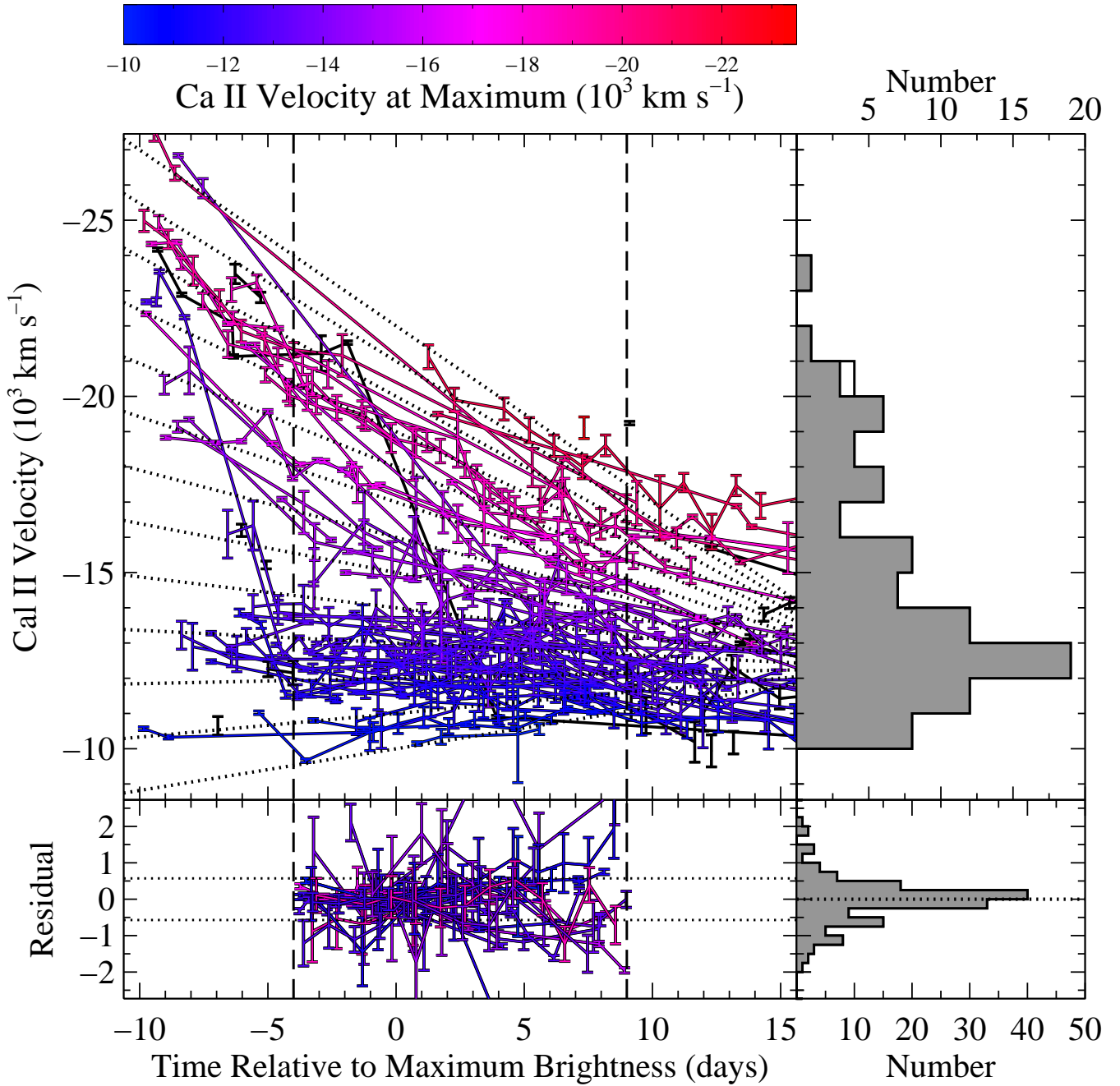}}
\caption{Same as Figure~\ref{f:vel_grad}, except for $v_{\rm
Ca~H\&K}$.}\label{f:ca_vel_grad}
\end{center}
\end{figure*}

In Figure~\ref{f:ca_vel_grad}, the red \ion{Ca}{2} velocity is shown
as a function of phase relative to $B$ maximum for the 685 spectra of
109 SNe~Ia with a \ion{Ca}{2} H\&K measurement, a time of maximum, and
$1 \le \Delta m_{15} (B) \le 1.5$~mag.  Only spectra which have passed
our previous cuts are shown.  The velocity evolution of \ion{Ca}{2} is
similar to that of \ion{Si}{2}, but \ion{Ca}{2} is typically at a
higher velocity at a given epoch (see Figure~\ref{f:ca_rb}).

\begin{figure}
\begin{center}
\epsscale{1.}
\rotatebox{90}{
\plotone{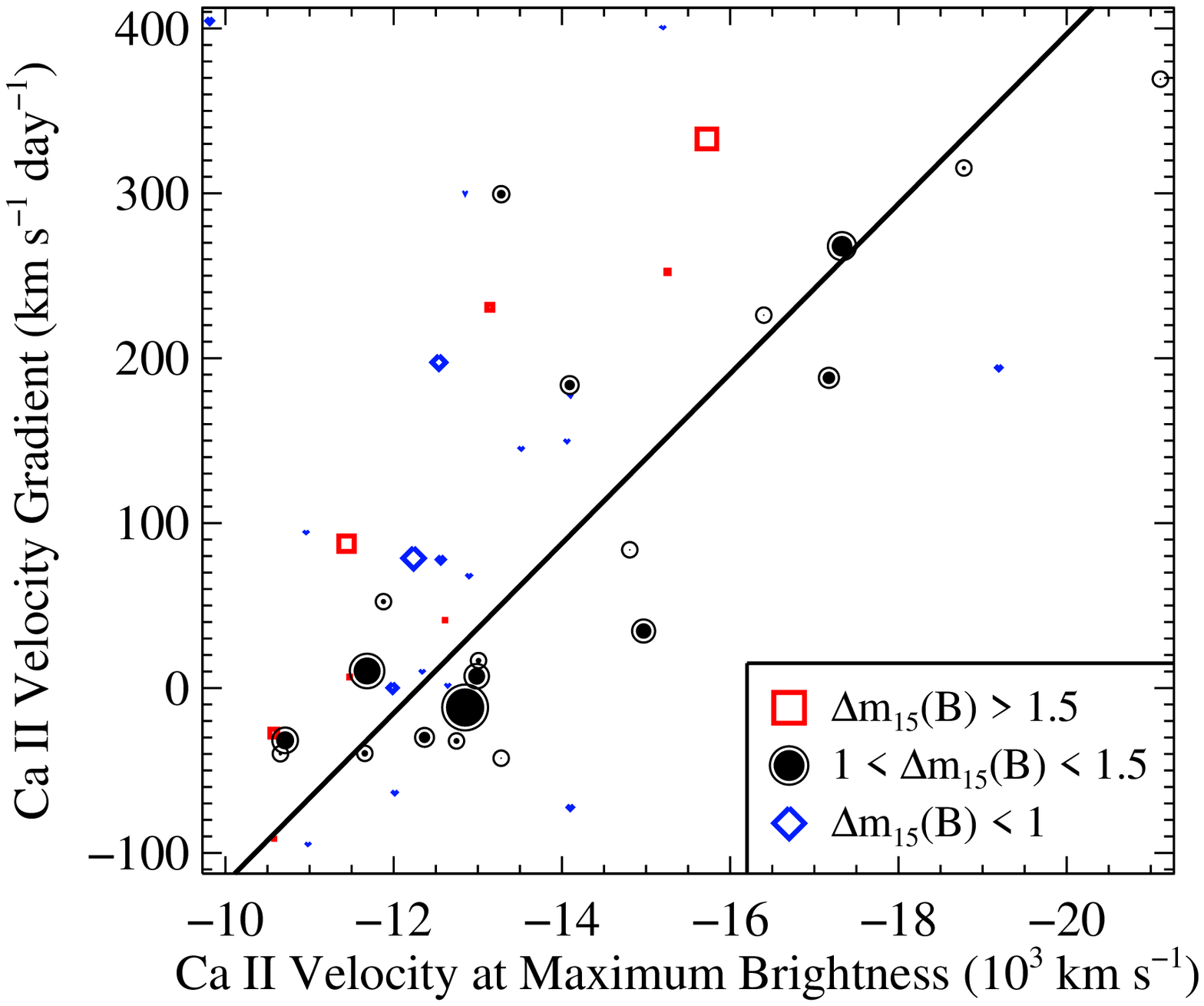}}
\caption{Same as Figure~\ref{f:family}, except for \ion{Ca}{2}
H\&K.}\label{f:ca_family}
\end{center}
\end{figure}

There are fewer spectra and fewer SNe with a $v_{\rm Ca~H\&K}$
measurement than for $v_{\rm Si~II}$.  This is because (1) the CfA
FAST spectra extend only to \about3500~\AA\ in the blue, and the
\ion{Ca}{2} H\&K absorption minimum is often blueward of these spectra,
and (2) the above procedure rejected many spectra.  Regardless, the
same trend that higher velocity SNe have higher velocity gradients is
clear.  This is also shown in Figure~\ref{f:ca_family}, which is
similar to Figure~\ref{f:family}.

It appears that early-time ($t < -6$~d) \ion{Ca}{2} H\&K data might be
a cleaner diagnostic of maximum-light velocity and velocity gradient
than for \ion{Si}{2} $\lambda 6355$, but the small number of spectra
at these phases prevents a definitive statement.  This might be the
result of inspecting and rejecting individual spectra for \ion{Ca}{2}
H\&K and not for \ion{Si}{2} $\lambda 6355$, which might also have
multiple components at early phases.  SNe with high-velocity
\ion{Ca}{2} H\&K features continue to have higher velocity than their
lower velocity counterparts to later phases than is seen with
\ion{Si}{2} $\lambda 6355$.  Despite this, the linear velocity
evolution does not maintain at these later phases (there appears to be
a plateau for most SNe).  As a result of a paucity of early-time data
and the changing velocity gradient at later times, we define the phase
range of $-4 < t < 9$~d to be the range over which $v_{\rm
Ca~H\&K}^{0}$ can be cleanly measured.

Performing the same procedure as above (again, ignoring SN~2004dt,
which also has peculiar \ion{Ca}{2} H\&K evolution --- SN~2003W only
has a \ion{Ca}{2} H\&K velocity for our $t = -11$~d spectrum, so it is
also not included) using $v_{\rm Ca~H\&K}$ for SNe~Ia with $1 \le
\Delta m_{15} (B) \le 1.5$~mag, we determine the relation
\begin{equation}\label{e:cagrad}
  v_{\rm Ca~H\&K} \left ( v_{\rm Ca~H\&K}^{0}, t \right ) = v_{\rm Ca~H\&K}^{0} (1 - 0.0515 t) - 0.633 t.
\end{equation}
Examples of this family of functions are shown in
Figure~\ref{f:ca_vel_grad}.

The residuals (as defined in Section~\ref{ss:sivel}), have a larger
scatter ($\sigma = 570$~\kms) than that of \ion{Si}{2} $\lambda 6355$,
but the residuals are not that large considering other factors
contributing to the total uncertainty in this measurement.  It is
unclear if the larger scatter is the result of a poorer representation
of the velocity gradient because of less data, less linear velocity
evolution of \ion{Ca}{2} H\&K, a less linear relation between
velocity at maximum brightness and velocity gradient, interloper
measurements, and/or the relation has larger intrinsic scatter.
There are also more ``catastrophic'' outliers (residuals more than
1000~\kms), which are likely the remaining absorption velocities that
overlap with the blue cloud.

Similar to what was done with Equation~\ref{e:grad} and \ion{Si}{2}
$\lambda 6355$, $v_{\rm Ca~H\&K}^{0}$ can be found using
Equation~\ref{e:cagrad},
\begin{equation}\label{e:cav0}
  v_{\rm Ca~H\&K}^{0} = (v_{\rm Ca~H\&K} + 0.631 t) / (1 - 0.0513 t).
\end{equation}

Table~\ref{t:obj} lists $v_{\rm Ca~H\&K}^{0}$ for each SN in the
F11 sample with a measurement in $-4 \le t \le 12$~d.  The value given
is for the measurement closest to $t = 0$~d.


\section{Pseudo-Equivalent Width}\label{s:pew}

Although \citetalias{Foley11:vel} showed that ejecta velocity and
intrinsic color were connected and we show in Section~\ref{s:col} that
the relation is robust, ejecta velocity may not be the best
indicator of intrinsic color.  \citetalias{Foley11:vel} postulated
that the redder colors in high-velocity SNe may be the result of
broader lines causing increased line blanketing in the near-UV.  If
this explanation is correct, one might expect line widths or line
strength to be more strongly correlated with intrinsic color.

FWHM and pEW measurements have observational advantages over velocity
measurements.  Specifically, even if the redshift of a SN~Ia is poorly
known (e.g., no host galaxy redshift for high-$z$ SNe), the FWHM and
pEW is still robust.  Unfortunately, pEW is sensitive to galaxy light
contamination \citep[e.g.,][]{Foley08:comp}.  If FWHM is measured as
part of fitting a Gaussian profile to a spectral feature, then
deviations from that profile will cause the measured FWHM to lack a
physical connection to the data.

For the F11 sample, the measured \ion{Si}{2} $\lambda 6355$ FWHMs
assume a Gaussian profile.  This has not been attempted for the
\ion{Ca}{2} H\&K feature since it rarely has a Gaussian profile.  We
performed the below analysis for the FWHM(\ion{Si}{2}) measurements,
but found pEW(\ion{Si}{2}) measurements to yield a more robust result.
We therefore only report the pEW results below.

Several groups have examined pEWs of \ion{Ca}{2} H\&K and \ion{Si}{2}
$\lambda 6355$ at both low and high $z$ \citep[e.g.,][]{Folatelli04,
Hachinger06, Garavini07, Bronder08, Foley08:comp, Balland09, Branch09,
Blondin11, Konishi11, Nordin11:sdss, Walker11}.  \citet{Branch06} used
the pEW of \ion{Si}{2} $\lambda 6355$, along with the pEW of
\ion{Si}{2} $\lambda 5972$ to subclassify SNe~Ia.  His classifications
match up well with the \citet{Benetti05} classifications based on
velocity gradient.

\subsection{\ion{Si}{2} $\lambda 6355$ Pseudo-Equivalent Width}\label{ss:si_pew}

\begin{figure}
\begin{center}
\epsscale{0.95}
\rotatebox{90}{
\plotone{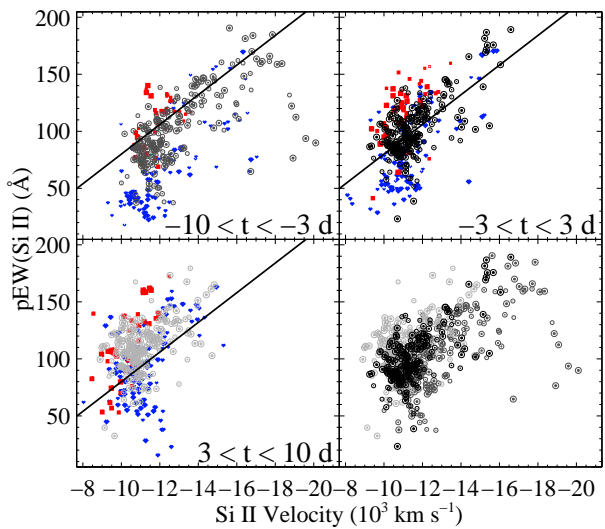}}
\caption{pEW(\ion{Si}{2}) as a function of $v_{\rm Si~II}$ for
different phase and $\Delta m_{15} (B)$ ranges.  The panels (clockwise
from lower-left) show a phase range of $3 < t < 10$, $-10 < t < -3$,
$-3 < t < 3$, and $-10 < t < 10$~d, respectively.  The blue diamonds
and red squares represent SNe~Ia with $\Delta m_{15} (B) < 1$ and
$\Delta m_{15} (B) > 1.5$~mag, respectively.  The dark grey, black,
and light grey circles represent SNe~Ia with $1 < \Delta m_{15} (B) <
1.5$~mag and $-10 < t < -3$, $-3 < t < 3$, and $3 < t < 10$~d,
respectively.  The size of the symbol corresponds to the inverse of
the uncertainty.  A solid line is plotted to correspond roughly to the
relation between $v_{\rm Si~II}$ and pEW(\ion{Si}{2}) at early
times and high velocity.  The same line is plotted in the upper-right
and lower-left panels to aid the eye in comparing the different
phases.  The lower-right panel shows the moderate decliners ($1 <
\Delta m_{15} (B) < 1.5$~mag) for the three phase
ranges.}\label{f:pew_si2_ev}
\end{center}
\end{figure}

As noted above, several studies have examined pEWs for several
features, including \ion{Si}{2}.  In Figure~\ref{f:pew_si2_ev}, we
show how pEW(\ion{Si}{2}) is related to the spectral phase, $\Delta
m_{15} (B)$, and $v_{\rm Si~II}$.  There is a wealth of information in
these data, but we will only briefly discuss them, leaving further
analysis to future work.

Examining Figure~\ref{f:pew_si2_ev}, we note several observational
facts.  First, as expected, higher-velocity SNe tend to have stronger
lines.  Second, slower-declining SNe tend to have weaker lines.
Third, lines tend to get stronger over the phase range shown here
($-10 < t < 10$~d).  We also note a generic trend where SNe~Ia with
$v_{\rm Si~II} \gtrsim -12,000$~\kms\ can have a large range of pEW
that appears to be completely unrelated to the measured velocity.
However, above this velocity, there appears to be a strong
relation between pEW(\ion{Si}{2}) and $v_{\rm Si~II}$ at all
phases shown here.  This is likely the result of the line saturating.
This effect may also be connected to the fact that SNe~Ia with $v_{\rm
Si~II}^{0} \gtrsim -12,000$~\kms\ have a small scatter in intrinsic
color, while higher velocity SNe~Ia have a larger range in intrinsic
color (\citetalias{Foley11:vel}; Section~\ref{s:col}).

\begin{figure*}
\begin{center}
\epsscale{1.18}
\rotatebox{90}{
\plotone{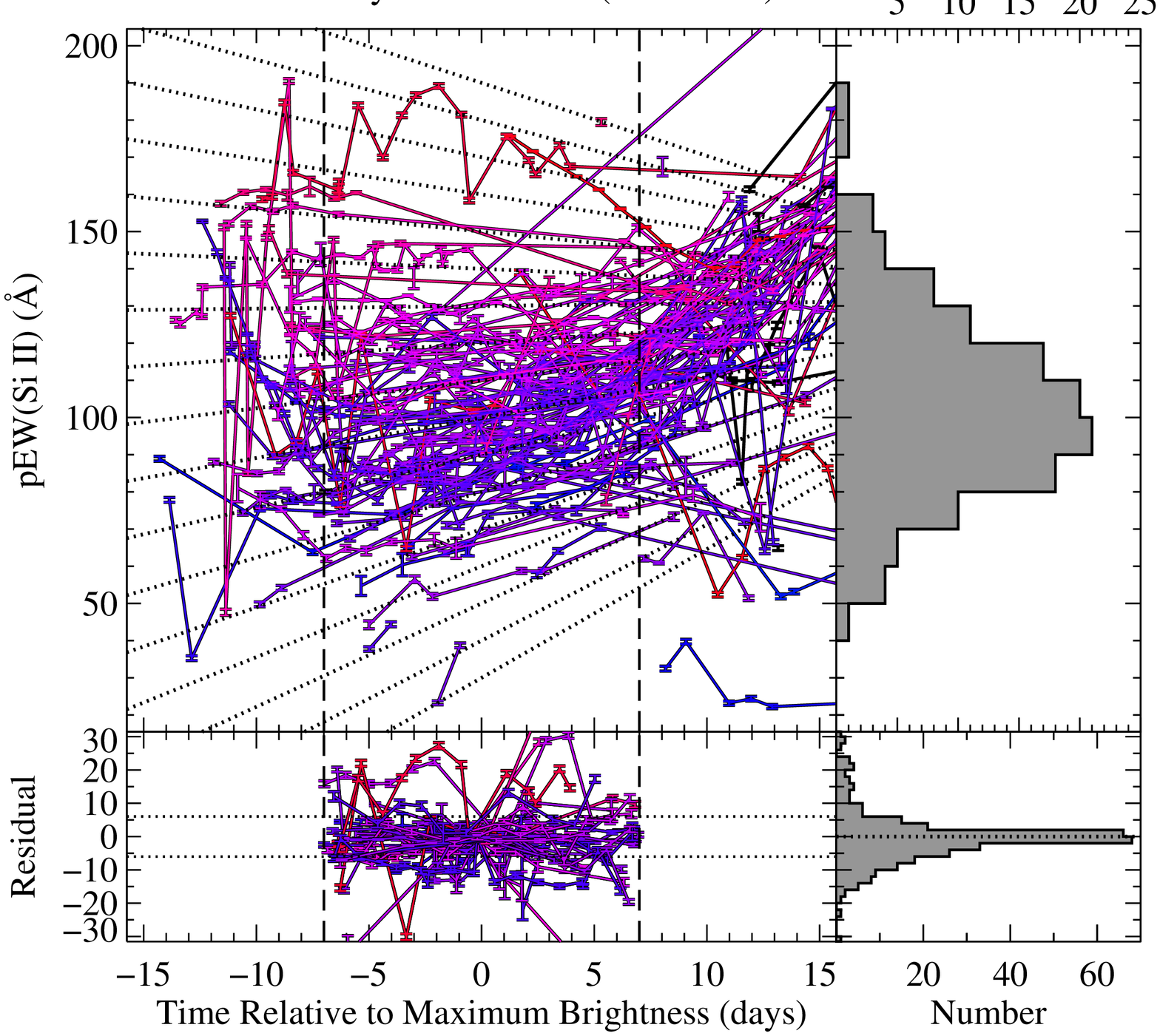}}
\caption{Same as Figure~\ref{f:vel_grad}, except for
pEW(\ion{Si}{2}).}\label{f:pew_grad}
\end{center}
\end{figure*}

Figure~\ref{f:pew_grad} shows the pEW evolution as a function of phase
for individual SNe.  Similar to the method outlined in
Section~\ref{s:vgrad}, we find a family of functions which describe
the pEW evolution as a function of time.  Using the same process as
before, we can determine pEW($t = 0$) = pEW$_{0}$ from a single pEW
measurement in the appropriate phase range ($-7 < t < 7$~d for
pEW(\ion{Si}{2})).  We find
\begin{equation}\label{e:pew0}
  {\rm pEW}_{0}({\rm Si~II}) = ({\rm pEW}({\rm Si~II}) + 4.48 t) / (1 - 0.0339 t).
\end{equation}

In the phase range $-7 < t < 7$~d, the evolution is relatively linear.
For most SNe, the pEW changes $<$20\% over this phase range;
therefore, knowing the exact evolution of this feature with time is
not critical.  Unlike $v_{\rm Si~II}^{0}$, pEW$_{0}$(\ion{Si}{2}) has
a relatively Gaussian distribution for our sample.

Table~\ref{t:obj} lists pEW$_{0}$(\ion{Si}{2}) for each SN in the
F11 sample with a measurement in $-7 \le t \le 7$~d.  The value given
is for the measurement closest to $t = 0$~d.

\subsection{\ion{Ca}{2} H\&K Pseudo-Equivalent Width}

Unlike $v_{\rm Ca~H\&K}$, pEW(\ion{Ca}{2}) is not systematically
affected by multiple absorption components.  Although incorrect
definition of the borders of the feature can produce incorrect
measurements, misidentifying a blue absorption component as a red
absorption component does not change any measurements.

\begin{figure*}
\begin{center}
\epsscale{1.18}
\rotatebox{90}{
\plotone{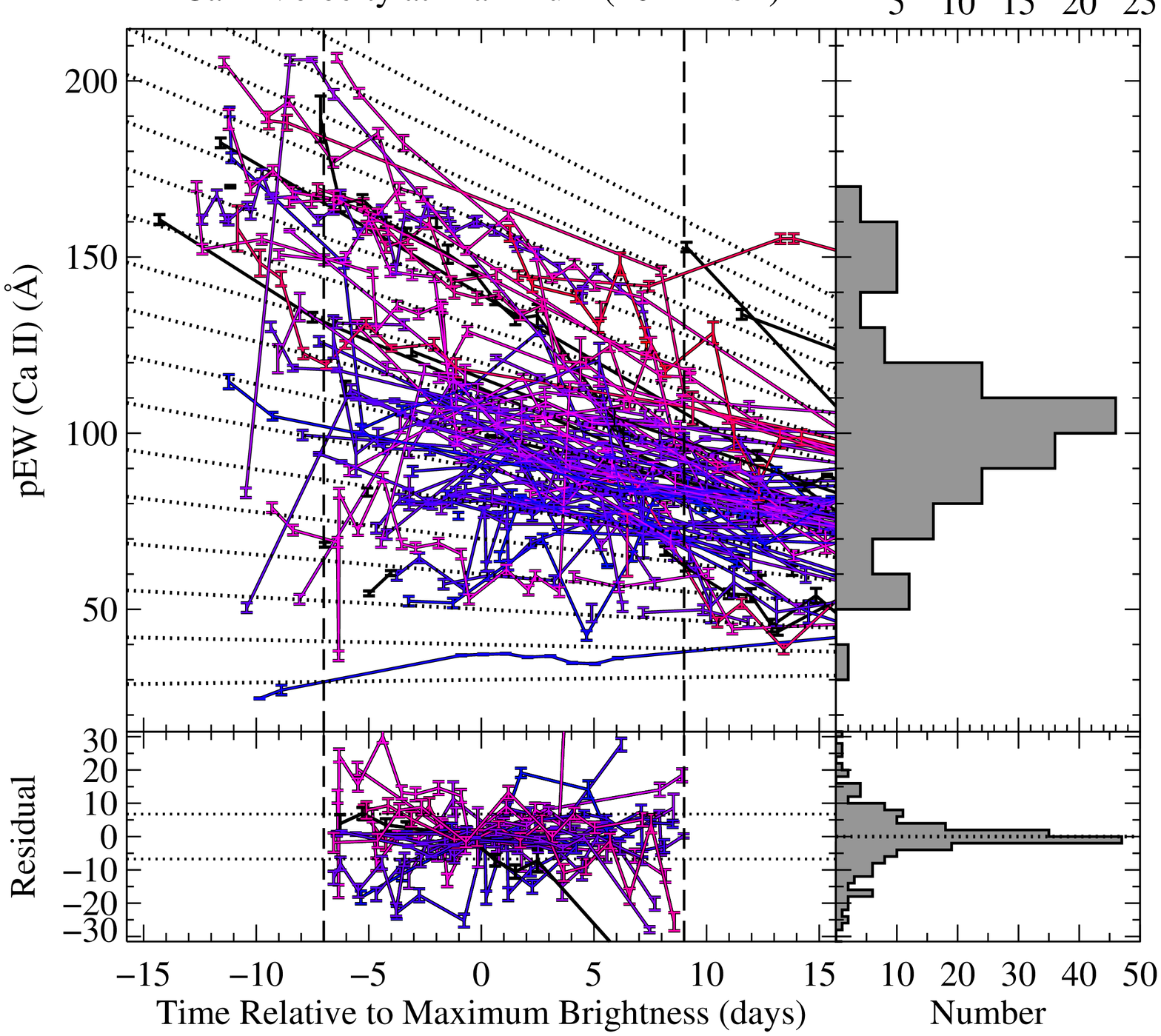}}
\caption{Same as Figure~\ref{f:vel_grad}, except for
pEW(\ion{Ca}{2}).}\label{f:ca_pew_grad}
\end{center}
\end{figure*}

Figure~\ref{f:ca_pew_grad} shows the \ion{Ca}{2} H\&K pEW evolution as
a function of phase for individual SNe.  Using the same method
outlined in Section~\ref{ss:si_pew}, we fit a family of functions to
the near-maximum ($-7 < t < 9$~d) temporal evolution of
pEW(\ion{Ca}{2}), which provides a way to convert pEW measurements in
that phase range to a maximum-light measurement, pEW$_{0}$.  We find
\begin{equation}\label{e:capew0}
  {\rm pEW}_{0}({\rm Ca~II}) = ({\rm pEW}({\rm Ca~II}) + 0.708 t) / (1 - 0.0210 t).
\end{equation}

Similar to \ion{Si}{2} $\lambda 6355$, \ion{Ca}{2} H\&K has a
relatively Gaussian pEW$_{0}$ distribution for our sample.

Table~\ref{t:obj} lists pEW$_{0}$(\ion{Ca}{2}) for each SN in the F11
sample with a measurement in $-7 \le t \le 9$~d.  The value given is
for the measurement closest to $t = 0$~d.


\section{Intrinsic Color}\label{s:col}

\subsection{Matching the F11 Sample to the W09 Sample}

The \citetalias{Wang09:2pop} SN~Ia sample is large.  It is further
useful for various studies since it is separated into two groups by
ejecta velocity.  However, it lacks published velocity measurements.
We are able to provide velocity measurements for 59 of the 121 SNe~Ia
in the \citetalias{Wang09:2pop} sample (after applying the same cuts
as \citetalias{Foley11:vel}).

\citetalias{Wang09:2pop} used the \ion{Si}{2} $\lambda 6355$ velocity
evolution of 10 well-observed low-velocity gradient SNe~Ia to
determine an average velocity as a function of phase and a dispersion
around that average.  SNe which had velocity measurements
$>$3$\sigma$ above the average in the time interval $-7 \le t \le 7$~d
were considered ``High-Velocity'' SNe, while all others with
measurements within that time interval were considered ``Normal.''
Since there are intermediate-velocity SNe and since SN velocities at
later times are similar, this method produces a coarse assessment of
the SNe with potential ambiguity.

\begin{figure}
\begin{center}
\epsscale{1.}
\rotatebox{90}{
\plotone{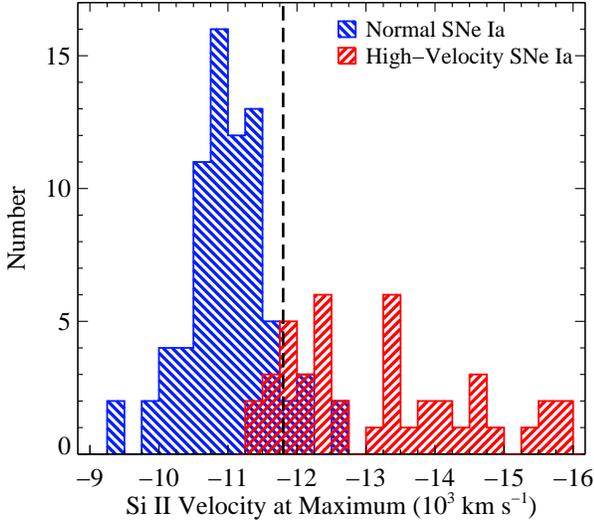}}
\caption{Histograms of $v_{\rm Si~II}^{0}$ measured from the F11
spectra.  The blue and red histograms represent SNe classified as
``Normal'' and ``High Velocity,'' respectively, by
\citetalias{Wang09:2pop}.  The dashed line represents the value of
$v_{\rm Si~II}^{0}$ that \citetalias{Wang09:2pop} nominally used to
separate the two subsamples.}\label{f:wang_hist}
\end{center}
\end{figure}

Figure~\ref{f:wang_hist} shows histograms of $v_{\rm Si~II}^{0}$ for
the F11 sample of SNe~Ia that overlap with the
\citetalias{Wang09:2pop} sample.  The two histograms represent SNe
classified as ``Normal'' and ``High-Velocity'' by
\citetalias{Wang09:2pop}.  The two subsamples are relatively separate,
but there is a significant amount of overlap for $-11,000 \gtrsim
v_{\rm Si~II}^{0} \gtrsim -12,000$~\kms.  The standard deviation for
our relation that measures $v_{\rm Si~II}^{0}$ is 220~km~s$^{-1}$,
the typical measurement uncertainty is $\lesssim$100~\kms. (Velocity
uncertainty from host-galaxy rotation does not affect this direct
comparison.)  Given the range of overlap of the subsamples, the data
sets and/or methodology must be responsible for some of the overlap.
This gives additional motivation for using individual velocity
measurements to describe the ejecta velocity of a SN~Ia.

\subsection{Intrinsic Color for the W09 Sample}\label{ss:w09}

\begin{figure}
\begin{center}
\epsscale{1.15}
\rotatebox{90}{
\plotone{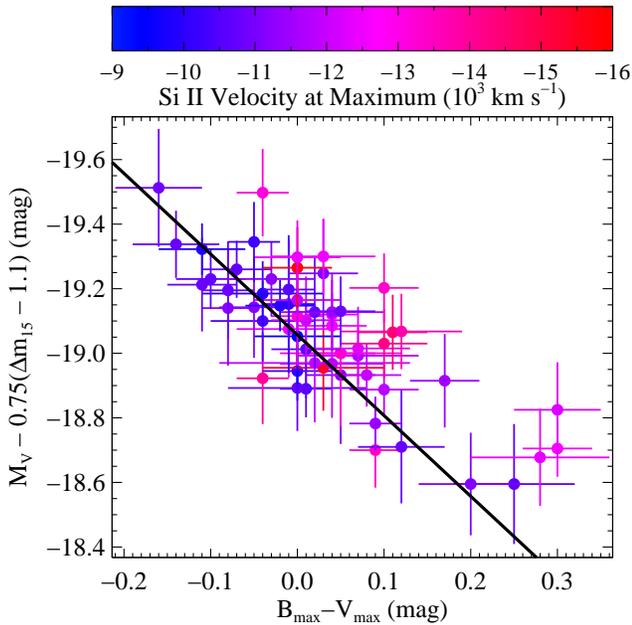}}
\caption{The light-curve shape corrected peak absolute $V$ brightness
as a function of $B_{\rm max} - V_{\rm max}$.  The color of the symbol
corresponds to the SN's $v_{\rm Si~II}^{0}$, with the color bar at the
top of the figure displaying the correspondence.  The solid black line
represents the host-galaxy dust reddening of a zero-color SN~Ia.
Deviations from this line in color is the intrinsic color of the
SN.}\label{f:wang}
\end{center}
\end{figure}

In Figure~\ref{f:wang}, we plot the light-curve (but not host-galaxy
reddening) corrected peak absolute magnitude of the
\citetalias{Wang09:2pop} matched sample as a function of their $B_{\rm
max} - V_{\rm max}$ pseudo-color.  This is similar to Figure~4 from
\citetalias{Foley11:vel}.  The symbols representing each SN are
color-coded by $v_{\rm Si~II}^{0}$.  As noted by
\citetalias{Foley11:vel}, the higher-velocity SNe tend to be
redder (brighter) for a given peak magnitude (color).

Using Equation~1 from \citetalias{Foley11:vel}, we have a relation
between measured $E(B-V)$ and the peak magnitude.  The
\citetalias{Wang09:2pop} $E(B-V)$ and the F11 $B_{\rm max} - V_{\rm
max}$ values are highly correlated.  Fitting a line to the data, we
find a simple offset of $B_{\rm max} - V_{\rm max} = E(B-V) -
0.081$~mag.  This is consistent with the offset ($-0.070 \pm 0.012$)
found by \citet{Phillips99}, especially considering the F11 $B_{\rm
max} - V_{\rm max}$ values include a Milky Way correction.  Applying
this offset to the best-fit values of Equation~1 of
\citetalias{Foley11:vel} for the ``Normal'' SNe, we obtain the solid
line in Figure~\ref{f:wang}.  This line represents the dust reddening
for a SN~Ia with an ``intrinsic'' color of zero.  Although this may
not be the true intrinsic color, deviations from this line in the
color direction represent deviations from a nominal color that is
corrected for host-galaxy reddening.  Specifically, we define
\begin{align}
  \left ( B_{\rm max} - V_{\rm max} \right )_{0} &=
    \left ( B_{\rm max} - V_{\rm max} \right ) \notag \\
    &- \left ( M_{\rm max,~corr}^{V} - M_{\rm zp} \right )/R_{V} - C,
\end{align}
where $M_{\rm max,~corr}^{V} = M_{\rm max}^{V} - \alpha (\Delta m_{15}
(B) - 1.1)$, $M_{\rm zp} = -19.26$~mag, $\alpha = 0.75$, $R_{V} =
2.5$, and $C = 0.081$~mag \citepalias{Foley11:vel}.  This measurement
may be offset from the true intrinsic color, but the same offset
should apply to all SNe, making the measurement useful for directly
comparing SNe.

We restrict the following analysis to SNe~Ia with $B_{\rm max} -
V_{\rm max} < 0.4 - 0.081 = 0.319$~mag, corresponding to the $E(B-V) <
0.4$~mag cut used by \citetalias{Foley11:vel}.  This is the color
range where low and high-velocity SNe~Ia appear to have the same
reddening law \citepalias{Foley11:vel}.

\begin{figure}
\begin{center}
\epsscale{1.}
\rotatebox{90}{
\plotone{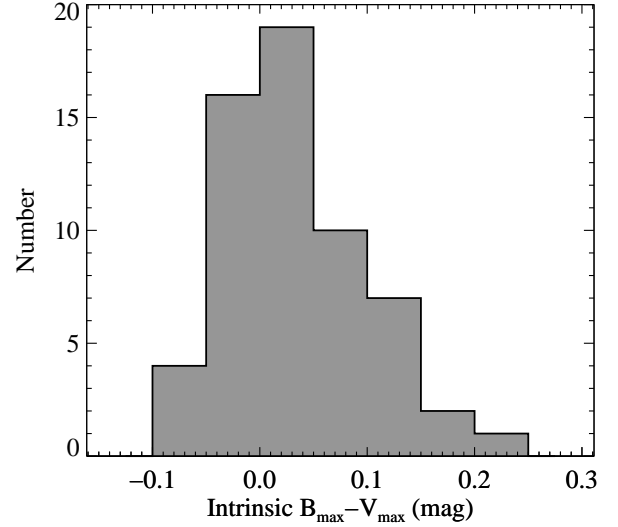}}
\caption{Distribution of intrinsic $B_{\rm max} - V_{\rm max}$ for
the \citetalias{Wang09:2pop}-F11 matched sample of SNe~Ia.}\label{f:col_hist}
\end{center}
\end{figure}

Figure~\ref{f:col_hist} presents a histogram of the intrinsic $B_{\rm
max} - V_{\rm max}$ for the sample.  The shape is slightly skewed to
red colors, with a skewness of 0.48.  The shape of the color
distribution is similar to the color distribution expected from the
asymmetric explosion models of \citet{Kasen07:asym} when accounting
for the frequency with which a particular viewing angle is observed.

\begin{figure}
\begin{center}
\epsscale{1.}
\rotatebox{90}{
\plotone{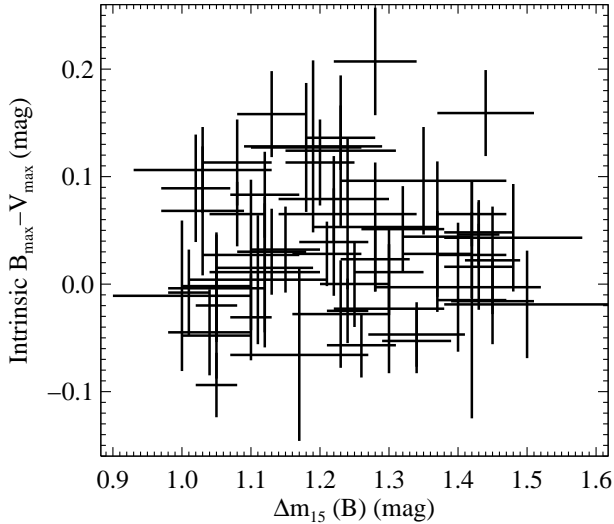}}
\caption{The intrinsic $B_{\rm max} - V_{\rm max}$ pseudo-color as a
function of $\Delta m_{15} (B)$ for the \citetalias{Wang09:2pop}-F11
sample.  The correlation coefficient is 0.04.}\label{f:col_dm15}
\end{center}
\end{figure}

SN~Ia color is correlated with light-curve shape
\citep[e.g.,][]{Riess96, Tripp98}, with faster declining SNe having
redder colors.  \citet{Maeda11} corrected the observed colors of SNe
in their sample using the relation between $B_{\rm max} - V_{\rm
max}$ and $\Delta m_{15} (B)$ found by \citet{Folatelli10}.  Over the
full $\Delta m_{15} (B)$ range used for this analysis, the
\citet{Folatelli10} relation predicts a correction of $0.060 \pm
0.025$~mag.  Although there is only a slight correlation between
light-curve shape and ejecta velocity for the range of $\Delta m_{15}
(B)$ examined here (e.g., Figure~\ref{f:vel_dm15}), there may still be
a significant correlation between $\Delta m_{15} (B)$ and intrinsic
color for the \citetalias{Wang09:2pop}-F11 matched sample.
Figure~\ref{f:col_dm15} shows the distribution of color and
light-curve shape for the \citetalias{Wang09:2pop}-F11 matched sample.
The two quantities displayed ($B_{\rm max} - V_{\rm max}$ and $\Delta
m_{15} (B)$) are not correlated for the sample, having a correlation
coefficient of 0.04.  Therefore, the intrinsic color is not
significantly affected by light-curve shape for the sample.  Given the
uncertainty in the \citet{Folatelli10} relation, the fact that it
was derived using a combination of low and high-velocity SNe~Ia, the
small correction expected for SNe~Ia in this analysis, and the lack of
a correlation between light-curve shape and intrinsic color for the
sample, we do not apply this correction.

\begin{figure*}
\begin{center}
\epsscale{0.7}
\rotatebox{90}{
\plotone{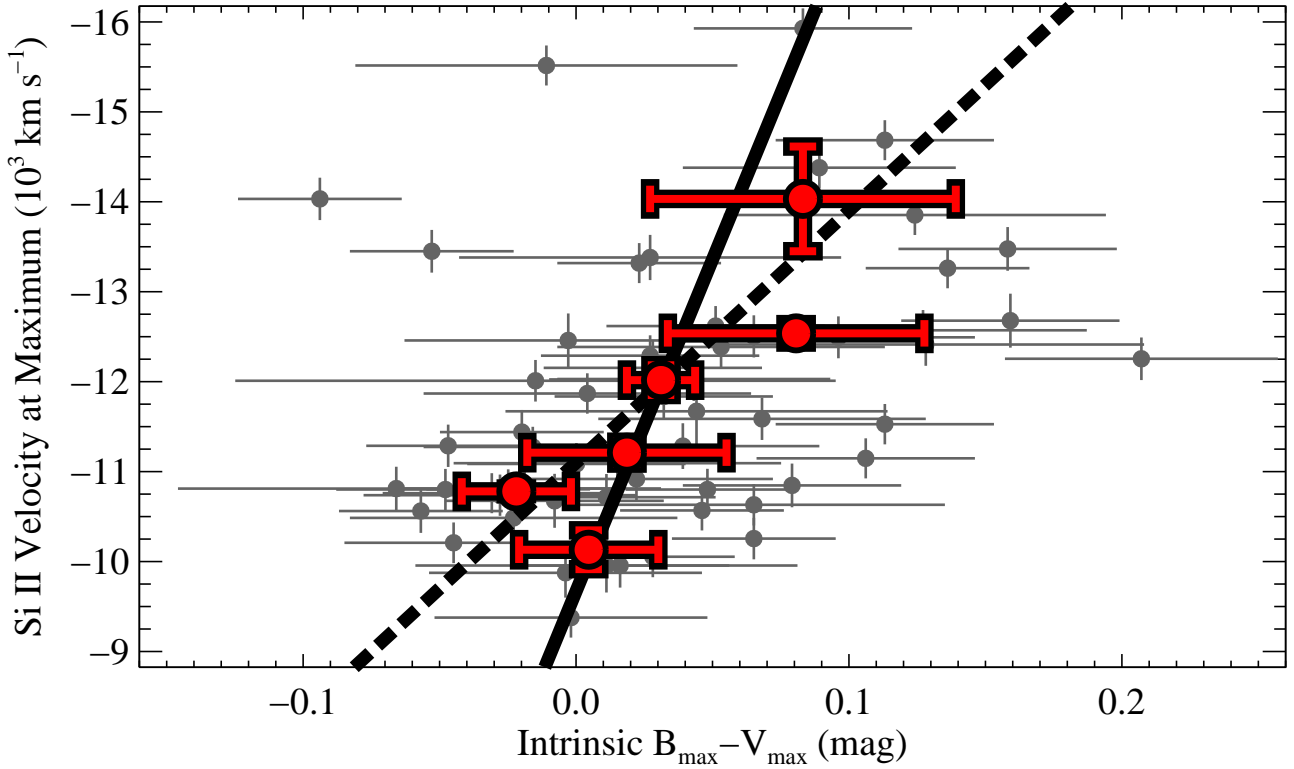}}
\caption{Maximum-light \ion{Si}{2} $\lambda 6355$ velocity ($v_{\rm
Si~II}^{0}$) as a function of intrinsic $B_{\rm max} - V_{\rm max}$
pseudo-color for the \citetalias{Wang09:2pop} sample of SNe~Ia.  The
red circles represent the median values of equal-numbered velocity
bins with the error bars being the median absolute deviation.  The
dashed line is the relation between velocity and color for the
\citet{Kasen07:asym} models (Equation~3 of \citetalias{Foley11:vel}).
The solid line is the best-fit linear model for the
data.}\label{f:vel_col}
\end{center}
\end{figure*}

In Figure~\ref{f:vel_col}, we compare the intrinsic color of each SN
to its $v_{\rm Si~II}^{0}$.  There is a trend where SNe with redder
intrinsic colors systematically have higher ejecta velocity.  The
correlation between the two quantities is 0.28, and a linear
least-squares fit results in a slope that is significant at the
3.4-$\sigma$ level.  Performing a Bayesian Monte-Carlo linear
regression on the data \citep{Kelly07}, we find that 99.142\% of the
realizations have a negative slope.  The trend is more apparent when
separating the sample into equal-numbered groups based on $v_{\rm
Si~II}^{0}$ and examining their median values.  The trend is also
qualitatively similar to that found by \citetalias{Foley11:vel} for
the synthetic spectra of \citet{Kasen07:asym}.

From the \citetalias{Wang09:2pop} sample, it is unclear if the trend
is linear or if there is a simple color offset.  There is a larger
intrinsic color scatter for the higher velocity SNe than for the lower
velocity SNe.  Splitting the sample by $v_{\rm Si~II}^{0} =
-11,800$~\kms, the intrinsic color scatter is 0.070 and 0.047~mag for
the high and low-velocity subsamples, respectively.  This was
postulated by \citetalias{Foley11:vel} as the reason that the
``Normal'' SNe~Ia produced a smaller scatter in Hubble residuals than
the ``High-Velocity'' SNe~Ia.  However, even after removing a linear
trend between velocity and intrinsic color, the higher-velocity SNe~Ia
continue to have a higher intrinsic color scatter (0.071 vs.\
0.046~mag).  This suggests that SNe~Ia with lower ejecta velocity may
be intrinsically better distance indicators, and with enough SNe~Ia,
it may be prudent to reject high-velocity SNe~Ia from cosmological
samples.

\subsection{Intrinsic Color for the F11 Sample}

As described in Section~\ref{ss:phot}, there are some drawbacks to
using the \citetalias{Wang09:2pop} data.  Here we repeat the above
analysis to determine the intrinsic color of SNe~Ia, but use the F11
sample.  Specifically, derived photometric values have been compiled
from CfA3 \citep{Hicken09:lc} and LOSS \citep{Ganeshalingam10}.  The
CfA3 peak magnitudes were corrected to match the LOSS system.  See
Section~\ref{ss:phot} for details.  We derive intrinsic $B_{\rm max} -
V_{\rm max}$ pseudo-colors for each SN in the F11 sample.  Using the
measurements of $v_{\rm Si~II}$, $v_{\rm Ca~H\&K}$, pEW(\ion{Si}{2}),
and pEW(\ion{Ca}{2}) from CfA and literature spectra near maximum
brightness and the time evolution of these quantities as defined by
Equations~\ref{e:v0}, \ref{e:cav0}, \ref{e:pew0}, and \ref{e:capew0},
we derive $v_{\rm Si~II}^{0}$, $v_{\rm CaH\&K}^{0}$,
pEW$_{0}$(\ion{Si}{2}), and pEW$_{0}$(\ion{Ca}{2}), respectively, for
individual SNe.  Below, we will compare these values to their
intrinsic colors.  The F11 sample takes no data directly from
\citetalias{Wang09:2pop}, but the samples have some SNe in common and
may share derived light-curve values.  Furthermore, the $v_{\rm
Si~II}^{0}$ values used in the above analysis are identical to those
presented here for the SNe in both samples.

Similar to what was done in Section~\ref{ss:w09}, we restrict the
following analysis to SNe~Ia with $B_{\rm max} - V_{\rm max} <
0.319$~mag, the color range where low and high-velocity SNe~Ia appear
to have the same reddening law \citepalias{Foley11:vel}.  We exclude
SNe~Ia with $z < 0.01$, except for those with Cepheid distances from
\citet{Riess11}.  There are 65, 59, 42, and 49 SNe~Ia with measured
maximum-brightness light-curve parameters, $1 \le \Delta m_{15} (B)
\le 1.5$~mag, and a maximum-brightness spectral parameter for
$v_{\rm Si~II}^{0}$, pEW$_{0}$(\ion{Si}{2}), $v_{\rm CaH\&K}^{0}$, and
pEW$_{0}$(\ion{Ca}{2}), respectively.

\begin{figure}
\begin{center}
\epsscale{1.15}
\rotatebox{90}{
\plotone{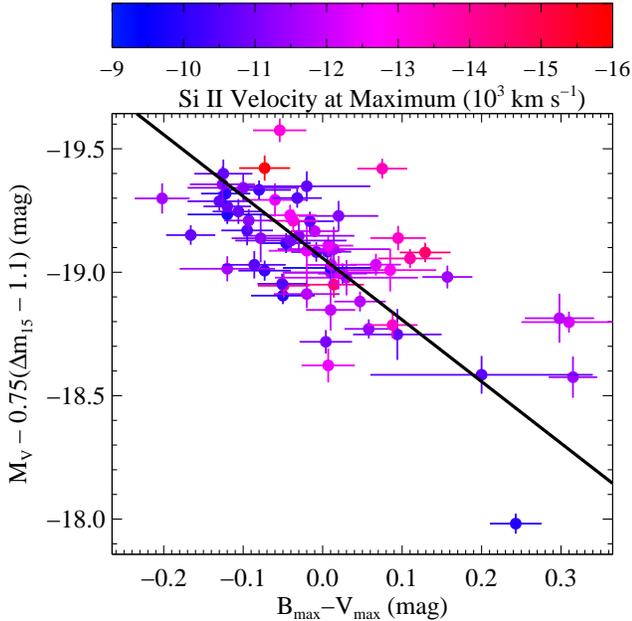}}
\caption{Same as Figure~\ref{f:wang}, except for the F11
sample.}\label{f:wang2}
\end{center}
\end{figure}

Using the $v_{\rm Si~II}^{0}$ data, Figure~\ref{f:wang2} shows the
light-curve (but not host-galaxy reddening) corrected peak absolute
magnitude of the F11 sample as a function of their $B_{\rm max} -
V_{\rm max}$ pseudo-color.  This is similar to Figure~\ref{f:wang}.

\begin{figure}
\begin{center}
\epsscale{2.8}
\rotatebox{90}{
\plotone{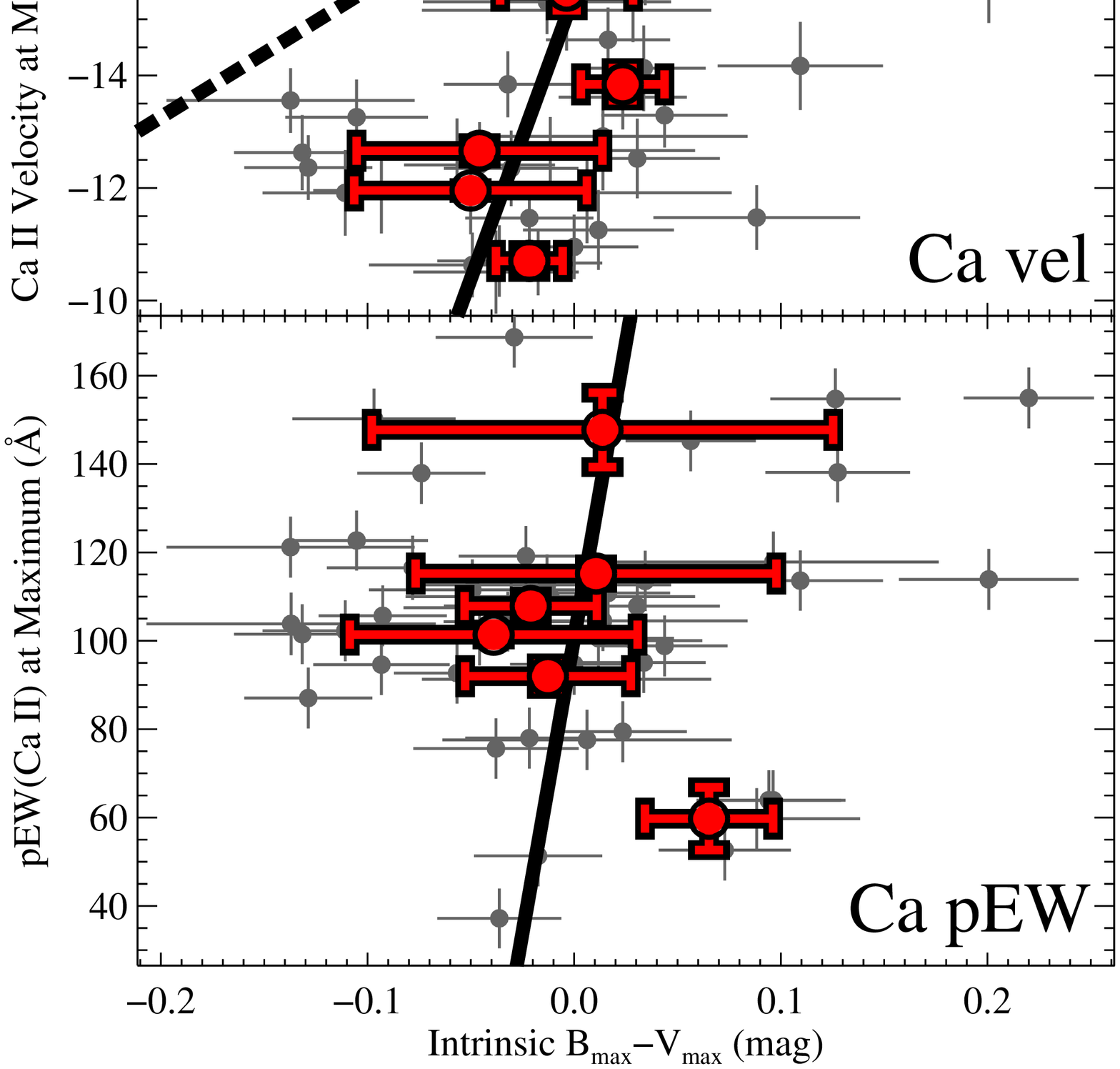}}
\caption{Maximum-light \ion{Si}{2} $\lambda 6355$ velocity,
pEW(\ion{Si}{2}), \ion{Ca}{2} H\&K velocity, and pEW(\ion{Ca}{2}) as a
function of intrinsic $B_{\rm max} - V_{\rm max}$ pseudo-color (top to
bottom panel, respectively) for the F11 sample of SNe~Ia.  The top
panel is similar to Figure~\ref{f:vel_col}, except for the F11 sample.
The red circles represent the median values of equal-numbered velocity
or pEW bins with the error bars representing the median absolute
deviation.  The dashed lines in the top and third panels are the
relation between velocity and color for the \citet{Kasen07:asym}
models (Equations~3 and 4 of \citetalias{Foley11:vel}, respectively).
The theoretical relations are strikingly similar and particularly poor
for the the \ion{Si}{2} $\lambda 6355$ and \ion{Ca}{2} H\&K data,
respectively.  The solid lines are the best-fit linear model for the
data.}\label{f:col}
\end{center}
\end{figure}

As was seen in the \citetalias{Wang09:2pop} sample, the higher
velocity SNe~Ia in the F11 sample tend to be redder than lower
velocity SNe~Ia.  Using the procedure outlined above, we again
calculate the intrinsic color for the sample.  Figure~\ref{f:col}
compares the intrinsic color of each SN to its ejecta velocity (as
derived from \ion{Si}{2} $\lambda 6355$ and \ion{Ca}{2} H\&K) and pEW
values.

As with the \citetalias{Wang09:2pop} sample, the F11 sample shows a
strong trend between $v_{\rm Si~II}^{0}$ and intrinsic color such that
intrinsically redder SNe~Ia tend to have higher velocity ejecta.
Unlike the \citetalias{Wang09:2pop} sample, the F11 sample has an
obvious linear relation between these quantities --- not just an
offset, corresponding to
\begin{align}
  (B_{\rm max} - V_{\rm max})_{0} &= (-0.39 \pm 0.04) - (0.033 \pm 0.004) \notag  \\
    &\times (v_{\rm Si~II}^{0} / 1000 {\rm ~km~s}^{-1}) {\rm ~mag}.
\end{align}
Performing a linear least-squares fit results in a non-zero slope that
is significant at the 8.8-$\sigma$ level.  The correlation between the
two quantities is $-0.39$.  Performing a Bayesian analysis of
Monte-Carlo linear regressions on the data \citep{Kelly07}, we find
that 99.949\% of the realizations have a negative slope and the median
correlation coefficient for the realizations is $-0.47$.  There is a
reasonable amount of scatter to the relation, but on average, the
trend is quite clear.  Although slightly offset, the linear
relation is strikingly similar to the theoretical trend found by
\citetalias{Foley11:vel}.

Similar to what was found with the \citetalias{Wang09:2pop} sample,
the intrinsic color scatter is larger for higher-velocity (as
determined by $v_{\rm Si~II}^{0}$) SNe~Ia.  The intrinsic color
scatter for SNe~Ia with $v_{\rm Si~II}^{0}$ higher/lower than
$-11,800$~\kms\ is 0.095 and 0.072~mag, respectively.  After
correcting for the linear trend, the scatter does not change much,
resulting in a scatter of 0.093 and 0.070~mag for the subsamples,
respectively.  However, correcting for the linear trend reduces the
overall scatter from $\sigma = 0.087$ to 0.080~mag.

Comparing pEW$_{0}$(\ion{Si}{2}) to the intrinsic $B_{\rm max} -
V_{\rm max}$ pseudo-color (Figure~\ref{f:col}), there is a reasonable
correlation ($\rho = 0.28$), with a linear relation defined by
\begin{align}
  (B_{\rm max} - V_{\rm max})_{0} &= (-0.16 \pm 0.03) - (0.0016 \pm 0.0003) \notag  \\
    &\times {\rm pEW}_{0}({\rm Si~II}) {\rm ~mag}.
\end{align}
The slope of the best-fit line for the two quantities is 6.3-$\sigma$
from zero.  Again, looking at medians (now for bins of equal number in
pEW space), the trend is clear.  Performing a Bayesian analysis of
Monte-Carlo linear regressions on the data \citep{Kelly07}, we find
that 98.6\% of the realizations have a positive slope and a median
correlation coefficient of 0.34.  The intrinsic color of SNe~Ia can be
derived from both the velocity and pEW of \ion{Si}{2} $\lambda 6355$
near maximum brightness.

The \ion{Ca}{2} H\&K measurements produce less robust results.
Intrinsic color is correlated with $v_{\rm Ca~H\&K}^{0}$, where
(similar to \ion{Si}{2} $\lambda 6355$) SNe~Ia with higher velocity
ejecta tend to be intrinsically redder.  A linear least-squares fit to
the data results in a slope that is 4.7-$\sigma$ from zero.  However,
the data have a large scatter and the correlation is not apparent by
eye ($\rho = -0.24$).  A Bayesian Monte-Carlo analysis of the data
results in 96.1\% of the realizations having a negative slope and a
median correlation coefficient of $-0.33$.  Nonetheless,
lower-velocity SNe~Ia, as determined from \ion{Ca}{2} H\&K, appear to
have less intrinsic color scatter than higher-velocity SNe~Ia.
Splitting the sample at $v_{\rm Ca~H\&K}^{0} = -14,000$~\kms, we find
that the higher/lower velocity SNe~Ia have intrinsic color scatter of
0.095 and 0.058~mag, respectively.  Therefore, a velocity cut may
improve cosmological results.  The linear fit to the data is
significantly different from that found by \citetalias{Foley11:vel}
for the \citet{Kasen07:asym} models, and detrending the data by the
best-fit line slightly {\it increases} (\about 0.003~mag) the scatter
of the velocity subsamples (but decreases the scatter of the full
sample by \about 0.002~mag).

The pEW(\ion{Ca}{2}) measurements do not show a correlation with
intrinsic color.  The correlation coefficient is 0.03, and there is no
significant linear trend as found by a fit to the data (performing the
similar linear fits provided a non-zero slope at the 1.8-$\sigma$
level and positive slopes for 65.1\% of the realizations).  There is
no clear way to make a cut on pEW(\ion{Ca}{2}) to reduce the intrinsic
color scatter.  Two competing effects drive the value of
pEW(\ion{Ca}{2}).  First, higher velocity ejecta correspond to broader
lines (as clearly seen for \ion{Si}{2} $\lambda 6355$;
Figure~\ref{f:pew_si2_ev}), so SNe~Ia with higher ejecta velocity,
which have redder intrinsic colors, should have higher pEW.  However,
SNe~Ia with redder intrinsic $B-V$ colors, which have higher ejecta
velocity, will have a depressed UV continuum, and should have a lower
pEW.  These competing effects may reduce any correlation for
pEW(\ion{Ca}{2}).


\section{Discussion \& Conclusions}\label{s:conc}

SN~Ia kinematics (as measured by the minimum of the absorption or the
width of the absorption of spectral features) are related to the
intrinsic color of a SN~Ia.  The correlation between SN~Ia ejecta
velocity and intrinsic color found by \citetalias{Foley11:vel} has
been further investigated here by examining additional kinematic
probes: velocity and pEW measurements for both \ion{Si}{2} $\lambda
6355$ and \ion{Ca}{2} H\&K.  Using spectral series of many SNe~Ia,
families of functions, which provide a measurement of velocity and pEW
at maximum brightness given a measurement near maximum brightness,
were constructed.  Both features show a strong correlation between
their velocity gradient and velocity at maximum brightness such that
SNe~Ia with larger velocity gradients tend to have higher velocities
at maximum brightness.  Similar to high velocity-gradient SNe
\citep{Hachinger06}, higher-velocity SNe also tend to have broader
features (see also \citetalias{Wang09:2pop}).  Restricting the sample
to $1 \le \Delta m_{15} (B) \le 1.5$~mag, we find no correlation
between host-galaxy morphology and $v_{\rm Si~II}^{0}$.  Using the
families of functions, SNe~Ia with spectra at different phases can be
directly compared.  These functions provide a way to directly compare
SNe~Ia using a single near-maximum spectrum, providing an easy way to
categorize SNe~Ia in a manner similar to how $\Delta m_{15}$ has been
used for photometry.  Using the photometric parameters of
\citetalias{Wang09:2pop} as well as \citet{Hicken09:lc} and
\citet{Ganeshalingam10} along with the reddening law derived in
\citetalias{Foley11:vel}, we are able to determine the intrinsic color
of both the \citetalias{Wang09:2pop} and F11 samples, respectively.

Comparing the maximum-light velocity and pEW measurements with the
intrinsic color data for individual SNe, we find correlations between
intrinsic color and $v_{\rm Si~II}^{0}$, $v_{\rm Ca~H\&K}^{0}$, and
pEW$_{0}$(\ion{Si}{2}), although the correlation with $v_{\rm
Ca~H\&K}^{0}$ is less robust.  We find no correlation between
intrinsic color and pEW$_{0}$(\ion{Ca}{2}), and speculate that the
competing effects of higher-velocity SNe~Ia having broader lines and
depressed UV continua balance each other to some extent, negating any
correlation.  The intrinsic color of a SN~Ia varies linearly with both
$v_{\rm Si~II}^{0}$ and pEW$_{0}$(\ion{Si}{2}), although there is
significant scatter.

\begin{figure*}
\begin{center}
\epsscale{1.2}
\rotatebox{90}{
\plotone{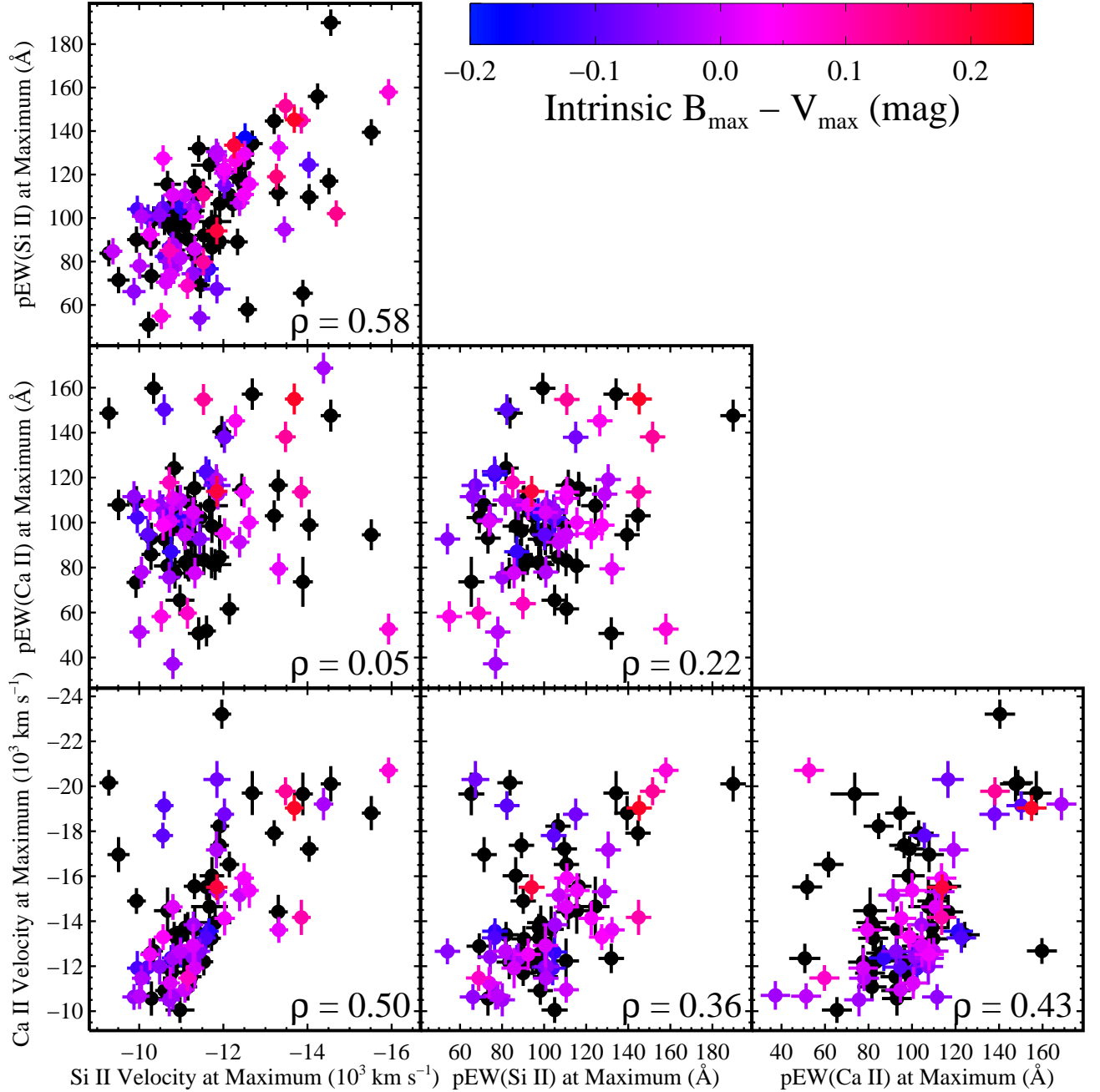}}
\caption{Maximum-light \ion{Si}{2} $\lambda 6355$ velocity,
pEW(\ion{Si}{2}), \ion{Ca}{2} H\&K velocity, and pEW(\ion{Ca}{2}) for
the F11 sample.  SNe with no intrinsic color information are black,
while those with intrinsic color information are color-coded by their
intrinsic color with the mapping represented by the color
bar.  The Pearson correlation coefficient for each set of parameters
is noted in each panel.}\label{f:comp}
\end{center}
\end{figure*}

Thus far, we have only discussed certain aspects of how our measured
parameters correlate with each other.  Specifically, we find that
$v_{\rm Si~II}$ and $v_{\rm Ca~H\&K}$ are highly correlated
(Figure~\ref{f:ca_rb}).  We also find that $v_{\rm Si~II}$ and
pEW(\ion{Si}{2}) are highly correlated for the highest velocities, but
uncorrelated at lower velocities (Figure~\ref{f:pew_si2_ev}).  In
Figure~\ref{f:comp}, $v_{\rm Si~II}^{0}$, pEW$_{0}$(\ion{Si}{2}),
$v_{\rm Ca~H\&K}^{0}$, and pEW$_{0}$(\ion{Ca}{2}) are plotted against
each other for the F11 sample.  SNe with intrinsic color information
are color-coded by their intrinsic color.  Most parameters are highly
correlated with each other (the absolute Pearson correlation
coefficient for each combination is presented in Figure~\ref{f:comp}).
The exceptions are $v_{\rm Si~II}^{0}$ and pEW$_{0}$(\ion{Ca}{2}), and
pEW$_{0}$(\ion{Si}{2}) and pEW$_{0}$(\ion{Ca}{2}).  This is not
unexpected since $v_{\rm Si~II}^{0}$ and pEW$_{0}$(\ion{Si}{2})
correlate with intrinsic color while pEW$_{0}$(\ion{Ca}{2}) does not.

There are a handful of outliers for each relation between the spectral
parameters, but the outliers do not appear to have a strong
relation with intrinsic color.  Therefore, measuring multiple
parameters may not directly improve our knowledge of intrinsic color.
Moreover, it is not clear if rejecting SNe based on their placement in
this multidimensional parameter space would improve any of the
previously determined relations.

Explosion models combined with radiative transfer models must account
for the relation between ejecta kinematics and intrinsic color.
Accurate SN~Ia models should reproduce both the general trends seen in
our data as well as the scatter.  We find that high-velocity (as
defined by either $v_{\rm Si~II}^{0}$ or $v_{\rm Ca~H\&K}^{0}$) SNe~Ia
have a larger scatter in their intrinsic color even after accounting
for the velocity-color relation.  \citetalias{Foley11:vel} suggested
that the larger Hubble scatter seen for ``High-Velocity'' SNe~Ia is
caused primarily by the larger scatter in intrinsic color, but also
suggested that this is primarily driven by the larger velocity range
for those SNe.  This appears to be correct, but the residual scatter
after correcting for velocity indicates lower-velocity SNe~Ia are
better standard crayons than higher-velocity SNe~Ia.

To reduce the scatter and potential bias \citepalias{Foley11:vel} of
their samples, future cosmological analyses with SNe~Ia may elect to
reject higher velocity SNe~Ia, along with or instead of fitting a
relation between velocity and color.  Furthermore, the most
precise and accurate SN~Ia distances must account for this effect.
\citetalias{Foley11:vel} outlined how ignoring this effect will
increase the scatter associated with SN~Ia distances and may bias
those distances, resulting in biased measurements of cosmological
parameters.

Although we provide a prescription for determining the intrinsic color
of SNe~Ia with a rest-frame optical spectrum near maximum brightness,
this relation is not robust for high-velocity SNe~Ia if the
spectrum does not cover \ion{Si}{2} $\lambda 6355$.  For SNe~Ia at $z
\gtrsim 0.4$, this feature is rarely observed.  Alternatively, a
relatively homogeneous data set can be obtained by making a cut based
on \ion{Ca}{2} H\&K velocity.  SNe~Ia with $v_{\rm Ca~H\&K}^{0}
\gtrsim -14,000$~\kms\ have a relatively small range of intrinsic
$B_{\rm max} - V_{\rm max}$ pseudo-color ($\sigma = 0.06$~mag).  To
avoid biases and to produce the most precise and accurate SN~Ia
distances, all future SN~Ia cosmology surveys should strongly consider
obtaining spectroscopy near maximum light for their primary
cosmological sample.  Measuring \ion{Si}{2} $\lambda 6355$ appears to
have some advantages over the easier to observe \ion{Ca}{2} H\&K.
Therefore, surveys targeting SNe~Ia at $z > 0.4$ may want to explore
NIR spectroscopy.  Even \ion{Ca}{2} H\&K is redshifted out of the
optical window for $z \gtrsim 1.2$, making NIR spectroscopy necessary
for the highest-redshift SNe~Ia.  We note that if one has a detector
that is sensitive to 2~$\mu$m, one can in principle measure
\ion{Si}{2} $\lambda 6355$ and \ion{Ca}{2} H\&K to $z \approx 2.2$ and
$z \approx 4$, respectively.

We could not have performed the analysis presented here without the
large CfA data set.  Although further analysis with an expanded sample
is necessary to strengthen the claims presented here and examine the
potential for combining different measurements to improve our
knowledge of the intrinsic color, the current sample has provided
insight and direction for these future studies.  To expand the current
sample, our group will soon release its CfA4 sample of SN~Ia light
curves.  On a similar time scale, the Carnegie Supernova Project (CSP)
will provide its next sample of light curves.  For spectroscopy, the
Berkeley Supernova Ia Program (BSNIP) will soon release its sample of
SN~Ia spectra, including all spectra used in the
\citetalias{Wang09:2pop} analysis.  Similarly, the CSP will soon release
their spectroscopic data set.  While the BSNIP sample will not have
the well-sampled spectral series of the CfA and CSP samples, it does
have spectra of many SNe~Ia near maximum brightness, and its spectra
typically cover bluer and redder wavelengths than the CfA spectra.
Therefore, it should provide many additional SNe for this particular
type of analysis.  On a slightly longer time scale, the Palomar
Transient Factory \citep{Rau09} and Supernova Factory
\citep{Aldering02} will provide light curves and
spectra of thousands of SNe~Ia.  With these combined data sets, it
should be possible to further refine and extend the results presented
here.

\begin{acknowledgments} 

\bigskip
R.J.F.\ is supported by a Clay Fellowship.  N.E.S.\ is supported by
the National Science Foundation through a Graduate Research
Fellowship.  Supernova studies at the Harvard College Observatory are
supported by NSF grant AST09-07903.

This work would not have been possible without the tireless work of
S.\ Blondin, who reduced the CfA spectra and made the spectral
measurements.  His comments have also significantly improved this
work.  P.\ Berlind and M.\ Calkins spent countless nights observing at
the 1.5~m telescope and obtained most of the spectra used for this
study.  We would like to thank them for their dedication.  We would
like to thank the many observers who contributed spectra to our
Literature sample and especially thank those that provided velocity
measurements in the circulars and telegrams.  R.J.F.\ would like to
thank S.\ Jha and K.\ Mandel, who influenced the direction of this
work.  M.\ Hicken and M.\ Ganeshalingam graciously contributed
information regarding the CfA3 and LOSS photometry necessary for this
study.  Some ideas presented here were first discussed at the Aspen
Center for Physics during the Summer 2010 workshop, ``Taking Supernova
Cosmology into the Next Decade.''

\end{acknowledgments}

\bibliographystyle{fapj}
\bibliography{../astro_refs}

\begin{thebibliography}{134}
\expandafter\ifx\csname natexlab\endcsname\relax\def\natexlab#1{#1}\fi

\bibitem[{{Aldering} {et~al.}(2002){Aldering}, {Adam}, {Antilogus}, {Astier},
  {Bacon}, {Bongard}, {Bonnaud}, {Copin}, {Hardin}, {Henault}, {Howell},
  {Lemonnier}, {Levy}, {Loken}, {Nugent}, {Pain}, {Pecontal}, {Pecontal},
  {Perlmutter}, {Quimby}, {Schahmaneche}, {Smadja}, \&
  {Wood-Vasey}}]{Aldering02}
{Aldering}, G., {et~al.} 2002, in Society of Photo-Optical Instrumentation
  Engineers (SPIE) Conference Series, Vol. 4836, Society of Photo-Optical
  Instrumentation Engineers (SPIE) Conference Series, ed. {J.~A.~Tyson \&
  S.~Wolff}, 61--72

\bibitem[{{Aldering} \& {Conley}(2000)}]{Aldering00}
{Aldering}, G., \& {Conley}, A. 2000, \iaucirc, 7413, 2

\bibitem[{{Altavilla} {et~al.}(2007){Altavilla}, {Stehle}, {Ruiz-Lapuente},
  {Mazzali}, {Pignata}, {Balastegui}, {Benetti}, {Blanc}, {Canal},
  {Elias-Rosa}, {Goobar}, {Harutyunyan}, {Pastorello}, {Patat}, {Rich},
  {Salvo}, {Schmidt}, {Stanishev}, {Taubenberger}, {Turatto}, \&
  {Hillebrandt}}]{Altavilla07}
{Altavilla}, G., {et~al.} 2007, \aap, 475, 585

\bibitem[{{Amanullah} {et~al.}(2010){Amanullah}, {Lidman}, {Rubin}, {Aldering},
  {Astier}, {Barbary}, {Burns}, {Conley}, {Dawson}, {Deustua}, {Doi}, {Fabbro},
  {Faccioli}, {Fakhouri}, {Folatelli}, {Fruchter}, {Furusawa}, {Garavini},
  {Goldhaber}, {Goobar}, {Groom}, {Hook}, {Howell}, {Kashikawa}, {Kim}, {Knop},
  {Kowalski}, {Linder}, {Meyers}, {Morokuma}, {Nobili}, {Nordin}, {Nugent},
  {{\"O}stman}, {Pain}, {Panagia}, {Perlmutter}, {Raux}, {Ruiz-Lapuente},
  {Spadafora}, {Strovink}, {Suzuki}, {Wang}, {Wood-Vasey}, {Yasuda}, \&
  {Supernova Cosmology Project}}]{Amanullah10}
{Amanullah}, R., {et~al.} 2010, \apj, 716, 712

\bibitem[{{Anupama} {et~al.}(2005){Anupama}, {Sahu}, \& {Jose}}]{Anupama05}
{Anupama}, G.~C., {Sahu}, D.~K., \& {Jose}, J. 2005, \aap, 429, 667

\bibitem[{{Bailey} {et~al.}(2009){Bailey}, {Aldering}, {Antilogus}, {Aragon},
  {Baltay}, {Bongard}, {Buton}, {Childress}, {Chotard}, {Copin}, {Gangler},
  {Loken}, {Nugent}, {Pain}, {Pecontal}, {Pereira}, {Perlmutter}, {Rabinowitz},
  {Rigaudier}, {Runge}, {Scalzo}, {Smadja}, {Swift}, {Tao}, {Thomas}, {Wu}, \&
  {The Nearby Supernova Factory}}]{Bailey09}
{Bailey}, S., {et~al.} 2009, \aap, 500, L17

\bibitem[{{Balland} {et~al.}(2009){Balland}, {Baumont}, {Basa}, {Mouchet},
  {Howell}, {Astier}, {Carlberg}, {Conley}, {Fouchez}, {Guy}, {Hardin}, {Hook},
  {Pain}, {Perrett}, {Pritchet}, {Regnault}, {Rich}, {Sullivan}, {Antilogus},
  {Arsenijevic}, {Le Du}, {Fabbro}, {Lidman}, {Mour{\~a}o},
  {Palanque-Delabrouille}, {P{\'e}contal}, \& {Ruhlmann-Kleider}}]{Balland09}
{Balland}, C., {et~al.} 2009, \aap, 507, 85

\bibitem[{{Benetti} {et~al.}(2005){Benetti}, {Cappellaro}, {Mazzali},
  {Turatto}, {Altavilla}, {Bufano}, {Elias-Rosa}, {Kotak}, {Pignata}, {Salvo},
  \& {Stanishev}}]{Benetti05}
{Benetti}, S., {et~al.} 2005, \apj, 623, 1011

\bibitem[{{Benetti} {et~al.}(2004){Benetti}, {Meikle}, {Stehle}, {Altavilla},
  {Desidera}, {Folatelli}, {Goobar}, {Mattila}, {Mendez}, {Navasardyan},
  {Pastorello}, {Patat}, {Riello}, {Ruiz-Lapuente}, {Tsvetkov}, {Turatto},
  {Mazzali}, \& {Hillebrandt}}]{Benetti04}
------. 2004, \mnras, 348, 261

\bibitem[{{Blondin} \& {Berlind}(2008)}]{Blondin08:08dt}
{Blondin}, S., \& {Berlind}, P. 2008, Central Bureau Electronic Telegrams,
  1424, 1

\bibitem[{{Blondin} {et~al.}(2006){Blondin}, {Dessart}, {Leibundgut}, {Branch},
  {H{\"o}flich}, {Tonry}, {Matheson}, {Foley}, {Chornock}, {Filippenko},
  {Sollerman}, {Spyromilio}, {Kirshner}, {Wood-Vasey}, {Clocchiatti},
  {Aguilera}, {Barris}, {Becker}, {Challis}, {Covarrubias}, {Davis},
  {Garnavich}, {Hicken}, {Jha}, {Krisciunas}, {Li}, {Miceli}, {Miknaitis},
  {Pignata}, {Prieto}, {Rest}, {Riess}, {Salvo}, {Schmidt}, {Smith}, {Stubbs},
  \& {Suntzeff}}]{Blondin06}
{Blondin}, S., {et~al.} 2006, \aj, 131, 1648

\bibitem[{{Blondin} {et~al.}(2011{\natexlab{a}}){Blondin}, {Kasen},
  {R{\"o}pke}, {Kirshner}, \& {Mandel}}]{Blondin11:2D}
{Blondin}, S., {Kasen}, D., {R{\"o}pke}, F.~K., {Kirshner}, R.~P., \& {Mandel},
  K.~S. 2011{\natexlab{a}}, \mnras, 1228

\bibitem[{{Blondin} {et~al.}(2011{\natexlab{b}}){Blondin}, {Mandel}, \&
  {Kirshner}}]{Blondin11}
{Blondin}, S., {Mandel}, K.~S., \& {Kirshner}, R.~P. 2011{\natexlab{b}}, \aap,
  526, A81+

\bibitem[{{Blondin} \& {Tonry}(2007)}]{Blondin07}
{Blondin}, S., \& {Tonry}, J.~L. 2007, \apj, 666, 1024

\bibitem[{{Branch}(1987)}]{Branch87}
{Branch}, D. 1987, \apjl, 316, L81

\bibitem[{{Branch} {et~al.}(2009){Branch}, {Dang}, \& {Baron}}]{Branch09}
{Branch}, D., {Dang}, L.~C., \& {Baron}, E. 2009, \pasp, 121, 238

\bibitem[{{Branch} {et~al.}(2006){Branch}, {Dang}, {Hall}, {Ketchum},
  {Melakayil}, {Parrent}, {Troxel}, {Casebeer}, {Jeffery}, \&
  {Baron}}]{Branch06}
{Branch}, D., {et~al.} 2006, \pasp, 118, 560

\bibitem[{{Branch} {et~al.}(1988){Branch}, {Drucker}, \& {Jeffery}}]{Branch88}
{Branch}, D., {Drucker}, W., \& {Jeffery}, D.~J. 1988, \apjl, 330, L117+

\bibitem[{{Branch} {et~al.}(2003){Branch}, {Garnavich}, {Matheson}, {Baron},
  {Thomas}, {Hatano}, {Challis}, {Jha}, \& {Kirshner}}]{Branch03}
{Branch}, D., {et~al.} 2003, \aj, 126, 1489

\bibitem[{{Bronder} {et~al.}(2008){Bronder}, {Hook}, {Astier}, {Balam},
  {Balland}, {Basa}, {Carlberg}, {Conley}, {Fouchez}, {Guy}, {Howell}, {Neill},
  {Pain}, {Perrett}, {Pritchet}, {Regnault}, {Sullivan}, {Baumont}, {Fabbro},
  {Filliol}, {Perlmutter}, \& {Ripoche}}]{Bronder08}
{Bronder}, T.~J., {et~al.} 2008, \aap, 477, 717

\bibitem[{{Conley} {et~al.}(2011){Conley}, {Guy}, {Sullivan}, {Regnault},
  {Astier}, {Balland}, {Basa}, {Carlberg}, {Fouchez}, {Hardin}, {Hook},
  {Howell}, {Pain}, {Palanque-Delabrouille}, {Perrett}, {Pritchet}, {Rich},
  {Ruhlmann-Kleider}, {Balam}, {Baumont}, {Ellis}, {Fabbro}, {Fakhouri},
  {Fourmanoit}, {Gonz{\'a}lez-Gait{\'a}n}, {Graham}, {Hudson}, {Hsiao},
  {Kronborg}, {Lidman}, {Mourao}, {Neill}, {Perlmutter}, {Ripoche}, {Suzuki},
  \& {Walker}}]{Conley11}
{Conley}, A., {et~al.} 2011, \apjs, 192, 1

\bibitem[{{Cristiani} {et~al.}(1992){Cristiani}, {Cappellaro}, {Turatto},
  {Bergeron}, {Bues}, {Buson}, {Danziger}, {di Serego-Alighieri}, {Duerbeck},
  {Heydari-Malayeri}, {Krautter}, {Schmutz}, \&
  {Schulte-Ladbeck}}]{Cristiani92}
{Cristiani}, S., {et~al.} 1992, \aap, 259, 63

\bibitem[{{Elias-Rosa} {et~al.}(2006{\natexlab{a}}){Elias-Rosa}, {Benetti},
  {Cappellaro}, {Harutyunyan}, {Pastorello}, {Mazzali}, {Taubenberger}, \&
  {Andreuzzi}}]{Elias-Rosa06:06en}
{Elias-Rosa}, N., {Benetti}, S., {Cappellaro}, E., {Harutyunyan}, A.,
  {Pastorello}, A., {Mazzali}, P., {Taubenberger}, S., \& {Andreuzzi}, G.
  2006{\natexlab{a}}, Central Bureau Electronic Telegrams, 608, 1

\bibitem[{{Elias-Rosa} {et~al.}(2006{\natexlab{b}}){Elias-Rosa}, {Benetti},
  {Cappellaro}, {Turatto}, {Mazzali}, {Patat}, {Meikle}, {Stehle},
  {Pastorello}, {Pignata}, {Kotak}, {Harutyunyan}, {Altavilla}, {Navasardyan},
  {Qiu}, {Salvo}, \& {Hillebrandt}}]{Elias-Rosa06}
{Elias-Rosa}, N., {et~al.} 2006{\natexlab{b}}, \mnras, 369, 1880

\bibitem[{{Fabricant} {et~al.}(1998){Fabricant}, {Cheimets}, {Caldwell}, \&
  {Geary}}]{Fabricant98}
{Fabricant}, D., {Cheimets}, P., {Caldwell}, N., \& {Geary}, J. 1998, \pasp,
  110, 79

\bibitem[{{Filippenko} {et~al.}(2003){Filippenko}, {Foley}, \&
  {Desroches}}]{Filippenko03:03gt}
{Filippenko}, A.~V., {Foley}, R.~J., \& {Desroches}, L. 2003, \iaucirc, 8175, 2

\bibitem[{{Filippenko} {et~al.}(1992{\natexlab{a}}){Filippenko}, {Richmond},
  {Branch}, {Gaskell}, {Herbst}, {Ford}, {Treffers}, {Matheson}, {Ho}, {Dey},
  {Sargent}, {Small}, \& {van Breugel}}]{Filippenko92:91bg}
{Filippenko}, A.~V., {et~al.} 1992{\natexlab{a}}, \aj, 104, 1543

\bibitem[{{Filippenko} {et~al.}(1992{\natexlab{b}}){Filippenko}, {Richmond},
  {Matheson}, {Shields}, {Burbidge}, {Cohen}, {Dickinson}, {Malkan}, {Nelson},
  {Pietz}, {Schlegel}, {Schmeer}, {Spinrad}, {Steidel}, {Tran}, \&
  {Wren}}]{Filippenko92:91T}
------. 1992{\natexlab{b}}, \apjl, 384, L15

\bibitem[{{Folatelli}(2004)}]{Folatelli04}
{Folatelli}, G. 2004, \nar, 48, 623

\bibitem[{{Folatelli} {et~al.}(2010){Folatelli}, {Phillips}, {Burns},
  {Contreras}, {Hamuy}, {Freedman}, {Persson}, {Stritzinger}, {Suntzeff},
  {Krisciunas}, {Boldt}, {Gonz{\'a}lez}, {Krzeminski}, {Morrell}, {Roth},
  {Salgado}, {Madore}, {Murphy}, {Wyatt}, {Li}, {Filippenko}, \&
  {Miller}}]{Folatelli10}
{Folatelli}, G., {et~al.} 2010, \aj, 139, 120

\bibitem[{{Foley} {et~al.}(2008{\natexlab{a}}){Foley}, {Filippenko},
  {Aguilera}, {Becker}, {Blondin}, {Challis}, {Clocchiatti}, {Covarrubias},
  {Davis}, {Garnavich}, {Jha}, {Kirshner}, {Krisciunas}, {Leibundgut}, {Li},
  {Matheson}, {Miceli}, {Miknaitis}, {Pignata}, {Rest}, {Riess}, {Schmidt},
  {Smith}, {Sollerman}, {Spyromilio}, {Stubbs}, {Suntzeff}, {Tonry},
  {Wood-Vasey}, \& {Zenteno}}]{Foley08:comp}
{Foley}, R.~J., {et~al.} 2008{\natexlab{a}}, \apj, 684, 68

\bibitem[{{Foley} {et~al.}(2008{\natexlab{b}}){Foley}, {Filippenko}, \&
  {Jha}}]{Foley08:uv}
{Foley}, R.~J., {Filippenko}, A.~V., \& {Jha}, S.~W. 2008{\natexlab{b}}, \apj,
  686, 117

\bibitem[{{Foley} \& {Kasen}(2011)}]{Foley11:vel}
{Foley}, R.~J., \& {Kasen}, D. 2011, \apj, 729, 55

\bibitem[{{Foley} {et~al.}(2009){Foley}, {Matheson}, {Blondin}, {Chornock},
  {Silverman}, {Challis}, {Clocchiatti}, {Filippenko}, {Kirshner},
  {Leibundgut}, {Sollerman}, {Spyromilio}, {Tonry}, {Davis}, {Garnavich},
  {Jha}, {Krisciunas}, {Li}, {Pignata}, {Rest}, {Riess}, {Schmidt}, {Smith},
  {Stubbs}, {Tucker}, \& {Wood-Vasey}}]{Foley09:year4}
{Foley}, R.~J., {et~al.} 2009, \aj, 137, 3731

\bibitem[{{Foley} {et~al.}(2010){Foley}, {Narayan}, {Challis}, {Filippenko},
  {Kirshner}, {Silverman}, \& {Steele}}]{Foley10:06bt}
{Foley}, R.~J., {Narayan}, G., {Challis}, P.~J., {Filippenko}, A.~V.,
  {Kirshner}, R.~P., {Silverman}, J.~M., \& {Steele}, T.~N. 2010, \apj, 708,
  1748

\bibitem[{{Foley} {et~al.}(2005){Foley}, {Perley}, {Bloom}, \&
  {Prochaska}}]{Foley05:05de}
{Foley}, R.~J., {Perley}, D., {Bloom}, J.~S., \& {Prochaska}, J.~X. 2005,
  \iaucirc, 8581, 3

\bibitem[{{Foley} {et~al.}(2006){Foley}, {Silverman}, {Moore}, \&
  {Filippenko}}]{Foley06:06ef}
{Foley}, R.~J., {Silverman}, J.~M., {Moore}, M., \& {Filippenko}, A.~V. 2006,
  Central Bureau Electronic Telegrams, 604, 1

\bibitem[{{Ganeshalingam} {et~al.}(2010){Ganeshalingam}, {Li}, {Filippenko},
  {Anderson}, {Foster}, {Gates}, {Griffith}, {Grigsby}, {Joubert}, {Leja},
  {Lowe}, {Macomber}, {Pritchard}, {Thrasher}, \& {Winslow}}]{Ganeshalingam10}
{Ganeshalingam}, M., {et~al.} 2010, \apjs, 190, 418

\bibitem[{{Garavini} {et~al.}(2005){Garavini}, {Aldering}, {Amadon},
  {Amanullah}, {Astier}, {Balland}, {Blanc}, {Conley}, {Dahl{\'e}n}, {Deustua},
  {Ellis}, {Fabbro}, {Fadeyev}, {Fan}, {Folatelli}, {Frye}, {Gates}, {Gibbons},
  {Goldhaber}, {Goldman}, {Goobar}, {Groom}, {Haissinski}, {Hardin}, {Hook},
  {Howell}, {Kent}, {Kim}, {Knop}, {Kowalski}, {Kuznetsova}, {Lee}, {Lidman},
  {Mendez}, {Miller}, {Moniez}, {Mouchet}, {Mour{\~a}o}, {Newberg}, {Nobili},
  {Nugent}, {Pain}, {Perdereau}, {Perlmutter}, {Quimby}, {Regnault}, {Rich},
  {Richards}, {Ruiz-Lapuente}, {Schaefer}, {Schahmaneche}, {Smith},
  {Spadafora}, {Stanishev}, {Thomas}, {Walton}, {Wang}, \&
  {Wood-Vasey}}]{Garavini05}
{Garavini}, G., {et~al.} 2005, \aj, 130, 2278

\bibitem[{{Garavini} {et~al.}(2004){Garavini}, {Folatelli}, {Goobar}, {Nobili},
  {Aldering}, {Amadon}, {Amanullah}, {Astier}, {Balland}, {Blanc}, {Burns},
  {Conley}, {Dahl{\'e}n}, {Deustua}, {Ellis}, {Fabbro}, {Fan}, {Frye}, {Gates},
  {Gibbons}, {Goldhaber}, {Goldman}, {Groom}, {Haissinski}, {Hardin}, {Hook},
  {Howell}, {Kasen}, {Kent}, {Kim}, {Knop}, {Lee}, {Lidman}, {Mendez},
  {Miller}, {Moniez}, {Mour{\~a}o}, {Newberg}, {Nugent}, {Pain}, {Perdereau},
  {Perlmutter}, {Prasad}, {Quimby}, {Raux}, {Regnault}, {Rich}, {Richards},
  {Ruiz-Lapuente}, {Sainton}, {Schaefer}, {Schahmaneche}, {Smith}, {Spadafora},
  {Stanishev}, {Walton}, {Wang}, \& {Wood-Vasey}}]{Garavini04}
------. 2004, \aj, 128, 387

\bibitem[{{Garavini} {et~al.}(2007{\natexlab{a}}){Garavini}, {Folatelli},
  {Nobili}, {Aldering}, {Amanullah}, {Antilogus}, {Astier}, {Blanc}, {Bronder},
  {Burns}, {Conley}, {Deustua}, {Doi}, {Fabbro}, {Fadeyev}, {Gibbons},
  {Goldhaber}, {Goobar}, {Groom}, {Hook}, {Howell}, {Kashikawa}, {Kim},
  {Kowalski}, {Kuznetsova}, {Lee}, {Lidman}, {Mendez}, {Morokuma}, {Motohara},
  {Nugent}, {Pain}, {Perlmutter}, {Quimby}, {Raux}, {Regnault},
  {Ruiz-Lapuente}, {Sainton}, {Schahmaneche}, {Smith}, {Spadafora},
  {Stanishev}, {Thomas}, {Walton}, {Wang}, {Wood-Vasey}, \&
  {Yasuda}}]{Garavini07}
------. 2007{\natexlab{a}}, \aap, 470, 411

\bibitem[{{Garavini} {et~al.}(2007{\natexlab{b}}){Garavini}, {Nobili},
  {Taubenberger}, {Pastorello}, {Elias-Rosa}, {Stanishev}, {Blanc}, {Benetti},
  {Goobar}, {Mazzali}, {Sanchez}, {Salvo}, {Schmidt}, \&
  {Hillebrandt}}]{Garavini07:05cf}
------. 2007{\natexlab{b}}, \aap, 471, 527

\bibitem[{{Garnavich} {et~al.}(2004){Garnavich}, {Bonanos}, {Krisciunas},
  {Jha}, {Kirshner}, {Schlegel}, {Challis}, {Macri}, {Hatano}, {Branch},
  {Bothun}, \& {Freedman}}]{Garnavich04}
{Garnavich}, P.~M., {et~al.} 2004, \apj, 613, 1120

\bibitem[{{G{\'o}mez} \& {L{\'o}pez}(1998)}]{Gomez98}
{G{\'o}mez}, G., \& {L{\'o}pez}, R. 1998, \aj, 115, 1096

\bibitem[{{Gurugubelli} {et~al.}(2006){Gurugubelli}, {Anupama}, \&
  {Sahu}}]{Gurugubelli06}
{Gurugubelli}, U.~K., {Anupama}, G.~C., \& {Sahu}, D.~K. 2006, Central Bureau
  Electronic Telegrams, 790, 1

\bibitem[{{Hachinger} {et~al.}(2006){Hachinger}, {Mazzali}, \&
  {Benetti}}]{Hachinger06}
{Hachinger}, S., {Mazzali}, P.~A., \& {Benetti}, S. 2006, \mnras, 370, 299

\bibitem[{{Hamuy} {et~al.}(2002){Hamuy}, {Maza}, {Pinto}, {Phillips},
  {Suntzeff}, {Blum}, {Olsen}, {Pinfield}, {Ivanov}, {Augusteijn}, {Brillant},
  {Chadid}, {Cuby}, {Doublier}, {Hainaut}, {Le Floc'h}, {Lidman},
  {Petr-Gotzens}, {Pompei}, \& {Vanzi}}]{Hamuy02}
{Hamuy}, M., {et~al.} 2002, \aj, 124, 417

\bibitem[{{Hamuy} {et~al.}(1996{\natexlab{a}}){Hamuy}, {Phillips}, {Suntzeff},
  {Schommer}, {Maza}, {Antezan}, {Wischnjewsky}, {Valladares}, {Muena},
  {Gonzales}, {Aviles}, {Wells}, {Smith}, {Navarrete}, {Covarrubias},
  {Williger}, {Walker}, {Layden}, {Elias}, {Baldwin}, {Hernandez}, {Tirado},
  {Ugarte}, {Elston}, {Saavedra}, {Barrientos}, {Costa}, {Lira}, {Ruiz},
  {Anguita}, {Gomez}, {Ortiz}, {della Valle}, {Danziger}, {Storm}, {Kim},
  {Bailyn}, {Rubenstein}, {Tucker}, {Cersosimo}, {Mendez}, {Siciliano},
  {Sherry}, {Chaboyer}, {Koopmann}, {Geisler}, {Sarajedini}, {Dey}, {Tyson},
  {Rich}, {Gal}, {Lamontagne}, {Caldwell}, {Guhathakurta}, {Phillips},
  {Szkody}, {Prosser}, {Ho}, {McMahan}, {Baggley}, {Cheng}, {Havlen},
  {Wakamatsu}, {Janes}, {Malkan}, {Baganoff}, {Seitzer}, {Shara}, {Sturch},
  {Hesser}, {Hartig}, {Hughes}, {Welch}, {Williams}, {Ferguson}, {Francis},
  {French}, {Bolte}, {Roth}, {Odewahn}, {Howell}, \& {Krzeminski}}]{Hamuy96:lc}
------. 1996{\natexlab{a}}, \aj, 112, 2408

\bibitem[{{Hamuy} {et~al.}(1996{\natexlab{b}}){Hamuy}, {Phillips}, {Suntzeff},
  {Schommer}, {Maza}, \& {Aviles}}]{Hamuy96:lum}
{Hamuy}, M., {Phillips}, M.~M., {Suntzeff}, N.~B., {Schommer}, R.~A., {Maza},
  J., \& {Aviles}, R. 1996{\natexlab{b}}, \aj, 112, 2391

\bibitem[{{Harutyunyan} {et~al.}(2008){Harutyunyan}, {Benetti}, {Fiorenzano},
  \& {Stanishev}}]{Harutyunyan08}
{Harutyunyan}, A., {Benetti}, S., {Fiorenzano}, A., \& {Stanishev}, V. 2008,
  Central Bureau Electronic Telegrams, 1438, 1

\bibitem[{{Hicken} {et~al.}(2009{\natexlab{a}}){Hicken}, {Challis}, {Jha},
  {Kirshner}, {Matheson}, {Modjaz}, {Rest}, {Michael Wood-Vasey}, {Bakos},
  {Barton}, {Berlind}, {Bragg}, {Brice{\~n}o}, {Brown}, {Caldwell}, {Calkins},
  {Cho}, {Ciupik}, {Contreras}, {Dendy}, {Dosaj}, {Durham}, {Eriksen},
  {Esquerdo}, {Everett}, {Falco}, {Fernandez}, {Gaba}, {Garnavich}, {Graves},
  {Green}, {Groner}, {Hergenrother}, {Holman}, {Hradecky}, {Huchra},
  {Hutchison}, {Jerius}, {Jordan}, {Kilgard}, {Krauss}, {Luhman}, {Macri},
  {Marrone}, {McDowell}, {McIntosh}, {McNamara}, {Megeath}, {Mochejska},
  {Munoz}, {Muzerolle}, {Naranjo}, {Narayan}, {Pahre}, {Peters}, {Peterson},
  {Rines}, {Ripman}, {Roussanova}, {Schild}, {Sicilia-Aguilar}, {Sokoloski},
  {Smalley}, {Smith}, {Spahr}, {Stanek}, {Barmby}, {Blondin}, {Stubbs},
  {Szentgyorgyi}, {Torres}, {Vaz}, {Vikhlinin}, {Wang}, {Westover}, {Woods}, \&
  {Zhao}}]{Hicken09:lc}
{Hicken}, M., {et~al.} 2009{\natexlab{a}}, \apj, 700, 331

\bibitem[{{Hicken} {et~al.}(2009{\natexlab{b}}){Hicken}, {Wood-Vasey},
  {Blondin}, {Challis}, {Jha}, {Kelly}, {Rest}, \& {Kirshner}}]{Hicken09:de}
{Hicken}, M., {Wood-Vasey}, W.~M., {Blondin}, S., {Challis}, P., {Jha}, S.,
  {Kelly}, P.~L., {Rest}, A., \& {Kirshner}, R.~P. 2009{\natexlab{b}}, \apj,
  700, 1097

\bibitem[{{H\"{o}flich} {et~al.}(1996){H\"{o}flich}, {Khokhlov}, {Wheeler},
  {Phillips}, {Suntzeff}, \& {Hamuy}}]{Hoflich96}
{H\"{o}flich}, P., {Khokhlov}, A., {Wheeler}, J.~C., {Phillips}, M.~M.,
  {Suntzeff}, N.~B., \& {Hamuy}, M. 1996, \apjl, 472, L81+

\bibitem[{{Hook} {et~al.}(2005){Hook}, {Howell}, {Aldering}, {Amanullah},
  {Burns}, {Conley}, {Deustua}, {Ellis}, {Fabbro}, {Fadeyev}, {Folatelli},
  {Garavini}, {Gibbons}, {Goldhaber}, {Goobar}, {Groom}, {Kim}, {Knop},
  {Kowalski}, {Lidman}, {Nobili}, {Nugent}, {Pain}, {Pennypacker},
  {Perlmutter}, {Ruiz-Lapuente}, {Sainton}, {Schaefer}, {Smith}, {Spadafora},
  {Stanishev}, {Thomas}, {Walton}, {Wang}, \& {Wood-Vasey}}]{Hook05}
{Hook}, I.~M., {et~al.} 2005, \aj, 130, 2788

\bibitem[{{Howell}(2001)}]{Howell01}
{Howell}, D.~A. 2001, \apjl, 554, L193

\bibitem[{{Jeffery} {et~al.}(1992){Jeffery}, {Leibundgut}, {Kirshner},
  {Benetti}, {Branch}, \& {Sonneborn}}]{Jeffery92}
{Jeffery}, D.~J., {Leibundgut}, B., {Kirshner}, R.~P., {Benetti}, S., {Branch},
  D., \& {Sonneborn}, G. 1992, \apj, 397, 304

\bibitem[{{Jha} {et~al.}(1999{\natexlab{a}}){Jha}, {Garnavich}, {Challis},
  {Kirshner}, {Calkins}, {Filippenko}, \& {Stern}}]{Jha99}
{Jha}, S., {Garnavich}, P., {Challis}, P., {Kirshner}, R., {Calkins}, M.,
  {Filippenko}, A.~V., \& {Stern}, D. 1999{\natexlab{a}}, \iaucirc, 7206, 1

\bibitem[{{Jha} {et~al.}(1999{\natexlab{b}}){Jha}, {Garnavich}, {Kirshner},
  {Challis}, {Soderberg}, {Macri}, {Huchra}, {Barmby}, {Barton}, {Berlind},
  {Brown}, {Caldwell}, {Calkins}, {Kannappan}, {Koranyi}, {Pahre}, {Rines},
  {Stanek}, {Stefanik}, {Szentgyorgyi}, {V{\"a}is{\"a}nen}, {Wang}, {Zajac},
  {Riess}, {Filippenko}, {Li}, {Modjaz}, {Treffers}, {Hergenrother}, {Grebel},
  {Seitzer}, {Jacoby}, {Benson}, {Rizvi}, {Marschall}, {Goldader}, {Beasley},
  {Vacca}, {Leibundgut}, {Spyromilio}, {Schmidt}, \& {Wood}}]{Jha99:98bu}
{Jha}, S., {et~al.} 1999{\natexlab{b}}, \apjs, 125, 73

\bibitem[{{Jha} {et~al.}(2006){Jha}, {Kirshner}, {Challis}, {Garnavich},
  {Matheson}, {Soderberg}, {Graves}, {Hicken}, {Alves}, {Arce}, {Balog},
  {Barmby}, {Barton}, {Berlind}, {Bragg}, {Brice{\~n}o}, {Brown}, {Buckley},
  {Caldwell}, {Calkins}, {Carter}, {Concannon}, {Donnelly}, {Eriksen},
  {Fabricant}, {Falco}, {Fiore}, {Garcia}, {G{\'o}mez}, {Grogin}, {Groner},
  {Groot}, {Haisch}, {Hartmann}, {Hergenrother}, {Holman}, {Huchra},
  {Jayawardhana}, {Jerius}, {Kannappan}, {Kim}, {Kleyna}, {Kochanek},
  {Koranyi}, {Krockenberger}, {Lada}, {Luhman}, {Luu}, {Macri}, {Mader},
  {Mahdavi}, {Marengo}, {Marsden}, {McLeod}, {McNamara}, {Megeath}, {Moraru},
  {Mossman}, {Muench}, {Mu{\~n}oz}, {Muzerolle}, {Naranjo}, {Nelson-Patel},
  {Pahre}, {Patten}, {Peters}, {Peters}, {Raymond}, {Rines}, {Schild},
  {Sobczak}, {Spahr}, {Stauffer}, {Stefanik}, {Szentgyorgyi}, {Tollestrup},
  {V{\"a}is{\"a}nen}, {Vikhlinin}, {Wang}, {Willner}, {Wolk}, {Zajac}, {Zhao},
  \& {Stanek}}]{Jha06:lc}
------. 2006, \aj, 131, 527

\bibitem[{{Kasen} \& {Plewa}(2007)}]{Kasen07:asym}
{Kasen}, D., \& {Plewa}, T. 2007, \apj, 662, 459

\bibitem[{{Kasen} \& {Woosley}(2007)}]{Kasen07:wlr}
{Kasen}, D., \& {Woosley}, S.~E. 2007, \apj, 656, 661

\bibitem[{{Kasliwal} {et~al.}(2008){Kasliwal}, {Ofek}, {Gal-Yam}, {Rau},
  {Brown}, {Cenko}, {Cameron}, {Quimby}, {Kulkarni}, {Bildsten}, {Milne}, \&
  {Bryngelson}}]{Kasliwal08}
{Kasliwal}, M.~M., {et~al.} 2008, \apjl, 683, L29

\bibitem[{{Kelly}(2007)}]{Kelly07}
{Kelly}, B.~C. 2007, \apj, 665, 1489

\bibitem[{{Kessler} {et~al.}(2009){Kessler}, {Becker}, {Cinabro}, {Vanderplas},
  {Frieman}, {Marriner}, {Davis}, {Dilday}, {Holtzman}, {Jha}, {Lampeitl},
  {Sako}, {Smith}, {Zheng}, {Nichol}, {Bassett}, {Bender}, {Depoy}, {Doi},
  {Elson}, {Filippenko}, {Foley}, {Garnavich}, {Hopp}, {Ihara}, {Ketzeback},
  {Kollatschny}, {Konishi}, {Marshall}, {McMillan}, {Miknaitis}, {Morokuma},
  {M{\"o}rtsell}, {Pan}, {Prieto}, {Richmond}, {Riess}, {Romani}, {Schneider},
  {Sollerman}, {Takanashi}, {Tokita}, {van der Heyden}, {Wheeler}, {Yasuda}, \&
  {York}}]{Kessler09}
{Kessler}, R., {et~al.} 2009, \apjs, 185, 32

\bibitem[{{Kirshner} {et~al.}(1993){Kirshner}, {Jeffery}, {Leibundgut},
  {Challis}, {Sonneborn}, {Phillips}, {Suntzeff}, {Smith}, {Winkler}, {Winge},
  {Hamuy}, {Hunter}, {Roth}, {Blades}, {Branch}, {Chevalier}, {Fransson},
  {Panagia}, {Wagoner}, {Wheeler}, \& {Harkness}}]{Kirshner93}
{Kirshner}, R.~P., {et~al.} 1993, \apj, 415, 589

\bibitem[{{Konishi} {et~al.}(2011){Konishi}, {Frieman}, {Goobar}, {Marriner},
  {Nordin}, {{\"O}stman}, {Sako}, {Schneider}, \& {Yasuda}}]{Konishi11}
{Konishi}, K., {et~al.} 2011, ArXiv e-prints, 1103.2497

\bibitem[{{Kotak} {et~al.}(2005){Kotak}, {Meikle}, {Pignata}, {Stehle},
  {Smartt}, {Benetti}, {Hillebrandt}, {Lennon}, {Mazzali}, {Patat}, \&
  {Turatto}}]{Kotak05}
{Kotak}, R., {et~al.} 2005, \aap, 436, 1021

\bibitem[{{Krisciunas} {et~al.}(2007){Krisciunas}, {Garnavich}, {Stanishev},
  {Suntzeff}, {Prieto}, {Espinoza}, {Gonzalez}, {Salvo}, {Elias de la Rosa},
  {Smartt}, {Maund}, \& {Kudritzki}}]{Krisciunas07}
{Krisciunas}, K., {et~al.} 2007, \aj, 133, 58

\bibitem[{{Krisciunas} {et~al.}(2011){Krisciunas}, {Li}, {Matheson}, {Howell},
  {Stritzinger}, {Aldering}, {Berlind}, {Calkins}, {Challis}, {Chornock},
  {Conley}, {Filippenko}, {Ganeshalingam}, {Germany}, {Gonz{\'a}lez},
  {Gooding}, {Hsiao}, {Kasen}, {Kirshner}, {Howie Marion}, {Muena}, {Nugent},
  {Phelps}, {Phillips}, {Qiu}, {Quimby}, {Rines}, {Silverman}, {Suntzeff},
  {Thomas}, \& {Wang}}]{Krisciunas11}
------. 2011, \aj, 142, 74

\bibitem[{{Leibundgut} {et~al.}(1991){Leibundgut}, {Kirshner}, {Filippenko},
  {Shields}, {Foltz}, {Phillips}, \& {Sonneborn}}]{Leibundgut91}
{Leibundgut}, B., {Kirshner}, R.~P., {Filippenko}, A.~V., {Shields}, J.~C.,
  {Foltz}, C.~B., {Phillips}, M.~M., \& {Sonneborn}, G. 1991, \apjl, 371, L23

\bibitem[{{Leibundgut} {et~al.}(1993){Leibundgut}, {Kirshner}, {Phillips},
  {Wells}, {Suntzeff}, {Hamuy}, {Schommer}, {Walker}, {Gonzalez}, {Ugarte},
  {Williams}, {Williger}, {Gomez}, {Marzke}, {Schmidt}, {Whitney}, {Coldwell},
  {Peters}, {Chaffee}, {Foltz}, {Rehner}, {Siciliano}, {Barnes}, {Cheng},
  {Hintzen}, {Kim}, {Maza}, {Parker}, {Porter}, {Schmidtke}, \&
  {Sonneborn}}]{Leibundgut93}
{Leibundgut}, B., {et~al.} 1993, \aj, 105, 301

\bibitem[{{Leloudas} {et~al.}(2009){Leloudas}, {Stritzinger}, {Sollerman},
  {Burns}, {Kozma}, {Krisciunas}, {Maund}, {Milne}, {Filippenko}, {Fransson},
  {Ganeshalingam}, {Hamuy}, {Li}, {Phillips}, {Schmidt}, {Skottfelt},
  {Taubenberger}, {Boldt}, {Fynbo}, {Gonzalez}, {Salvo}, \&
  {Thomas-Osip}}]{Leloudas09}
{Leloudas}, G., {et~al.} 2009, \aap, 505, 265

\bibitem[{{Leonard}(2005)}]{Leonard05:05ms}
{Leonard}, D.~C. 2005, Central Bureau Electronic Telegrams, 345, 1

\bibitem[{{Leonard} {et~al.}(2005){Leonard}, {Li}, {Filippenko}, {Foley}, \&
  {Chornock}}]{Leonard05}
{Leonard}, D.~C., {Li}, W., {Filippenko}, A.~V., {Foley}, R.~J., \& {Chornock},
  R. 2005, \apj, 632, 450

\bibitem[{{Li} {et~al.}(1999){Li}, {Qiu}, {Qiao}, {Zhu}, {Hu}, {Richmond},
  {Filippenko}, {Treffers}, {Peng}, \& {Leonard}}]{Li99}
{Li}, W.~D., {et~al.} 1999, \aj, 117, 2709

\bibitem[{{Maeda} {et~al.}(2010){Maeda}, {Benetti}, {Stritzinger}, {R{\"o}pke},
  {Folatelli}, {Sollerman}, {Taubenberger}, {Nomoto}, {Leloudas}, {Hamuy},
  {Tanaka}, {Mazzali}, \& {Elias-Rosa}}]{Maeda10:asym}
{Maeda}, K., {et~al.} 2010, \nat, 466, 82

\bibitem[{{Maeda} {et~al.}(2011){Maeda}, {Leloudas}, {Taubenberger},
  {Stritzinger}, {Sollerman}, {Elias-Rosa}, {Benetti}, {Hamuy}, {Folatelli}, \&
  {Mazzali}}]{Maeda11}
------. 2011, \mnras, 413, 3075

\bibitem[{{Mandel} {et~al.}(2011){Mandel}, {Narayan}, \& {Kirshner}}]{Mandel11}
{Mandel}, K.~S., {Narayan}, G., \& {Kirshner}, R.~P. 2011, \apj, 731, 120

\bibitem[{{Mandel} {et~al.}(2009){Mandel}, {Wood-Vasey}, {Friedman}, \&
  {Kirshner}}]{Mandel09}
{Mandel}, K.~S., {Wood-Vasey}, W.~M., {Friedman}, A.~S., \& {Kirshner}, R.~P.
  2009, \apj, 704, 629

\bibitem[{{Matheson} {et~al.}(2008){Matheson}, {Kirshner}, {Challis}, {Jha},
  {Garnavich}, {Berlind}, {Calkins}, {Blondin}, {Balog}, {Bragg}, {Caldwell},
  {Dendy Concannon}, {Falco}, {Graves}, {Huchra}, {Kuraszkiewicz}, {Mader},
  {Mahdavi}, {Phelps}, {Rines}, {Song}, \& {Wilkes}}]{Matheson08}
{Matheson}, T., {et~al.} 2008, \aj, 135, 1598

\bibitem[{{Maund} {et~al.}(2010){Maund}, {H{\"o}flich}, {Patat}, {Wheeler},
  {Zelaya}, {Baade}, {Wang}, {Clocchiatti}, \& {Quinn}}]{Maund10:asym}
{Maund}, J.~R., {et~al.} 2010, \apjl, 725, L167

\bibitem[{{Maza} {et~al.}(1992){Maza}, {Hamuy}, {Wischnjewsky}, {Wells},
  {Wakamatsu}, {Malkan}, \& {Aviles}}]{Maza92}
{Maza}, J., {Hamuy}, M., {Wischnjewsky}, M., {Wells}, L., {Wakamatsu}, K.,
  {Malkan}, M., \& {Aviles}, R. 1992, \iaucirc, 5555, 1

\bibitem[{{Mazzali} {et~al.}(1995){Mazzali}, {Danziger}, \&
  {Turatto}}]{Mazzali95}
{Mazzali}, P.~A., {Danziger}, I.~J., \& {Turatto}, M. 1995, \aap, 297, 509

\bibitem[{{Mazzali} {et~al.}(1993){Mazzali}, {Lucy}, {Danziger}, {Gouiffes},
  {Cappellaro}, \& {Turatto}}]{Mazzali93}
{Mazzali}, P.~A., {Lucy}, L.~B., {Danziger}, I.~J., {Gouiffes}, C.,
  {Cappellaro}, E., \& {Turatto}, M. 1993, \aap, 269, 423

\bibitem[{{McNaught} {et~al.}(1992){McNaught}, {Parker}, {della Valle},
  {Lorenz}, {Phillips}, \& {Alonso}}]{McNaught92}
{McNaught}, R.~H., {Parker}, Q.~A., {della Valle}, M., {Lorenz}, H.,
  {Phillips}, M.~M., \& {Alonso}, A. 1992, \iaucirc, 5569, 1

\bibitem[{{Nordin} {et~al.}(2011{\natexlab{a}}){Nordin}, {{\"O}stman},
  {Goobar}, {Amanullah}, {Nichol}, {Smith}, {Sollerman}, {Bassett}, {Frieman},
  {Garnavich}, {Leloudas}, {Sako}, \& {Schneider}}]{Nordin11:sdss}
{Nordin}, J., {et~al.} 2011{\natexlab{a}}, \aap, 526, A119+

\bibitem[{{Nordin} {et~al.}(2011{\natexlab{b}}){Nordin}, {{\"O}stman},
  {Goobar}, {Balland}, {Lampeitl}, {Nichol}, {Sako}, {Schneider}, {Smith},
  {Sollerman}, \& {Wheeler}}]{Nordin11:Si4000}
------. 2011{\natexlab{b}}, \apj, 734, 42

\bibitem[{{Nugent} {et~al.}(1995){Nugent}, {Phillips}, {Baron}, {Branch}, \&
  {Hauschildt}}]{Nugent95}
{Nugent}, P., {Phillips}, M., {Baron}, E., {Branch}, D., \& {Hauschildt}, P.
  1995, \apjl, 455, L147

\bibitem[{{Nugent} \& {Wang}(2001)}]{Nugent01}
{Nugent}, P., \& {Wang}, L. 2001, \iaucirc, 7614, 3

\bibitem[{{Pastorello} {et~al.}(2007){Pastorello}, {Mazzali}, {Pignata},
  {Benetti}, {Cappellaro}, {Filippenko}, {Li}, {Meikle}, {Arkharov}, {Blanc},
  {Bufano}, {Derekas}, {Dolci}, {Elias-Rosa}, {Foley}, {Ganeshalingam},
  {Harutyunyan}, {Kiss}, {Kotak}, {Larionov}, {Lucey}, {Napoleone},
  {Navasardyan}, {Patat}, {Rich}, {Ryder}, {Salvo}, {Schmidt}, {Stanishev},
  {Sz{\'e}kely}, {Taubenberger}, {Temporin}, {Turatto}, \&
  {Hillebrandt}}]{Pastorello07:04eo}
{Pastorello}, A., {et~al.} 2007, \mnras, 377, 1531

\bibitem[{{Patat} {et~al.}(1996){Patat}, {Benetti}, {Cappellaro}, {Danziger},
  {della Valle}, {Mazzali}, \& {Turatto}}]{Patat96}
{Patat}, F., {Benetti}, S., {Cappellaro}, E., {Danziger}, I.~J., {della Valle},
  M., {Mazzali}, P.~A., \& {Turatto}, M. 1996, \mnras, 278, 111

\bibitem[{{Patat} {et~al.}(2001){Patat}, {Contreras}, {Prieto}, {Altavilla},
  {Benetti}, {Cappellaro}, {Pastorello}, \& {Turatto}}]{Patat01}
{Patat}, F., {Contreras}, C., {Prieto}, J., {Altavilla}, G., {Benetti}, S.,
  {Cappellaro}, E., {Pastorello}, A., \& {Turatto}, M. 2001, \iaucirc, 7680, 1

\bibitem[{{Peek} \& {Graves}(2010)}]{Peek10}
{Peek}, J.~E.~G., \& {Graves}, G.~J. 2010, \apj, 719, 415

\bibitem[{{Perlmutter} {et~al.}(1999){Perlmutter}, {Aldering}, {Goldhaber},
  {Knop}, {Nugent}, {Castro}, {Deustua}, {Fabbro}, {Goobar}, {Groom}, {Hook},
  {Kim}, {Kim}, {Lee}, {Nunes}, {Pain}, {Pennypacker}, {Quimby}, {Lidman},
  {Ellis}, {Irwin}, {McMahon}, {Ruiz-Lapuente}, {Walton}, {Schaefer}, {Boyle},
  {Filippenko}, {Matheson}, {Fruchter}, {Panagia}, {Newberg}, \&
  {Couch}}]{Perlmutter99}
{Perlmutter}, S., {et~al.} 1999, \apj, 517, 565

\bibitem[{{Phillips}(1993)}]{Phillips93}
{Phillips}, M.~M. 1993, \apjl, 413, L105

\bibitem[{{Phillips} {et~al.}(1999){Phillips}, {Lira}, {Suntzeff}, {Schommer},
  {Hamuy}, \& {Maza}}]{Phillips99}
{Phillips}, M.~M., {Lira}, P., {Suntzeff}, N.~B., {Schommer}, R.~A., {Hamuy},
  M., \& {Maza}, J. 1999, \aj, 118, 1766

\bibitem[{{Pignata} {et~al.}(2008){Pignata}, {Benetti}, {Mazzali}, {Kotak},
  {Patat}, {Meikle}, {Stehle}, {Leibundgut}, {Suntzeff}, {Buson}, {Cappellaro},
  {Clocchiatti}, {Hamuy}, {Maza}, {Mendez}, {Ruiz-Lapuente}, {Salvo},
  {Schmidt}, {Turatto}, \& {Hillebrandt}}]{Pignata08:02dj}
{Pignata}, G., {et~al.} 2008, \mnras, 388, 971

\bibitem[{{Pinto} \& {Eastman}(2001)}]{Pinto01}
{Pinto}, P.~A., \& {Eastman}, R.~G. 2001, \na, 6, 307

\bibitem[{{Quimby} {et~al.}(2006{\natexlab{a}}){Quimby}, {Castro}, {Edelmann},
  \& {Riley}}]{Quimby06:06os}
{Quimby}, R., {Castro}, F., {Edelmann}, H., \& {Riley}, V. 2006{\natexlab{a}},
  Central Bureau Electronic Telegrams, 751, 1

\bibitem[{{Quimby} {et~al.}(2006{\natexlab{b}}){Quimby}, {H{\"o}flich},
  {Kannappan}, {Rykoff}, {Rujopakarn}, {Akerlof}, {Gerardy}, \&
  {Wheeler}}]{Quimby06:05cg}
{Quimby}, R., {H{\"o}flich}, P., {Kannappan}, S.~J., {Rykoff}, E.,
  {Rujopakarn}, W., {Akerlof}, C.~W., {Gerardy}, C.~L., \& {Wheeler}, J.~C.
  2006{\natexlab{b}}, \apj, 636, 400

\bibitem[{{Rau} {et~al.}(2009){Rau}, {Kulkarni}, {Law}, {Bloom}, {Ciardi},
  {Djorgovski}, {Fox}, {Gal-Yam}, {Grillmair}, {Kasliwal}, {Nugent}, {Ofek},
  {Quimby}, {Reach}, {Shara}, {Bildsten}, {Cenko}, {Drake}, {Filippenko},
  {Helfand}, {Helou}, {Howell}, {Poznanski}, \& {Sullivan}}]{Rau09}
{Rau}, A., {et~al.} 2009, \pasp, 121, 1334

\bibitem[{{Reindl} {et~al.}(2005){Reindl}, {Tammann}, {Sandage}, \&
  {Saha}}]{Reindl05}
{Reindl}, B., {Tammann}, G.~A., {Sandage}, A., \& {Saha}, A. 2005, \apj, 624,
  532

\bibitem[{{Riess} {et~al.}(1998){Riess}, {Filippenko}, {Challis},
  {Clocchiatti}, {Diercks}, {Garnavich}, {Gilliland}, {Hogan}, {Jha},
  {Kirshner}, {Leibundgut}, {Phillips}, {Reiss}, {Schmidt}, {Schommer},
  {Smith}, {Spyromilio}, {Stubbs}, {Suntzeff}, \& {Tonry}}]{Riess98:Lambda}
{Riess}, A.~G., {et~al.} 1998, \aj, 116, 1009

\bibitem[{{Riess} {et~al.}(1999){Riess}, {Kirshner}, {Schmidt}, {Jha},
  {Challis}, {Garnavich}, {Esin}, {Carpenter}, {Grashius}, {Schild}, {Berlind},
  {Huchra}, {Prosser}, {Falco}, {Benson}, {Brice{\~n}o}, {Brown}, {Caldwell},
  {dell'Antonio}, {Filippenko}, {Goodman}, {Grogin}, {Groner}, {Hughes},
  {Green}, {Jansen}, {Kleyna}, {Luu}, {Macri}, {McLeod}, {McLeod}, {McNamara},
  {McLean}, {Milone}, {Mohr}, {Moraru}, {Peng}, {Peters}, {Prestwich},
  {Stanek}, {Szentgyorgyi}, \& {Zhao}}]{Riess99:lc}
------. 1999, \aj, 117, 707

\bibitem[{{Riess} {et~al.}(2011){Riess}, {Macri}, {Casertano}, {Lampeitl},
  {Ferguson}, {Filippenko}, {Jha}, {Li}, {Chornock}, \& {Silverman}}]{Riess11}
------. 2011, \apj, 730, 119

\bibitem[{{Riess} {et~al.}(1996){Riess}, {Press}, \& {Kirshner}}]{Riess96}
{Riess}, A.~G., {Press}, W.~H., \& {Kirshner}, R.~P. 1996, \apj, 473, 88

\bibitem[{{Riess} {et~al.}(2007){Riess}, {Strolger}, {Casertano}, {Ferguson},
  {Mobasher}, {Gold}, {Challis}, {Filippenko}, {Jha}, {Li}, {Tonry}, {Foley},
  {Kirshner}, {Dickinson}, {MacDonald}, {Eisenstein}, {Livio}, {Younger}, {Xu},
  {Dahl{\'e}n}, \& {Stern}}]{Riess07}
{Riess}, A.~G., {et~al.} 2007, \apj, 659, 98

\bibitem[{{Salvo} {et~al.}(1999){Salvo}, {Benetti}, {Kjaergaard}, \&
  {Greve}}]{Salvo99}
{Salvo}, M.~E., {Benetti}, S., {Kjaergaard}, P., \& {Greve}, T.~R. 1999,
  \iaucirc, 7238, 1

\bibitem[{{Salvo} {et~al.}(2001){Salvo}, {Cappellaro}, {Mazzali}, {Benetti},
  {Danziger}, {Patat}, \& {Turatto}}]{Salvo01}
{Salvo}, M.~E., {Cappellaro}, E., {Mazzali}, P.~A., {Benetti}, S., {Danziger},
  I.~J., {Patat}, F., \& {Turatto}, M. 2001, \mnras, 321, 254

\bibitem[{{Sauer} {et~al.}(2008){Sauer}, {Mazzali}, {Blondin}, {Stehle},
  {Benetti}, {Challis}, {Filippenko}, {Kirshner}, {Li}, \&
  {Matheson}}]{Sauer08}
{Sauer}, D.~N., {et~al.} 2008, \mnras, 391, 1605

\bibitem[{{Schlegel} {et~al.}(1998){Schlegel}, {Finkbeiner}, \&
  {Davis}}]{Schlegel98}
{Schlegel}, D.~J., {Finkbeiner}, D.~P., \& {Davis}, M. 1998, \apj, 500, 525

\bibitem[{{Selj} {et~al.}(2006){Selj}, {Sharapov}, {Somero}, {Oestman},
  {Sollerman}, \& {Blondin}}]{Selj06}
{Selj}, J., {Sharapov}, D., {Somero}, A., {Oestman}, L., {Sollerman}, J., \&
  {Blondin}, S. 2006, Central Bureau Electronic Telegrams, 570, 1

\bibitem[{{Silverman} {et~al.}(2007){Silverman}, {Foley}, \&
  {Filippenko}}]{Silverman07}
{Silverman}, J.~M., {Foley}, R.~J., \& {Filippenko}, A.~V. 2007, Central Bureau
  Electronic Telegrams, 818, 3

\bibitem[{{Silverman} {et~al.}(2006){Silverman}, {Wong}, {Filippenko}, \&
  {Chornock}}]{Silverman06}
{Silverman}, J.~M., {Wong}, D., {Filippenko}, A.~V., \& {Chornock}, R. 2006,
  Central Bureau Electronic Telegrams, 765, 1

\bibitem[{{Stanishev} {et~al.}(2007){Stanishev}, {Goobar}, {Benetti}, {Kotak},
  {Pignata}, {Navasardyan}, {Mazzali}, {Amanullah}, {Garavini}, {Nobili},
  {Qiu}, {Elias-Rosa}, {Ruiz-Lapuente}, {Mendez}, {Meikle}, {Patat},
  {Pastorello}, {Altavilla}, {Gustafsson}, {Harutyunyan}, {Iijima},
  {Jakobsson}, {Kichizhieva}, {Lundqvist}, {Mattila}, {Melinder}, {Pavlenko},
  {Pavlyuk}, {Sollerman}, {Tsvetkov}, {Turatto}, \&
  {Hillebrandt}}]{Stanishev07:03du}
{Stanishev}, V., {et~al.} 2007, \aap, 469, 645

\bibitem[{{Stritzinger} {et~al.}(2002){Stritzinger}, {Hamuy}, {Suntzeff},
  {Smith}, {Phillips}, {Maza}, {Strolger}, {Antezana}, {Gonz{\'a}lez},
  {Wischnjewsky}, {Candia}, {Espinoza}, {Gonz{\'a}lez}, {Stubbs}, {Becker},
  {Rubenstein}, \& {Galaz}}]{Stritzinger02}
{Stritzinger}, M., {et~al.} 2002, \aj, 124, 2100

\bibitem[{{Sullivan} {et~al.}(2011){Sullivan}, {Guy}, {Conley}, {Regnault},
  {Astier}, {Balland}, {Basa}, {Carlberg}, {Fouchez}, {Hardin}, {Hook},
  {Howell}, {Pain}, {Palanque-Delabrouille}, {Perrett}, {Pritchet}, {Rich},
  {Ruhlmann-Kleider}, {Balam}, {Baumont}, {Ellis}, {Fabbro}, {Fakhouri},
  {Fourmanoit}, {Gonz{\'a}lez-Gait{\'a}n}, {Graham}, {Hudson}, {Hsiao},
  {Kronborg}, {Lidman}, {Mourao}, {Neill}, {Perlmutter}, {Ripoche}, {Suzuki},
  \& {Walker}}]{Sullivan11}
{Sullivan}, M., {et~al.} 2011, \apj, 737, 102

\bibitem[{{Suntzeff} \& {Smith}(2000)}]{Suntzeff00}
{Suntzeff}, N.~B., \& {Smith}, R.~C. 2000, \iaucirc, 7506, 2

\bibitem[{{Taubenberger} {et~al.}(2008){Taubenberger}, {Hachinger}, {Pignata},
  {Mazzali}, {Contreras}, {Valenti}, {Pastorello}, {Elias-Rosa},
  {B{\"a}rnbantner}, {Barwig}, {Benetti}, {Dolci}, {Fliri}, {Folatelli},
  {Freedman}, {Gonzalez}, {Hamuy}, {Krzeminski}, {Morrell}, {Navasardyan},
  {Persson}, {Phillips}, {Ries}, {Roth}, {Suntzeff}, {Turatto}, \&
  {Hillebrandt}}]{Taubenberger08}
{Taubenberger}, S., {et~al.} 2008, \mnras, 385, 75

\bibitem[{{Tripp}(1998)}]{Tripp98}
{Tripp}, R. 1998, \aap, 331, 815

\bibitem[{{Turatto} {et~al.}(1996){Turatto}, {Benetti}, {Cappellaro},
  {Danziger}, {Della Valle}, {Gouiffes}, {Mazzali}, \& {Patat}}]{Turatto96}
{Turatto}, M., {Benetti}, S., {Cappellaro}, E., {Danziger}, I.~J., {Della
  Valle}, M., {Gouiffes}, C., {Mazzali}, P.~A., \& {Patat}, F. 1996, \mnras,
  283, 1

\bibitem[{{Turatto} {et~al.}(1998){Turatto}, {Piemonte}, {Benetti},
  {Cappellaro}, {Mazzali}, {Danziger}, \& {Patat}}]{Turatto98}
{Turatto}, M., {Piemonte}, A., {Benetti}, S., {Cappellaro}, E., {Mazzali},
  P.~A., {Danziger}, I.~J., \& {Patat}, F. 1998, \aj, 116, 2431

\bibitem[{{Umbriaco} {et~al.}(2007){Umbriaco}, {Pietrogrande}, {di Mille},
  {Agnoletto}, {Harutyunyan}, \& {Benetti}}]{Umbriaco07}
{Umbriaco}, G., {Pietrogrande}, T., {di Mille}, F., {Agnoletto}, I.,
  {Harutyunyan}, A., \& {Benetti}, S. 2007, Central Bureau Electronic
  Telegrams, 1174, 1

\bibitem[{{Valentini} {et~al.}(2003){Valentini}, {Di Carlo}, {Massi}, {Dolci},
  {Arkharov}, {Larionov}, {Pastorello}, {Di Paola}, {Benetti}, {Cappellaro},
  {Turatto}, {Pedichini}, {D'Alessio}, {Caratti o Garatti}, {Li Causi},
  {Speziali}, {Danziger}, \& {Tornamb{\'e}}}]{Valentini03}
{Valentini}, G., {et~al.} 2003, \apj, 595, 779

\bibitem[{{Walker} {et~al.}(2011){Walker}, {Hook}, {Sullivan}, {Howell},
  {Astier}, {Balland}, {Basa}, {Bronder}, {Carlberg}, {Conley}, {Fouchez},
  {Guy}, {Hardin}, {Pain}, {Perrett}, {Pritchet}, {Regnault}, {Rich},
  {Aldering}, {Fakhouri}, {Kronborg}, {Palanque-Delabrouille}, {Perlmutter},
  {Ruhlmann-Kleider}, \& {Zhang}}]{Walker11}
{Walker}, E.~S., {et~al.} 2011, \mnras, 410, 1262

\bibitem[{{Wang}(2001)}]{Wang01}
{Wang}, L. 2001, \iaucirc, 7640, 2

\bibitem[{{Wang} {et~al.}(2003){Wang}, {Baade}, {H{\"o}flich}, {Khokhlov},
  {Wheeler}, {Kasen}, {Nugent}, {Perlmutter}, {Fransson}, \&
  {Lundqvist}}]{Wang03}
{Wang}, L., {et~al.} 2003, \apj, 591, 1110

\bibitem[{{Wang} {et~al.}(2006){Wang}, {Baade}, {H{\"o}flich}, {Wheeler},
  {Kawabata}, {Khokhlov}, {Nomoto}, \& {Patat}}]{Wang06}
{Wang}, L., {Baade}, D., {H{\"o}flich}, P., {Wheeler}, J.~C., {Kawabata}, K.,
  {Khokhlov}, A., {Nomoto}, K., \& {Patat}, F. 2006, \apj, 653, 490

\bibitem[{{Wang} \& {Wheeler}(2008)}]{Wang08:specpol}
{Wang}, L., \& {Wheeler}, J.~C. 2008, \araa, 46, 433

\bibitem[{{Wang} {et~al.}(2009{\natexlab{a}}){Wang}, {Filippenko},
  {Ganeshalingam}, {Li}, {Silverman}, {Wang}, {Chornock}, {Foley}, {Gates},
  {Macomber}, {Serduke}, {Steele}, \& {Wong}}]{Wang09:2pop}
{Wang}, X., {et~al.} 2009{\natexlab{a}}, \apjl, 699, L139

\bibitem[{{Wang} {et~al.}(2009{\natexlab{b}}){Wang}, {Li}, {Filippenko},
  {Foley}, {Kirshner}, {Modjaz}, {Bloom}, {Brown}, {Carter}, {Friedman},
  {Gal-Yam}, {Ganeshalingam}, {Hicken}, {Krisciunas}, {Milne}, {Silverman},
  {Suntzeff}, {Wood-Vasey}, {Cenko}, {Challis}, {Fox}, {Kirkman}, {Li}, {Li},
  {Malkan}, {Moore}, {Reitzel}, {Rich}, {Serduke}, {Shang}, {Steele}, {Swift},
  {Tao}, {Wong}, \& {Zhang}}]{Wang09:05cf}
------. 2009{\natexlab{b}}, \apj, 697, 380

\bibitem[{{Wells} {et~al.}(1994){Wells}, {Phillips}, {Suntzeff}, {Heathcote},
  {Hamuy}, {Navarrete}, {Fernandez}, {Weller}, {Schommer}, {Kirshner},
  {Leibundgut}, {Willner}, {Peletier}, {Schlegel}, {Wheeler}, {Harkness},
  {Bell}, {Matthews}, {Filippenko}, {Shields}, {Richmond}, {Jewitt}, {Luu},
  {Tran}, {Appleton}, {Robson}, {Tyson}, {Guhathakurta}, {Eder}, {Bond},
  {Potter}, {Veilleux}, {Porter}, {Humphreys}, {Janes}, {Williams}, {Costa},
  {Ruiz}, {Lee}, {Lutz}, {Rich}, {Winkler}, \& {Tyson}}]{Wells94}
{Wells}, L.~A., {et~al.} 1994, \aj, 108, 2233

\bibitem[{{Wood-Vasey} {et~al.}(2007){Wood-Vasey}, {Miknaitis}, {Stubbs},
  {Jha}, {Riess}, {Garnavich}, {Kirshner}, {Aguilera}, {Becker}, {Blackman},
  {Blondin}, {Challis}, {Clocchiatti}, {Conley}, {Covarrubias}, {Davis},
  {Filippenko}, {Foley}, {Garg}, {Hicken}, {Krisciunas}, {Leibundgut}, {Li},
  {Matheson}, {Miceli}, {Narayan}, {Pignata}, {Prieto}, {Rest}, {Salvo},
  {Schmidt}, {Smith}, {Sollerman}, {Spyromilio}, {Tonry}, {Suntzeff}, \&
  {Zenteno}}]{Wood-Vasey07}
{Wood-Vasey}, W.~M., {et~al.} 2007, \apj, 666, 694

\bibitem[{{Zheng} {et~al.}(2008){Zheng}, {Romani}, {Sako}, {Marriner},
  {Bassett}, {Becker}, {Choi}, {Cinabro}, {DeJongh}, {Depoy}, {Dilday}, {Doi},
  {Frieman}, {Garnavich}, {Hogan}, {Holtzman}, {Im}, {Jha}, {Kessler},
  {Konishi}, {Lampeitl}, {Marshall}, {McGinnis}, {Miknaitis}, {Nichol},
  {Prieto}, {Riess}, {Richmond}, {Schneider}, {Smith}, {Takanashi}, {Tokita},
  {van der Heyden}, {Yasuda}, {Assef}, {Barentine}, {Bender}, {Blandford},
  {Bremer}, {Brewington}, {Collins}, {Crotts}, {Dembicky}, {Eastman}, {Edge},
  {Elson}, {Eyler}, {Filippenko}, {Foley}, {Frank}, {Goobar}, {Harvanek},
  {Hopp}, {Ihara}, {Kahn}, {Ketzeback}, {Kleinman}, {Kollatschny},
  {Krzesi{\'n}ski}, {Leloudas}, {Long}, {Lucey}, {Malanushenko},
  {Malanushenko}, {McMillan}, {Morgan}, {Morokuma}, {Nitta}, {Ostman}, {Pan},
  {Romer}, {Saurage}, {Schlesinger}, {Snedden}, {Sollerman}, {Stritzinger},
  {Watson}, {Watters}, {Wheeler}, \& {York}}]{Zheng08}
{Zheng}, C., {et~al.} 2008, \aj, 135, 1766

\end{thebibliography}


\end{document}